\newcommand{\exclude}[1]{}
\RequirePackage{etoolbox}

\PassOptionsToPackage{inline}{enumitem}

\newtoggle{arXiv}
\toggletrue{arXiv}

\iftoggle{arXiv}{
}{%
  \PassOptionsToClass{times}{elsarticle}
}

\documentclass[final,3p]{elsarticle}

\iftoggle{arXiv}{
  \AtEndOfClass{
    \RequirePackage{newtxtext}
    \RequirePackage[varg]{newtxmath}
  }
}{%
}

\usepackage{graphicx}
\graphicspath{{figures/}{}}

\usepackage{mathtools}
\usepackage{amsthm,amsmath,amssymb,amsfonts}
\usepackage{bm}
\usepackage{color}

\usepackage{epigraph-keys}

\AtEndPreamble{
  \RequirePackage{hyperref}
  \RequirePackage[nameinlink]{cleveref}

    \crefname{table}{Table}{Tables}
    \Crefname{table}{Table}{Tables}
    \crefname{figure}{Fig.}{Figs.}
    \Crefname{figure}{Fig.}{Figs.}
    \crefname{equation}{Eq.}{Eqs.}
    \Crefname{equation}{Eq.}{Eqs.}
    \crefname{section}{Sec.}{Secs.}
    \Crefname{section}{Sec.}{Secs.}
}

\usepackage{algorithm}
\usepackage[noend]{algpseudocode}

\usepackage{booktabs}

\RequirePackage{listings}
\RequirePackage{xcolor} 

\lstdefinestyle{Input}{
  language=Python,
  basicstyle=\ttfamily\small,
  keywordstyle=\color{blue}\bfseries,
  commentstyle=\color{gray}\itshape,
  stringstyle=\color{orange!70!black},
  numbers=left,
  numberstyle=\tiny\color{gray},
  stepnumber=1,
  numbersep=5pt,
  tabsize=4,
  showstringspaces=false,
  breaklines=true,
  frame=single,
}

\lstdefinestyle{Python}{
  language=Python,
  basicstyle=\ttfamily\small,
  keywordstyle=\color{blue}\bfseries,
  commentstyle=\color{gray}\itshape,
  stringstyle=\color{orange!70!black},
  numbers=left,
  numberstyle=\tiny\color{gray},
  stepnumber=1,
  numbersep=5pt,
  tabsize=4,
  showstringspaces=false,
  breaklines=true,
  frame=single,
}

\lstdefinestyle{Commands}{
  language=bash,
  basicstyle=\ttfamily\small,
  keywordstyle=\color{blue}\bfseries,
  commentstyle=\color{gray}\itshape,
  stringstyle=\color{orange!70!black},
  numbers=none,
  numberstyle=\tiny\color{gray},
  stepnumber=1,
  numbersep=5pt,
  tabsize=4,
  showstringspaces=false,
  breaklines=true,
  frame=single,
}

\newcounter{bla}

\journal{Computer Physics Communications}

\RequirePackage{enumitem}           
\newcommand{\mylabelformat}[1]{\normalfont\small\itshape#1:}
\newcommand{\mydescriptionformat}[1]{\bfseries#1:}

\setlist[description]{format=\mydescriptionformat}
\setlist[enumerate]{noitemsep}

\SetEnumitemKey{PSdescription}{
  rightmargin=1.9in,
  format=\mylabelformat,
  parsep=0pt,
  listparindent=\parindent}

\newlist{PSdescriptionA}{description}{1}
\setlist[PSdescriptionA]{PSdescription}

\newlist{PSdescriptionB}{description}{1}
\setlist[PSdescriptionB]{PSdescription, style=nextline}

\providecommand{\mainmatter}{}
\providecommand{\preprint}[1]{\footnote{#1}}

\makeatletter
\iftoggle{arXiv}{%
  \let\@afterindenttrue\@afterindentfalse
  \@afterindentfalse
}{%
}
\makeatother

\newcommand{\filename}[1]{\nolinkurl{#1}}
\newcommand{\email}[1]{\href{mailto:#1}{#1}}

\newcommand{\Effg}{E_{\textrm{FG}}}

\newcommand{\kF}{k_{F}}
\newcommand{\eF}{\varepsilon_{F}}
\newcommand{\rhoP}{\rho_+}                        
\renewcommand{\rhoP}{\rho}
\newcommand{\abs}[1]{\lvert{#1}\rvert}            
\newcommand{\avg}[1]{\langle #1 \rangle}
\newcommand{\Udiss}{V^{\textrm{(d)}}}
\newcommand{\Ddiss}{\Delta^{\textrm{(d)}}}
\newcommand{\Nreq}{N_{\textrm{req.}}}
\newcommand{\Cinf}{C^{(\infty)}}

\newcommand{\ginf}{g}
\newcommand{\Cc}{C}
\newcommand{\gc}{g^{\mathrm{reg.}}}
\newcommand{\nuc}{\nu^{\mathrm{reg.}}}

\renewcommand{\Cinf}{C^{(\infty)}}

\renewcommand{\ginf}{g}
\renewcommand{\Cc}{C^{\mathrm{reg.}}}
\renewcommand{\gc}{g^{\mathrm{reg.}}}
\renewcommand{\nuc}{\nu^{\mathrm{reg.}}}

\providecommand\vb{}
\renewcommand\vb[1]{\ensuremath{\boldsymbol{#1}}}

\newcommand{\I}{\mathrm{i}}
\let\Im\relax

\DeclareMathOperator{\Im}{Im}

\AtBeginDocument{
  
  \providecommand{\mysc}[1]{#1}
}
\newcommand{\fundref}[2]{\href{https://dx.doi.org/10.13039/#2}{#1}}

\newcommand{\fundrefNSF}[1][\gls{NSF}]{\fundref{#1}{100000166}}
\newcommand{\fundrefNSFPHY}[1][\gls{NSF}]{\fundref{#1}{100000086}}
\newcommand{\fundrefDOE}[1][\gls{DOE}]{\fundref{#1}{100006132}}
\newcommand{\supportfromNSFgrant}[2][PHY]{%
  support from the
  \ifthenelse{\equal{#1}{PHY}}{\fundrefNSFPHY}{\fundrefNSF}\ through
  Grant No.\ \href{https://www.nsf.gov/awardsearch/showAward?AWD_ID=#2}{\mysc{#1-#2}}}

\makeatletter
\newcommand\DEFINEALPHABETLOOP[3]{%
  \ifx\relax#3\expandafter\@gobble
  \else\expandafter\@firstofone
  \fi
  {
  \expandafter\newcommand\expandafter*\csname#1#3\endcsname{#2{#3}}%
   \DEFINEALPHABETLOOP{#1}{#2}
  }
}
\newcommand\definealphabet[2]{%
  \DEFINEALPHABETLOOP{#1}{#2}abcdefghijklmnopqrstuvwxyzABCDEFGHIJKLMNOPQRSTUVWXYZ\relax
}%
\makeatother

\definealphabet{cal}{\mathcal}
\definealphabet{rm}{\mathrm}
\definealphabet{bb}{\mathbb}
\definealphabet{frak}{\mathfrak}
\definealphabet{scr}{\mathscr}
\definealphabet{vb}{\vb}

\newcommand{\ddcc}{\ensuremath{\lambda}}
\newcommand{\Vext}[1]{V_{#1}^{\textrm{(ext)}}}
\newcommand{\Dext}{\Delta^{\textrm{(ext)}}}
\newcommand{\velext}[1]{\mathbf{v}_{#1}^{\textrm{(ext)}}}
\renewcommand{\vbA}{\vb{\mathrm{A}}}
\providecommand\oprime{}\renewcommand\oprime{\ensuremath{^\prime}}

\renewcommand\textvisiblespace[1][1.0em]{          
  \makebox[#1]{%
    \kern.07em     \vrule height.3ex     \hrulefill     \vrule height.3ex     \kern.07em}}

\usepackage[normalem]{ulem}

\newcommand{\respch}[1]{#1}

\begin{document}

\begin{frontmatter}



  \title{W-SLDA Toolkit: A simulation platform for ultracold Fermi gases}
  

\author[a,b]{Gabriel Wlaz\l{}owski\corref{author}}
\author[a,b]{Piotr Magierski}
\author[c,b]{Michael McNeil Forbes}
\author[b]{Aurel Bulgac}

\cortext[author] {Corresponding author.\\
  \textit{E-mail address:} gabriel.wlazlowski@pw.edu.pl}
\address[a]{Faculty of Physics, Warsaw University of Technology, Ulica Koszykowa 75, 00-662 Warsaw, Poland}
\address[b]{Department of Physics, University of Washington, Seattle, Washington 98195--1560, USA}
\address[c]{Department of Physics and Astronomy, Washington State University, Pullman, WA 99164, USA}

\begin{abstract}
  We present the W-SLDA Toolkit, a general-purpose software package for simulating ultracold Fermi gases within the framework of density functional theory and its time-dependent extensions.
  The toolkit enables fully microscopic studies of interacting superfluid systems across the BCS-BEC crossover, including spin-imbalanced configurations and arbitrary external geometries.
  It provides both static and time-dependent solvers capable of describing a broad range of phenomena in one-, two-, and three-dimensional settings.
  In addition, the toolkit incorporates functionality for solving the standard Bogoliubov-de~Gennes equations for fermions, extending its applicability to other physical systems such as superconductors.
  The code is implemented in C with GPU acceleration and is optimized for hybrid CPU/GPU execution on modern high-performance computing platforms.
  It ensures scalability on leadership-class supercomputers, enabling fully three-dimensional simulations with large atom numbers, and allows for direct benchmarks of ultracold-atom experimental setups.
  Its modular architecture facilitates straightforward extensions, user customization, and seamless interoperability with other scientific software frameworks.
  Furthermore, an extensive collection of practical usage examples is provided through the integrated \textit{reproducibility packs} functionality, ensuring transparency and reproducibility of computational results.
 \preprint{INT-PUB-26-002}
\end{abstract}


\end{frontmatter}


\section*{Program summary}

\begin{small}
  \begin{PSdescriptionA}
  \item[Program Title] W-SLDA Toolkit
  \item[CPC Library link to program files] (to be added by Technical Editor)
  \item[Developer's repository link] \url{https://gitlab.fizyka.pw.edu.pl/wtools/wslda}
  \item[Code Ocean capsule] (to be added by Technical Editor)
  \item[Licensing provisions] GPLv3
  \item[Programming language] C, CUDA, MPI, Python
  \item[Supplementary material]
  \end{PSdescriptionA}
  \begin{PSdescriptionB}
  \item[Nature of problem]
    The simulation of superfluid or superconducting Fermi systems presents a major computational challenge, as both the normal and anomalous (pairing) densities must be treated self-consistently and on an equal footing.
    The W-SLDA Toolkit addresses this challenge by implementing a fully microscopic framework based on the Superfluid Local Density Approximation (SLDA) and its time-dependent extensions (TDSLDA).
    This approach generalizes density functional theory to superfluid systems, enabling modeling of both equilibrium and non-equilibrium phenomena in ultracold Fermi gases.

    The toolkit provides a unified computational environment for determining ground states, excitation spectra, and real-time dynamics in arbitrary external geometries, across the entire BCS-BEC crossover, and for both balanced and spin-imbalanced configurations.
    In addition, the capability to solve the standard Bogoliubov-de~Gennes equations extends its applicability to conventional superconductors, making the toolkit a versatile platform for exploring a broad range of correlated fermionic systems.

  \item[Solution method] The toolkit solves HFB / Bogoliubov-de~Gennes – type equations derived from superfluid density functionals using lattice discretization.
    Static problems are formulated as nonlinear fixed-point equations and solved self-consistently using linear mixing or Broyden schemes, with support for both fixed particle-number and fixed chemical-potential.
    Time-dependent simulations employ multistep predictor-corrector integrators of the Adams-Bashforth-Moulton type, which ensure stability and accuracy in large-scale three-dimensional systems.

    Fast Fourier transforms are used to evaluate kinetic-energy operators, while hybrid CPU/GPU parallelism with optimized FFT libraries provides high computational efficiency.
    The implementation supports one-, two-, and three-dimensional geometries, with significant performance gains for systems exhibiting translational symmetry.
    Regularization of the contact interaction guarantees ultraviolet convergence.
    
    The framework further incorporates built-in convergence and stability diagnostics, along with an automated reproducibility system that archives all parameters, input files, and results, ensuring transparent, verifiable, and repeatable computational workflows.
  \item[Additional comments including restrictions and unusual features] The W-SLDA Toolkit is an actively maintained, open-source,  high-performance computing package written in~C with CUDA and Python interfaces, distributed under the GPLv3 license.
    It is optimized for leadership-class supercomputing environments and routinely scales to thousands of GPUs, enabling fully three-dimensional simulations involving up to $10^5$ atoms, comparable to atom numbers realized in current ultracold-atom experiments.
    For smaller-scale problems, such as \respch{quasi-}1D or \respch{quasi-}2D systems, the code can also be executed efficiently on university clusters or even high-end workstations.

    The toolkit supports several superfluid density functionals (BdG, SLDA, ASLDA, SLDAE) and provides a flexible interface for defining custom energy density functionals.
    Although the implemented regularization scheme is formally derived for three-dimensional systems, extensions to lower dimensions are possible.
    Its modular design enables simulations of a wide range of fermionic systems, including superconductors, cold atoms, and neutron-star matter.
    Notably, the W-SLDA Toolkit also serves as the computational foundation for the W-BSk Toolkit~\cite{pecak2024WBSkMeff}, which targets applications in nuclear-matter physics.
  \end{PSdescriptionB}
\end{small}

\mainmatter

\section{Introduction}
The achievement of Bose-Einstein condensation (BEC) in dilute atomic gases by E.~A.~Cornell, W.~Ketterle, and C.~E.~Wieman at the end of the 20th century, for which they were awarded the Nobel Prize in Physics in 2001~\cite{RevModPhys.74.1131}, opened a new era in the exploration of quantum many-body systems.  Since then, ultracold atomic gases have become an unparalleled platform for exploring superfluidity and other emergent phenomena in correlated matter~\cite{RevModPhys.80.1215,Bloch2008,Bloch2012}.  A major milestone in this field was the experimental realization of the Bardeen-Cooper-Schrieffer (BCS) to Bose-Einstein condensate (BEC) crossover in two-component Fermi gases~\cite{Jin2004}, which made it possible to study the smooth evolution between two distinct superfluid regimes: the fermionic superfluid of loosely bound Cooper pairs and the bosonic condensate of tightly bound dimers.  The ability to control interparticle interactions through Feshbach resonances has turned ultracold Fermi gases into ideal laboratories for investigating strongly interacting superfluids under highly tunable conditions~\cite{Zwerger2012}.

Simulating such systems poses a formidable computational challenge.  Due to the Pauli exclusion principle, fermionic systems cannot be efficiently modeled with simplified approaches such as the Gross-Pitaevskii equation, which adequately describes weakly interacting Bose-Einstein condensates~\cite{Dalfovo1999}.  This limitation becomes particularly severe in the strongly interacting regime, where most experimentally relevant Fermi-gas phenomena occur.  Consequently, there has been a sustained effort to develop microscopic frameworks capable of describing strongly correlated superfluids in a controlled and quantitatively reliable manner.

A significant advance in this direction was made by A.~Bulgac, who formulated a density functional theory (DFT) for superfluid fermions, now known as the Superfluid Local Density Approximation (SLDA)~\cite{Bulgac2002,Bulgac2002a,Bulgac2007}.  This framework extends the Kohn-Sham formalism of DFT to include pairing correlations through the anomalous density, in accordance with general DFT theorems for superconductors~\cite{Oliveira1988}.
Unlike phenomenological models or qualitative approaches based on the Bogoliubov-de~Gennes (BdG) equations, the SLDA provides quantitatively accurate simulations of strongly interacting Fermi systems.  Its predictive power enables direct benchmarking against experiments, offering a rigorous platform for testing theoretical conjectures regarding the behavior of superfluid Fermi gases. 

A key strength of the SLDA lies in its natural extensibility to time-dependent phenomena:\ a domain where the distinctive properties of superfluids become most apparent.
This capability has become increasingly important with the advent of atomtronics:\ a rapidly developing field that seeks to engineer functional circuits and devices based on superfluid atomic flows~\cite{Amico2021,Amico2022,Pepino2021}.  In this context, numerical simulations play an essential role by guiding experimental design and predicting transport characteristics, stability conditions, and dissipation mechanisms in complex geometries.

The framework described in this work, the W-SLDA Toolkit, represents a mature implementation of the SLDA family of density functionals.
The name reflects both its scientific origin and design philosophy.
The acronym SLDA refers to the functional form that constitutes the core of the method.
The leading letter W acknowledges the institutions that played a key role in the development of this framework: the \underline{W}arsaw University of Technology (Poland), the University of \underline{W}ashington (Seattle, USA), and \underline{W}ashington State University (Pullman, USA).
The term Toolkit emphasizes that it is not a single monolithic code, but rather a comprehensive collection of tools for performing simulations and analyses of both static and time-dependent properties of superfluid Fermi gases.

The development of the W-SLDA Toolkit has a long history rooted in the pioneering applications of the SLDA to real-time dynamics.
The first demonstration of the applicability of the SLDA to time-dependent problems was reported in 2011~\cite{Bulgac2011}, where the real-time evolution of quantized vortices in a unitary Fermi gas was simulated.
Early implementations were written in Fortran and remained unavailable to the public.
The code was later ported to C and GPU-accelerated by G.~Wlazłowski.
The first scientific results from this accelerated version were published in 2014~\cite{Bulgac2014}, reporting simulations of quantized superfluid vortex rings.
This internal version served as a crucial testing ground and revealed the need for a more flexible workflow, unified data management, and standardized input/output formats suitable for large-scale parameter studies.
These insights motivated a comprehensive redesign in 2016, followed by a reimplementation from scratch, resulting in the current modular, toolkit-based architecture.
In this new design, the time-dependent implementation of the Asymmetric SLDA (ASLDA)~\cite{PhysRevLett.101.215301} functional was realized for the first time, enabling studies of dynamics in spin-imbalanced environments.
The modular structure facilitates straightforward code reuse and future extensions.
The first results obtained with the new framework were published in 2018 \cite{Wlazlowski2018}, demonstrating the dynamics of a spin-imbalanced superfluid Fermi gas.
In 2020, the W-SLDA Toolkit was publicly released for the first time~\cite{WSLDAToolkit}, marking its official name.
Since then, it has been under continuous development, with multiple releases each year that incorporate advances in numerical algorithms, data handling, and high-performance computing, ensuring scalability on leadership-class supercomputers.
Experience gained during the GPU porting of the ultracold-atom code also contributed to the development of the related nuclear-physics code LISE, developed by A.~Bulgac and collaborators~\cite{Jin2021}.
Although both projects share common algorithmic foundations, they have since evolved independently, addressing different physical regimes and adopting distinct implementation strategies.

\respch{This focus on dynamics distinguishes the development priorities of the W-SLDA Toolkit from those of many conventional DFT packages, which are designed primarily for normal systems. There, the central computational task is often the accurate determination of ground-state properties, and the static solver naturally becomes the main component of the software. In superfluid and superconducting systems, however, 
the majority of phenomena of interest -- such as vortex motion, phase slips, Josephson oscillations, pair breaking, collective modes, and other nonequilibrium processes -- are 
dynamical. For this reason, the W-SLDA Toolkit has been developed with particular emphasis on large-scale time evolution. In this workflow, static calculations remain essential, but primarily as a means of constructing controlled and physically meaningful initial states for subsequent real-time simulations. The numerical regime also differs from that of standard DFT calculations. In time-dependent BdG-type simulations, one typically evolves a number of quasiparticle states much larger than the number of particles -- often by an order of magnitude -- in order to ensure stable and accurate integration of equations of motion in the presence of pairing correlations. This requirement places the toolkit in a computational regime distinct from conventional DFT codes where only single-particle states below about the Fermi energy need to be occupied. This drives the emphasis of the toolkit on flexible state preparation, GPU-accelerated time evolution, and scalable handling of large quasiparticle bases.
}

Today, the W-SLDA Toolkit constitutes a robust and versatile platform for simulating Fermi superfluids.
Among its adjustable parameters are the interaction regimes (the BCS-BEC crossover, with an emphasis on the strongly interacting regime), population and mass imbalances, system dimensionality (ranging from 1D and \respch{quasi-}1D through 3D), arbitrary geometries of external potentials, transformations to moving or rotating reference frames, and finite-temperature formalisms.
The toolkit has been successfully applied to a broad range of ultracold-atom phenomena, including vortex dynamics, solitonic excitations, superfluid transport, and collective oscillations, providing indispensable numerical support for both experimental and theoretical research.
Moreover, it can be employed to solve standard Bogoliubov-de~Gennes equations, in either their static or time-dependent variants, enabling applications to superconducting systems as well.

\section{Theoretical framework}

\subsection{Units}
In the following sections, we use the metric system where $m=\hbar=k_B=1$. 

\subsection{General concepts}
The theoretical framework exploits the idea of Density Functional Theory (DFT)~\cite{Engel2011,Dreizler1990,Fiolhais2003,Koch2015,Parr1994,March1992}. \respch{In the case of ultracold Fermi gases, discussed here, we follow the superfluid formulation of the DFT framework~\cite{Oliveira1988}, which differs in many respects from the standard DFT framework.} 
We assume that the intrinsic energy of the system
\begin{align}\label{eq:local-functional}
	E_0 =& \int \calE[\rho_{\sigma}(\vbr),\tau_{\sigma}(\vbr),\vbj_\sigma,\nu(\vbr)]\,d\vbr
\end{align}
is a functional of the following densities:
\begin{subequations}
  \begin{description}[leftmargin=0pt]
  \item[Normal Density]
    \begin{equation}\label{eq:normal-density}
      \rho_{\sigma}(\vbr) = \sum_{\abs{E_n}<E_c}\abs{v_{n,\sigma}(\vbr)}^2 f_{\beta}(-E_n),
    \end{equation}
  \item[Kinetic Density]
    \begin{equation}\label{eq:kinetic-density}
      \tau_{\sigma}(\vbr) = \sum_{\abs{E_n}<E_c}\abs{\vb{\nabla} v_{n,\sigma}(\vbr)}^2 f_{\beta}(-E_n),
    \end{equation}
  \item[Current Density]
    \begin{equation}\label{eq:current-density}
      \vbj_\sigma(\vbr) = \sum_{\abs{E_n}<E_c} \textrm{Im}[v_{n,\sigma}(\vbr)\vb{\nabla} v_{n,\sigma}^*(\vbr)] f_{\beta}(-E_n),
    \end{equation}
  \item[Anomalous Density]
    \begin{equation}\label{eq:anomalous-density}
      \nu(\vbr) =  \dfrac{1}{2}\sum_{\abs{E_n}<E_c} \left [u_{n,a}(\vbr)v_{n,b}^{*}(\vbr)-u_{n,b}(\vbr)v_{n,a}^{*}(\vbr)\right ]f_{\beta}(-E_n).
    \end{equation}
  \end{description}
\end{subequations}
We adopt the convention
that $\nu(\vbr) =\langle c_{b}(\vbr) c_{a}(\vbr) \rangle$, where $c_{\sigma}(\vbr)$
denotes the annihilation operator.
The densities, except for the anomalous density, generally differ for the spin components $\sigma=\{a,b\}$.\footnote{Typically, the labels $(a,b)$ are associated with spin indices $(\uparrow,\downarrow)$. However, the framework presented here can also be applied to more general situations, where $a$ and $b$ may represent, for instance, two different hyperfine states or isotopes.} 
They are parametrized in terms of the Bogoliubov quasiparticle wave functions $\varphi_n(\vbr)=[u_{n,a}(\vbr),u_{n,b}(\vbr),v_{n,a}(\vbr),v_{n,b}(\vbr)]^{T}$, corresponding to quasienergies $E_n$. The Fermi-Dirac distribution function $f_{\beta}(E)=1/(\exp(\beta E)+1)$ is introduced to account for finite-temperature effects, with $T = 1/\beta$. This treatment follows the same approach as in the finite-temperature Hartree-Fock-Bogoliubov (HFB) theory~\cite{Goodman1981}. A cutoff energy scale $E_c$ is required to regularize the theory; see \cref{sec:regularization}.

The equations of motion for the quasiparticle orbitals $\varphi_n$ are derived by minimizing the functional~\cref{eq:local-functional}.
This procedure is typically carried out under external constraints, which are mathematically implemented through the generalized Lagrange multipliers.  The general expression subject to minimization takes the form
\begin{multline}
  \label{eq:E_tot}
  E =  E_0 
  -\sum_{\sigma}\int\left(\mu_{\sigma}-\Vext{\sigma}(\vbr)\right)\rho_{\sigma}(\vbr)\,d\vbr\\
  -\int\left(\Dext(\vbr)\nu^*(\vbr) + \textrm{h.c.}\right)d\vbr\\
  -\sum_{\sigma}\int {\velext{\sigma}}(\vbr)\cdot\vbj_{\sigma}(\vbr)\,d\vbr. 
\end{multline}
The Lagrange multipliers that couple to the densities have a well-defined physical interpretation:
\begin{description}[leftmargin=0pt]
\item[Chemical potential $\mu_\sigma$] controls total particle number $N_\sigma = \int \rho_{\sigma}(\vbr)\,d\vbr$.
\item[External potential $\Vext{\sigma}$] represents a confining potential.
\item[External pairing potential $\Dext$] models situations in which the superfluid gas is in contact with an external superfluid system, a scenario particularly relevant for applications to ultracold atomic systems.
\item[External velocity field $\velext{\sigma}$] enables the generation of solutions associated with flows.
  For instance, imposing $\velext{\sigma}(\vbr) = [v_0, 0, 0]$ corresponds to searching for solutions in a frame moving with velocity $v_0$ along the $x$-axis~\cite{Magierski2021}.
  Likewise, choosing $\velext{\sigma}(\vbr) = \vb\Omega \times \vbr$ leads to solving the problem in a rotating frame with angular velocity $\vb\Omega$~\cite{Kopycinski2021}.
  This Lagrange multiplier thus provides a convenient mechanism for formulating problems in various moving reference frames.
\end{description}

\subsection{Static formulation}\label{sec:st}
The minimization of quantity~\cref{eq:E_tot} is performed subject to the additional constraint $\int \varphi_n^{\dagger}(\vbr)\varphi_m(\vbr)\,d\vbr=\delta_{nm}$, where $\delta_{nm} = 1$ if $n=m$, and $0$ otherwise. This orthonormality condition for the quasiparticle orbitals reflects the Pauli principle, which must be satisfied by any fermionic many-body system.
The resulting equations of motion take the form of an eigenvalue problem, structurally equivalent to those encountered in the Hartree-Fock-Bogoliubov or Bogoliubov-de~Gennes (BdG) formalisms.  
They can be written as
\begin{subequations}
  \label{eq:st-Schrodinger}  
  \begin{equation}
    \calH
    \begin{pmatrix}u_{n,a}(\vbr) \\ 
      u_{n,b}(\vbr) \\ 
      v_{n,a}(\vbr) \\
      v_{n,b}(\vbr)
    \end{pmatrix}=E_n
    \begin{pmatrix}u_{n,a}(\vbr) \\ 
      u_{n,b}(\vbr) \\ 
      v_{n,a}(\vbr) \\
      v_{n,b}(\vbr)
    \end{pmatrix},
  \end{equation}
  with Hamiltonian matrix
  \begin{equation}
    \calH = 
    \begin{pmatrix}
      h_{a}(\vbr)-\mu_{a}  &  0 & 0 & \Delta(\vbr)+ \Delta_{\textrm{ext}}(\vbr) \\
      0 & h_{b}(\vbr)-\mu_{b}  & -\Delta(\vbr)-\Delta_{\textrm{ext}}(\vbr) & 0 \\
      0 & -\Delta^*(\vbr)- \Delta^*_{\textrm{ext}}(\vbr) & -h^*_{a}(\vbr)+\mu_{a} & 0 \\
      \Delta^*(\vbr)+ \Delta^*_{\textrm{ext}}(\vbr) & 0 & 0 & -h^*_{b}(\vbr)+\mu_{b}\label{eqn:HpsiEpsiFull}
    \end{pmatrix}.
  \end{equation}
  The single-particle Hamiltonian has a generic form
  \begin{equation}\label{eq:h_sigma}
    h_{\sigma} = -\dfrac{1}{2}\vb{\nabla}\alpha_{\sigma}(\vbr)\vb{\nabla}
    + V_{\sigma}(\vbr) + \Vext{\sigma}(\vbr)
    - \dfrac{\I}{2}\left\lbrace \vbA_{\sigma}(\vbr)
      -\velext{\sigma}(\vbr),\vb{\nabla} \right\rbrace,
  \end{equation}
  where $\{\; ,\,\}$ is the anticommutator.
\end{subequations}
The potentials that define the Hamiltonian matrix $\calH$ are expressed in terms of respective functional derivatives: 
\begin{description}
\item[Mean-field potential] ${ V}_{\sigma}=\dfrac{\delta{E}_{0}}{\delta \rho_{\sigma}}$, \\
\item[Pairing potential]    $\Delta=-\dfrac{\delta{E}_{0}}{\delta \nu^*}$,\\
\item[Current potential]    $\vbA_{\sigma}=\dfrac{\delta{E}_{0}}{\delta \vbj_{\sigma}}$,\\
\item[Inverse of the effective mass] $\alpha_\sigma=2\dfrac{\delta{E}_{0}}{\delta \tau_{\sigma}}$.
\end{description}
Their explicit forms depend on the choice of the energy functional, and will be discussed in \cref{sec:EDFs}. 

To significantly reduce the numerical cost, it is sufficient to solve the reduced problem
\begin{equation}
\begin{pmatrix}h_{a}(\vbr)-\mu_{a}  &  \Delta(\vbr)+ \Delta_{\textrm{ext}}(\vbr) \\\Delta^*(\vbr)+ \Delta^*_{\textrm{ext}}(\vbr) &  -h^*_{b}(\vbr)+\mu_{b}\end{pmatrix} \begin{pmatrix}u_{n,a}(\vbr) \\ v_{n,b}(\vbr)\end{pmatrix}= E_n\begin{pmatrix}u_{n,a}(\vbr) \\ v_{n,b}(\vbr)\end{pmatrix}\label{eqn:HpsiEpsi}
\end{equation}
which yields a subset of solutions of \cref{eq:st-Schrodinger} of the form $[u_{n,a}(\vbr),0,0,v_{n,b}(\vbr)]^T$.  The complementary branch of solutions, $[0,u_{n,b}(\vbr),v_{n,a}(\vbr),0]^T$, is obtained from
\begin{equation}
\begin{pmatrix}h_{b}(\vbr)-\mu_{b}  & -\Delta(\vbr)- \Delta_{\textrm{ext}}(\vbr) \\-\Delta^*(\vbr)-\Delta^*_{\textrm{ext}}(\vbr) &  -h^*_{a}(\vbr)+\mu_{a}\end{pmatrix} \begin{pmatrix}u_{n,b}(\vbr) \\ v_{n,a}(\vbr)\end{pmatrix}= E_n\begin{pmatrix}u_{n,b}(\vbr) \\ v_{n,a}(\vbr)\end{pmatrix}.\label{eqn:HpsiEpsi2}
\end{equation}
A symmetry relation exists between \cref{eqn:HpsiEpsi,eqn:HpsiEpsi2}.
Specifically, if $[u_{n,a},v_{n,b}]^T$ is a solution of \cref{eqn:HpsiEpsi} with eigenvalue $E_n$, then $[u_{n,b},v_{n,a}]^T=[v^*_{n,b},u^*_{n,a}]^T$ is a solution of \cref{eqn:HpsiEpsi2} corresponding to eigenvalue $-E_n$.
This symmetry ensures that solving \cref{eqn:HpsiEpsi} alone is sufficient to reconstruct the full set of solutions of the original problem \cref{eq:st-Schrodinger}.
The corresponding densities then are given by
\begin{subequations}
  \label{eqn:densities1}
  \begin{align}
    \rho_{a}(\vbr) &= \sum_{\abs{E_n}<E_c}\abs{u_{n,a}(\vbr)}^2 f_{\beta}(E_n)\,,\label{eq:na3D}\\
    \rho_{b}(\vbr) &= \sum_{\abs{E_n}<E_c}\abs{v_{n,b}(\vbr)}^2 f_{\beta}(-E_n)\,,\\
    \tau_{a}(\vbr) &= \sum_{\abs{E_n}<E_c}\abs{\vb{\nabla} u_{n,a}(\vbr)}^2 f_{\beta}(E_n)\,,\\
    \tau_{b}(\vbr) &= \sum_{\abs{E_n}<E_c}\abs{\vb{\nabla} v_{n,b}(\vbr)}^2 f_{\beta}(-E_n)\,,\\
    \nu(\vbr) &= \dfrac{1}{2}\sum_{\abs{E_n}<E_c} u_{n,a}(\vbr)v_{n,b}^{*}(\vbr)\left( f_{\beta}(-E_n)-f_{\beta}(E_n)\right)\,,\\
    \vb{j}_{a}(\vbr) &= -\sum_{\abs{E_n}<E_c} \Im[u_{n,a}(\vbr)\vb{\nabla} u_{n,a}^*(\vbr)] f_{\beta}(E_n)\,,\\
    \vb{j}_{b}(\vbr) &= \sum_{\abs{E_n}<E_c} \Im[v_{n,b}(\vbr)\vb{\nabla} v_{n,b}^*(\vbr)] f_{\beta}(-E_n)\,.
  \label{eq:jb3D}  
  \end{align}
\end{subequations}

In the case of spin-symmetric systems, where $h_{a}(\vbr)-\mu_{a}=h_{b}(\vbr)-\mu_{b}$ the computational procedure can be further simplified due to time-reversal symmetry.
It can be shown that if the vector $[ u_{n,a},v_{n,b}]^{T}$ is a solution of \cref{eqn:HpsiEpsi} with eigenvalue $E_n$, then the vector $[v_{n,b}^*,-u_{n,a}^*]^{T}$ is also a solution, corresponding to eigenvalue $-E_n$. Consequently, the computation can be restricted to solutions with positive eigenvalues $E_n$. For spin-symmetric systems, the densities are given by:
\begin{subequations}
  \begin{align}
    \rho_{a}(\vbr)=\rho_{b}(\vbr) &= \sum_{0<E_n<E_c}\left( \abs{v_{n,b}(\vbr)}^2 f_{\beta}(-E_n)+\abs{u_{n,a}(\vbr)}^2 f_{\beta}(E_n)\right)\,, \label{eq:nab3D}\\
    \tau_{a}(\vbr)=\tau_{b}(\vbr) &= \sum_{0<E_n<E_c}\left( \abs{\vb{\nabla} v_{n,b}(\vbr)}^2 f_{\beta}(-E_n) + \abs{\vb{\nabla} u_{n,a}(\vbr)}^2 f_{\beta}(E_n)\right)\,,\\
    \nu(\vbr) &= \sum_{0<E_n<E_c} u_{n,a}(\vbr)v_{n,b}^{*}(\vbr)\left( f_{\beta}(-E_n)-f_{\beta}(E_n)\right) ,\\
    \vb{j}_{a}(\vbr)=\vb{j}_{b}(\vbr) &= \sum_{0<E_n<E_c}\left(\Im[v_{n,b}(\vbr)\vb{\nabla} v_{n,b}^*(\vbr)] f_{\beta}(-E_n)-  \Im[u_{n,a}(\vbr)\vb{\nabla} u_{n,a}^*(\vbr)] f_{\beta}(E_n) \right)\,.\label{eq:jab3D}
  \end{align}
\end{subequations}

\respch{The number of quasiparticle states contributing to the densities directly impacts the computational cost. In contrast to standard DFT, where the number of states considered scales with the number of particles, in superfluid systems it is determined by the applied regularization scheme and typically scales with the size of the computational basis. This aspect is discussed in a separate sec.~\ref{sec:regularization}. }

\subsection{Energy density functional}\label{sec:EDFs}
We assume that the energy density takes the following general form:
\begin{subequations}
  \begin{gather}
    \label{eq:sldae-functional-generic}
    \calE  = \calE_0 + \calE_{\textrm{CM}},
  \end{gather}
  where $\calE_0$ denotes the Galilean-invariant~\cite{Engel:1975,Bender:2003,Bulgac2012} intrinsic energy density and $\calE_{\textrm{CM}}$ represents the Galilean-covariant contribution associated with the center-of-mass motion.
  The center-of-mass contribution is
  \begin{gather}
    \label{eq:sldae-functional-CM}
    \calE_{\textrm{CM}}  = \sum_\sigma \frac{\vbj^2_\sigma}{2\rho_{\sigma}}\,,
  \end{gather}
  while the Galilean-invariant intrinsic part is
  \begin{gather}
    \label{eq:sldae-functional}
    \calE_0 = \sum_\sigma \frac{A_{\sigma}}{2}
    \left(\tau_{\sigma}-\frac{\vbj^2_\sigma}{\rho_{\sigma}}\right)
    + \frac{3}{5}B \rhoP \eF(\rhoP)
    + \underbrace{\frac{\Cinf}{\rhoP^{1/3}}}_{\ginf}\abs{\nu}^2\,,
  \end{gather}
  where we have introduced $\rhoP$ as the total density, and $\eF(\rhoP)$ as the Fermi energy:
  \begin{align}
    \rhoP &= \rho_{a} + \rho_{b},
    &
    \eF &= \frac{\kF^2}{2} = \frac{(3\pi^2\rhoP)^{2/3}}{2},
  \end{align}
  to simplify the expression.
  
  The \emph{functional parameters} are dimensionless functions $\{A_{\sigma},B,C\}$ of the densities with the interpretation that $A_{\sigma}$ set the inverse effective masses,\footnote{Do not confuse the functional parameter $A_{\sigma}$ with the vector potential $\vbA_{\sigma}$, which is denoted using a bold symbol.
    Note that in this notation $A_{\sigma} = \alpha_\sigma$, which introduces redundancy in symbols.  However, we deliberately retain this distinction, as it facilitates a clearer and more systematic hierarchy at the implementation level.} $B$ sets the strength of normal-state interactions, and $C$ characterizes the strength of the pairing interaction through the coupling constant $g$.
  The kinetic and anomalous densities separately diverge.
  These divergences cancel when included in the specific combination appearing in $\calE_0$, but a regularization prescription is still needed so that the pairing field $\Delta = -g\nu$ remains finite.
  This means that the parameter $\Cinf \rightarrow 0$ formally vanishes in the continuum limit and must instead be expressed in terms of a finite cutoff-independent function $\Cc$ through the relationship \cref{eq:Creg}
  \begin{equation*}
    \underbrace{\frac{\rhoP^{1/3}}{\Cc}}_{1/\gc}
    = \underbrace{\frac{\rhoP^{1/3}}{\Cinf}}_{1/g} - \Lambda_c,
  \end{equation*}
  where the divergence in $1/\Cinf$ is canceled by the similarly-diverging regulator $\Lambda_c$.
  Details will be presented below in \cref{sec:regularization}.
\end{subequations}

We take the convention here that $g < 0$ for attractive interactions, and include an additional minus sign in the definition of the pairing potential
\begin{gather}
  \label{eq:Delta}
  \Delta = -\frac{\delta E_0}{\delta \nu^*} = -\underbrace{\frac{\Cinf}{\rhoP^{1/3}}}_{g}\nu
\end{gather}
to preserve consistency with this convention.

\begin{figure}[t]
	\centering
	\includegraphics[width=1.0\linewidth]{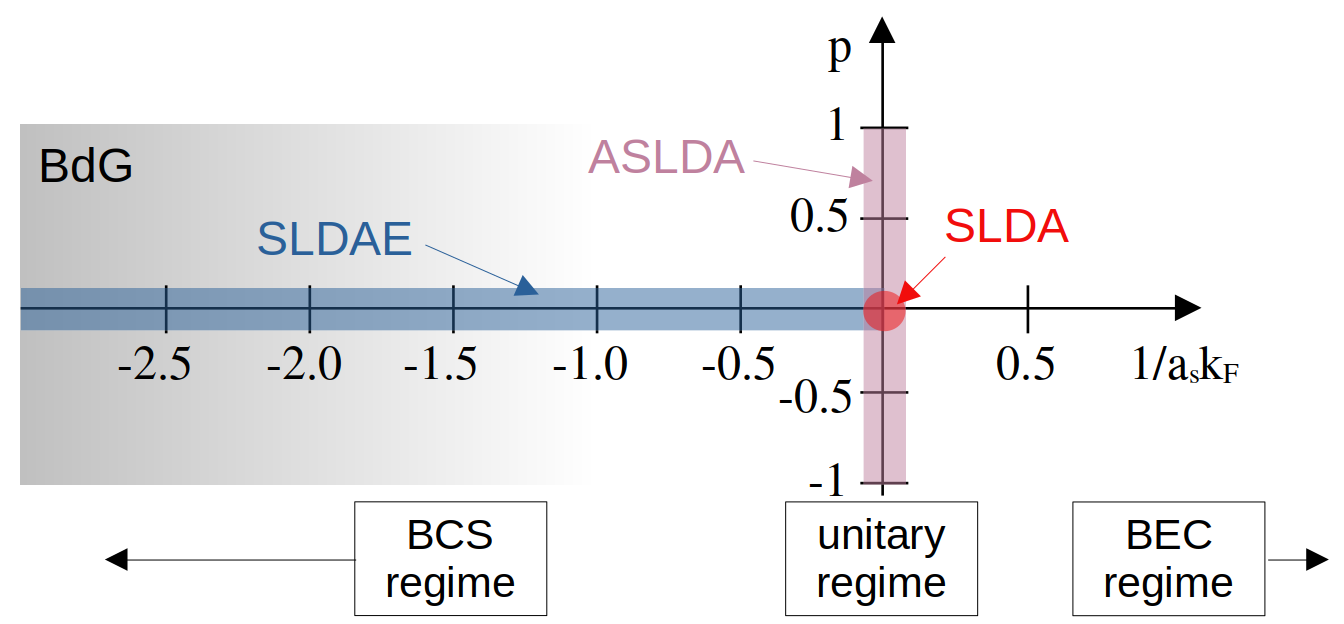}
	\caption{Range of applicability of the functionals implemented within the W-SLDA Toolkit.
    The horizontal axis represents the dimensionless interaction strength $1/(a_s\kF)$, while the vertical axis denotes the spin polarization $p = (\rho_a - \rho_b)/(\rho_a + \rho_b)$.
    The BdG functional is valid in the weak-coupling limit, but it is frequently employed for qualitative studies across the entire BCS--BEC crossover.
    The ASLDA functional~\cite{PhysRevLett.101.215301,Bulgac2012} is designed to accurately describe spin-imbalanced Fermi gases at unitarity; in the spin-symmetric limit, it reduces to the original SLDA functional~\cite{Bulgac2007}.
    The SLDAE functional~\cite{Boulet2022} extends this framework to cover the full range of coupling strengths from the BCS to the unitary regime.
    By construction, it reproduces the BdG limit in the weak-coupling regime and the SLDA limit at unitarity.}
	\label{fig:functionals-range}
\end{figure}
A functional of the generic form~\cref{eq:sldae-functional-generic} will be referred to as an \textit{SLDA-type} functional.  This class of functionals was first introduced in the context of the unitary Fermi gas by A.~Bulgac in Ref.~\cite{Bulgac2007}. 
The W-SLDA Toolkit implements the following functionals (see \cref{fig:functionals-range} for an overview of their respective ranges of applicability):

\begin{description}
\item[BdG] There always exists a functional which, upon minimization, yields equations identical to the mean-field equations.  In the context of superfluid Fermi gases or superconductors, these are commonly referred to in the literature as the Bogoliubov-de~Gennes (BdG) equations~\cite{Zhu2016}.  The BdG functional for a contact interaction is obtained by choosing $A_{\sigma} = 1$, $B = 0$, and $\tfrac{C}{\rhoP^{1/3}} = 4\pi a_s$, where $a_s$ denotes the $s$-wave scattering length.  After minimization, the resulting Hamiltonian is (for simplicity, we set $\Dext = 0$ and $\velext{\sigma} = 0$):
  \begin{equation}
    \calH^{\textrm{BdG}}=
    \begin{pmatrix}
      -\frac{1}{2}\nabla^2 + \Vext{a}(\vbr) - \mu_a & \Delta(\vbr) \\
      \Delta^*(\vbr) & \frac{1}{2}\nabla^2-\Vext{b}(\vbr) + \mu_b \\
    \end{pmatrix}.
    \label{eqn:HBdG}
  \end{equation}
  In this simplified approach, all interaction effects are modeled solely through the pairing term  $\Delta(\vbr)=-4\pi a_s \nuc(\vbr)$, which introduces nonlinearity into the problem.  The interaction must be attractive ($a_s < 0$) in order to induce the emergence of a superfluid state ($\nu \neq 0$); otherwise, the functional is minimized at $\nu = 0$, corresponding to the normal state of a free Fermi gas.  Formally, the BdG description is valid only for weak attractive interactions ($\abs{a_s\kF}<1$), Nevertheless, it is often employed at a qualitative level to study various phenomena throughout the entire BCS-BEC crossover~\cite{RevModPhys.80.1215,Spuntarelli2010,PhysRevB.89.054511,PhysRevLett.96.090403}. 
  The same formalism can also be applied to conventional superconductors, in which case $4\pi a_s$ is replaced by the coupling constant $g$, representing the effective pairing interaction due to electron-phonon coupling~\cite{De_Gennes_2018}.
\item[ASLDA] The functional known as the \textit{Asymmetric Superfluid Local Density Approximation} (ASLDA) was developed to describe the properties of the unitary Fermi gas ($\abs{a_s \kF} \rightarrow \infty$) with arbitrary spin polarization $P = (N_a - N_b)/(N_a + N_b)$~\cite{PhysRevLett.101.215301,Bulgac2012}.  The functional parameters are assumed to depend on the local spin polarization, $A_{\sigma}(p)$, $B(p)$, and $C(p)$, where $p(\vbr)=\frac{\rho_a(\vbr)-\rho_b(\vbr)}{\rho_a(\vbr)+\rho_b(\vbr)}$.  Their explicit forms are given in Ref.~\cite{Bulgac2012}, and read
  \begin{subequations}
    \begin{align} 
      A_a(p) = A_b(-p)&= 1.094 + 0.156 p \left(
                        1 - \frac{2p^2}{3} + \frac{p^4}{5}\right)
                        - 0.532 p^2\left(1 - p^2 + \frac{p^4}{3}\right),\\
      \frac{B(p)}{2^{2/3}}&=0.357 + 0.642p^2 - A_a(p)\left(\frac{1+p}{2}\right)^{5/3} - A_b(p)\left(\frac{1-p}{2}\right)^{5/3},\\
      C(p) &= -11.11.
    \end{align} 
    This parametrization reproduces quantum Monte Carlo results with high accuracy for both uniform systems, covering spin-symmetric and imbalanced cases~\cite{PhysRevLett.97.200403,PhysRevLett.95.060401,PhysRevLett.100.150403}, and trapped systems~\cite{PhysRevLett.99.233201,PhysRevA.78.013635,PhysRevA.76.021603}.
    In the spin-symmetric limit, the functional reduces to the SLDA functional originally proposed by A.~Bulgac in Ref.~\cite{Bulgac2007}.  In this case, the self-consistent solution yields
    \begin{equation}
      \frac{E(p=0)}{\Effg}=0.39(1),\qquad \frac{\Delta(p=0)}{\eF}=0.51(1),
    \end{equation}
    where $\Effg=\frac{3}{5}N\eF$ is the energy of a free Fermi gas consisting of $N=N_a+N_b$ particles.  
  \end{subequations}
\item[SLDAE] Many experiments are performed for interaction strengths that lie outside the applicability range of the BdG mean-field theory ($-a_s \kF < 1$) and away from the unitary limit ($1/\abs{a_s \kF} \approx 0$), where the (A)SLDA functional is valid.
  To bridge this gap, the \textit{SLDA Extended} (SLDAE) functional was recently introduced~\cite{Boulet2022}.
  In this framework, the functional parameters are expressed as functions of the dimensionless coupling constant, $A_{\sigma}(\lambda)$, $B(\lambda)$, and $C(\lambda)$, where $\lambda(\vbr)=\abs{a_s\kF(\vbr)}$.
  Since $\kF$ depends on the density, all couplings are effectively density-dependent.
  Ref.~\cite{Boulet2022} demonstrated that the coupling constants $A_{\sigma}$, $B$, and $C$ can be determined analytically from the density dependence of quasiparticle properties alone.
  Specifically, knowledge of the chemical potential, the effective mass, and the pairing gap function is sufficient to obtain the coupling functions.
  The SLDAE functional assumes the following forms for these quantities:
 \begin{subequations}
    \label{eq:APSfunctional}
	\begin{align}
		\frac{\mu}{\eF}
		= & 1 - \frac{16}{3\pi} \calS_\ddcc 
		      - \frac{16\ddcc}{15\pi} \calS_\ddcc\oprime,    \\
	\frac{\Delta_\ddcc}{\eF} =&  \frac{8}{e^2} \exp\left(-\frac{\pi}{2\ddcc}\right)\times  	\frac{1+\ddcc y}{1+  \ddcc y z},\\
		\alpha_\lambda
		=  & 1 + {\frac{5uw- 7x u v }{9\pi u^2}}
		        \lambda^2 \calS_\lambda\oprime
		      + {\frac{2w+7x v}{9\pi u^2}}
		        (1+\lambda v) \lambda^2 {\calS_\lambda\oprime}^2,
	\end{align}
  with the parametric function~\cite{Boulet2019a}
  \begin{align}
    \mathcal{S}_\lambda & = \arctan \left(\frac{\lambda u}{1 + \lambda v}\right).
  \end{align}
  
  The constants are: $u = 5/24$, $v = 6(11-2\ln 2)/35\pi$, $w = 24(1-7\ln 2)/35\pi$, $x \simeq -0.75$, $y = 4/5$, and $z = ({8}/{e^2})/0.46$.  This parametrization, known as the \textit{Arctangent Phase-Space} (APS) form, was derived in Refs.~\cite{Boulet2018,Boulet2019,Boulet2019a} and accurately reproduces both experimental and quantum Monte Carlo results across the BCS-unitary regime.
  For example, the associated equation of state, obtained from $\mu=\left. \frac{d\calE}{d\rhoP}\right |_{V}$, reads
  \begin{equation}
    \frac{E}{\Effg} = 1 - \frac{16}{3\pi} \calS_\ddcc, 
  \end{equation}
  yielding at unitarity $\frac{E}{\Effg}=\frac{\mu}{\eF}\simeq 0.36$.  In the weak-coupling limit, the expressions reduce to the predictions of BCS theory.
\end{subequations}  
\item[CUSTOM-EDF] The W-SLDA Toolkit also enables users to define \emph{custom energy density functionals}.
  This requires specifying the explicit analytical form of the energy density, along with the corresponding functional derivatives that determine potentials ($V_{\sigma}$, $\Delta$, $\vbA_{\sigma}$, $\alpha_\sigma$).
  This option is particularly valuable for developing and testing new functional forms.
  The main limitation is that the functional must be expressible within the general SLDA-type framework, ensuring compatibility with the toolkit’s numerical infrastructure.
\end{description}

The case of $A_{\sigma} = 1$ is a special one, as the framework simplifies considerably, resulting in a substantial computational speed-up.  In this limit, the functional no longer depends explicitly on the currents $\vbj_\sigma$, and the single-particle Hamiltonian~\cref{eq:h_sigma} takes the simplified form (assuming $\velext{\sigma} = 0$):
\begin{equation}\label{eq:h_sigma_A1}
h_\sigma = -\frac{1}{2}\nabla^2 + V_{\sigma}(\vbr) + \Vext{\sigma}(\vbr).
\end{equation}
In this case, only the Laplacian of the wave functions needs to be evaluated, in contrast to the general formulation, which requires the computation of both gradients and Laplacians.  The absence of gradient terms leads to a significant reduction in computational cost.
It turns out that, in most physically relevant cases, $A_{\sigma}$ remains close to unity, implying that the effective mass of the particles is nearly equal to the bare mass.
For this reason, the toolkit provides an option to explicitly enforce $A_{\sigma} = 1$, allowing users to trade a small decrease in functional accuracy for a substantial gain in computational efficiency.
In practice, this simplification has little impact on qualitative results and typically affects quantitative observables only marginally.
For example, in the supplementary material of Ref.~\cite{Bulgac2014}, it was demonstrated that the relative difference in total energy between the $A_{\sigma} \neq 1$ and $A_{\sigma} = 1$ cases does not exceed a few percent. 

\subsection{Time-dependent formulation}
The time-dependent formalism is obtained by promoting $E_n\rightarrow \I\tfrac{\partial}{\partial t}$ in \cref{eq:st-Schrodinger}\footnote{It can be formally derived from the stationarity condition of the action expressed in terms of the energy density functional (see Ref.~\cite{APPBMagierski2018}).}. Each quasiparticle orbital becomes both position- and time-dependent, $\varphi_n(\vbr)\rightarrow\varphi_n(\vbr,t)$, and consequently all densities [\crefrange{eq:normal-density}{eq:anomalous-density}] as well as the potentials ($V_\sigma$, $\Delta$, $\vbA_{\sigma}$ and $\alpha_\sigma$) acquire explicit position and time dependence. The equations governing the time evolution of the superfluid system take the form
\begin{equation}\label{eqn:dpsidt-Hpsi}
  \I\dfrac{\partial}{\partial t}\begin{pmatrix}u_{n,a}(\vbr,t) \\ v_{n,b}(\vbr,t)\end{pmatrix}=
\begin{pmatrix}h_{a}(\vbr,t)-\mu_{a}  &  \Delta(\vbr,t)+ \Delta_{\textrm{ext}}(\vbr,t) \\\Delta^*(\vbr,t)+ \Delta^*_{\textrm{ext}}(\vbr,t) &  -h^*_{b}(\vbr,t)+\mu_{b}\end{pmatrix} \begin{pmatrix}u_{n,a}(\vbr,t) \\ v_{n,b}(\vbr,t)\end{pmatrix},
\end{equation}
with $h_\sigma$ and $\Delta$ defined analogously to the static case.  Note that the external potentials may also depend on both position and time, e.g.\ $\Vext{\sigma}(\vbr) \rightarrow \Vext{\sigma}(\vbr,t)$, with analogous generalizations for the other external fields.

To initiate the integration of \cref{eqn:dpsidt-Hpsi}, an initial state must be specified.  In particular, a meaningful set of quasiparticle wave functions $\{ \varphi_n(\vbr,0) \}$ is required.  This set is typically chosen as the solution of the static equations \cref{eqn:HpsiEpsi}.  Hence, solving the time-dependent problem is generally preceded by solving the corresponding static problem.
The initial wave functions are orthonormal, as they are eigenvectors of a Hermitian matrix, and this property is preserved throughout the evolution:
\begin{equation}
\int \left[u_{n,a}^*(\vbr,t)u_{m,a}(\vbr,t) + v_{n,b}^*(\vbr,t)v_{m,b}(\vbr,t) \right]d\vbr = \delta_{nm}, 
\end{equation}
\begin{equation}
\int \left[u_{n,b}^*(\vbr,t)u_{m,b}(\vbr,t) + v_{n,a}^*(\vbr,t)v_{m,a}(\vbr,t) \right]d\vbr = \delta_{nm},
\end{equation}
where there is no summation of either $a$ or $b$ subscripts.

It is worth noting that, from the perspective of time-dependent density functional theory, the substitution $E_n\rightarrow \I\tfrac{\partial}{\partial t}$ represents an approximation known as the \textit{adiabatic approximation}; see, e.g., Refs.~\cite{Dreizler:1990lr,Marques:2004,Gross:2006,Gross:2012,Ullrich2013}.  In general, an energy density functional constructed for static calculations does not satisfy the Runge-Gross theorem~\cite{PhysRevLett.52.997}, which establishes a one-to-one correspondence between the time-dependent density $\rho(\vbr,t)$ and the time-dependent many-body wave function $\Psi(\vbr_1, \vbr_2, \dots, \vbr_N, t)$.  In principle, the functional at time $t$ may depend on the densities at all previous times $t^\prime\le t$ (so-called \textit{memory effects}), as well as on the initial state $\Psi_0$.  For example, one could write 
\begin{equation}
E_0(t) = \int_0^t dt^\prime \int d\vbr\, \calE[\rhoP(\vbr,t^\prime),\dots, \Psi_0].
\end{equation}
The adiabatic approximation neglects such memory effects, and as a result the time evolution is time-reversible.

The static formulation \cref{sec:st} also accounts for finite-temperature effects.
These are incorporated through the thermal weights $f_{\beta}$ appearing in the definitions of the densities \crefrange{eq:normal-density}{eq:anomalous-density}.
In the time-dependent formalism, however, the quasiparticle energies $E_n$ are no longer well-defined quantities.\footnote{In general, the temperature $T$ is a well-defined concept only for systems in (thermal) equilibrium.}
In the time-dependent implementation of the W-SLDA Toolkit, the thermal weights are kept frozen, fixed to the values obtained from the corresponding static calculation. This procedure is justified if the system remains close to equilibrium during the evolution or the time scale of the evolution is
sufficiently short; otherwise, it constitutes an uncontrolled approximation. 
\respch{A controlled treatment of memory effects and quasiparticle re-thermalization would require extensions beyond the framework implemented here and remains an active research problem. In the present version of the toolkit, these approximations should be understood as part of the standard adiabatic  
approximation, whose validity should be assessed using physical diagnostics such as energy and particle-number conservation, and comparison with relevant experimental or benchmark data, if available.}
\subsection{Regularization of the functional}
\label{sec:regularization}
The framework relies exclusively on local densities, a property that is essential for constructing an efficient solver.
Ensuring locality in the presence of pairing, however, leads to the complication that the kinetic ($\tau_{\sigma}$) and anomalous ($\nu$) densities each diverge when considered separately.
Although the combination appearing in the energy density functional $\mathcal{E}_0$ \cref{eq:sldae-functional} is finite~\cite{Bulgac2002, Bulgac2002a, Bulgac2007, Boulet2022}, we must still regularize the theory to render the formalism applicable -- notably to keep the pairing field $\Delta = -g\nu$ finite in \cref{eq:Delta}. 

The toolkit supports two regularization schemes (enabled through a flag in the configuration file \filename{predefines.h}) and allows one to define additional custom regularization schemes.

The first regularization scheme (the spherical cut-off scheme) introduces an energy cutoff scale $E_c$ as a separate, controllable algorithmic parameter.  All densities are computed by including only states with $\abs{E_n} < E_c$, so they become finite.  Next, we redefine the coupling constant
\begin{equation}
  \label{eq:Creg}
    \underbrace{\frac{\rhoP^{1/3}}{\Cc}}_{1/\gc}
    = \underbrace{\frac{\rhoP^{1/3}}{\Cinf}}_{1/g} - \Lambda_c 
\end{equation}
where $\Lambda_c$ is the so-called regularization function.
The regularization scheme is chosen so that physical quantities, like the pairing field $\Delta = -\gc\nu$ become independent of the cutoff as it becomes large and physically meaningful. 
While it works very well for static properties, using a cutoff in time-dependent simulations can pose problems. For example, \cite{Bulgac:2024a} demonstrates that using a cutoff like this can have significant qualitative effects, suppressing fission in the LISE~\cite{Jin2021} code, which uses a similar framework.

In our experience in describing cold gases, we have found it satisfactory to use a simple prescription outlined above.  The code, however, has the option to use the box regularization if needed.
This so-called box regularization yields finite quantities through a self-consistent formulation of the problem on both a spatial lattice and in a periodic box. This formulation can be made mathematically precise and consistent between both static and time-dependent problems, often justifying the added computational expense.  The box introduces a natural cutoff $k_c$ directly tied to the lattice spacing and, in this sense, is not an independent controllable algorithmic parameter.

\subsubsection{Spherical cutoff regularization scheme}

The choice of the regularization term $\Lambda$ is not unique. It must be constructed so as to cancel the local ultraviolet (UV) divergence of the anomalous density $\nu$. In continuous space, a natural choice is to introduce a spherical momentum cutoff $k<k_c$, and obtain a regularization scheme consistent with the standard two-body scattering problem:
\begin{equation} \label{eq:Lambda}
  \Lambda = \iiint_{|\bm{k}|<k_c}\frac{d^3{\bm{k}}}{(2\pi)^3} \frac{1}{k^2 + i 0^{+}} 
  = 4\pi\int_0^{k_c} dk \, \frac{k^2}{k^2 + i 0^{+}}
  = \frac{k_c}{2\pi^2}
  = R_0 k_c,
\end{equation}
where we introduced the regularization coefficient
\begin{equation}
    R_0=\frac{1}{2\pi^2}\approx 0.05066.
\end{equation}
This expression corresponds to a vacuum regularization, in which the subtraction matches the ultraviolet behavior of the two-body problem. Bulgac proposed an improved scheme~\cite{Bulgac2007} by shifting the pole of the integrand to $k_0\approx \kF$, thereby matching the asymptotic behavior of quasiparticles in the medium. The position of the pole is determined from
\begin{equation}
     A_{+}\frac{k_0^2}{2}+V_+ -
   \mu_+
    = 0,
\end{equation}
where quantities with a subscript $+$ denote averages over spin components, e.g.\ $X_+=(X_a+X_b)/2$. The regularization function then takes the form
\begin{align}\label{eq:spherical-Lambda-c}
  \Lambda_c(k_0,k_c) &= \frac{1}{A_{+}}\iiint_{|\bm{k}|<k_c}\frac{d^3{\bm{k}}}{(2\pi)^3} \frac{1}{k^2-k_0^2 + i 0^{+}}
  = \frac{4\pi}{A_{+}}\int_0^{k_c} dk \, \frac{k^2}{k^2-k_0^2 + i 0^{+}}\nonumber\\
  &= \frac{k_c}{2A_{+}\pi^2}\left(1 - \frac{k_0}{2k_c}
    \ln\frac{k_c + k_0}{k_c - k_0}\right)
  = \frac{R_0 k_c}{A_{+}}\left(1 - R_1\frac{k_0}{k_c}
    \ln\frac{k_c + k_0}{k_c - k_0}\right),
\end{align}
where we introduced the second regularization coefficient
\begin{equation}
    R_1=\frac{1}{2}=0.5.
\end{equation}
The momentum cutoff scale $k_c$ can be related to the energy cutoff scale via
\begin{equation}\label{eq:Ec}
E_c = A_{+}\frac{k_c^2}{2}+V_+ - \mu_+.
\end{equation}

This procedure is generalized to non-uniform systems through a local-density-type approximation, whereby the regularized coupling $\Cc(\vbr)$ is computed separately at each spatial point. The method has been extensively tested in the context of ultracold Fermi gases and has been shown to exhibit good convergence properties~\cite{Bulgac2007,Bulgac2012,Boulet2022}.  In computations performed on a discretized spatial grid, a natural energy cutoff arises:
\begin{equation}
    E_c = \frac{k_c^2}{2}, \quad k_c=\frac{\pi}{\Delta x},
\end{equation}
where $\Delta x$ is the spatial discretization step. This choice sets the spherical energy cutoff to the maximum allowed by the discretization scheme. 

The number of evolved quasiparticle states can be estimated by counting the number of momentum states inside a sphere of radius $k_c=\sqrt{2E_c}$. The volume of this sphere is $V_k=\frac{4}{3}\pi k_c^3$, while the volume occupied by a single momentum state is $(\Delta k)^3=(2\pi/L)^3=8\pi^3/L^3$, where $L^3$ is the volume of the simulation domain. Thus, \begin{equation} 
\label{eq:NstatesEc}
N_{\textrm{wf}} \approx \frac{V_k}{(\Delta k)^3} = \frac{\sqrt{2}}{3\pi^2}E_c^{3/2}L^3 . 
\end{equation} 
For an energy cutoff set by the lattice spacing, $E_c=k_c^2/2$ with $k_c=\pi/\Delta x$, this gives 
\begin{equation}
N_{\textrm{wf}} \approx \frac{\pi}{6}\frac{L^3}{(\Delta x)^3} \approx 0.5\,\frac{L^3}{(\Delta x)^3}. 
\end{equation} 
Thus, for the recommended cutoff, the number of retained single-particle momentum states is approximately $50\%$ of the total number of lattice points.

\subsubsection{Box regularization scheme}
\label{sec:cubic-cutoff-scheme}

In practical implementations on a spatial lattice, momentum space is naturally restricted to the first Brillouin zone, $k_c=\frac{\pi}{\Delta x}$, which has a cubic geometry. In this case, the regularization integral, Eq.~(\ref{eq:Lambda}), takes the form
\begin{align}\label{eq:box-Lambda}
  \Lambda  
  &= \int_{-k_c}^{k_c}\frac{dk_x}{2\pi}
     \int_{-k_c}^{k_c}\frac{dk_y}{2\pi}
     \int_{-k_c}^{k_c}\frac{dk_z}{2\pi} 
     \frac{1}{k^2 + i 0^{+}}
     \approx \frac{K}{4\pi^2} \frac{\pi}{dx}
     = \frac{K}{4\pi^2} k_c
     = R_0 k_c,
\end{align}
where $K\approx 2.44275$~\cite{Castin2004}. Importantly, this expression retains the same scaling behavior,
$
    \Lambda \sim k_c,
$
but with a modified numerical prefactor $R_0=\frac{K}{4\pi^2}\approx 0.06188$, which can be compared with the spherical value $1/(2\pi^2)\approx 0.05066$.
Thus, changing the integration domain from a sphere to the full Brillouin zone modifies only the coefficient, while preserving the overall scaling with $k_c$. In the box regularization scheme, all quasiparticle states represented on the lattice are included in the sums defining the densities, so the change of cutoff geometry affects not only the regularization coefficient but also the number of contributing states and, consequently, quantities such as the pairing field and kinetic density.

\begin{figure}[h!]
	\centering
	\includegraphics[width=0.6\linewidth]{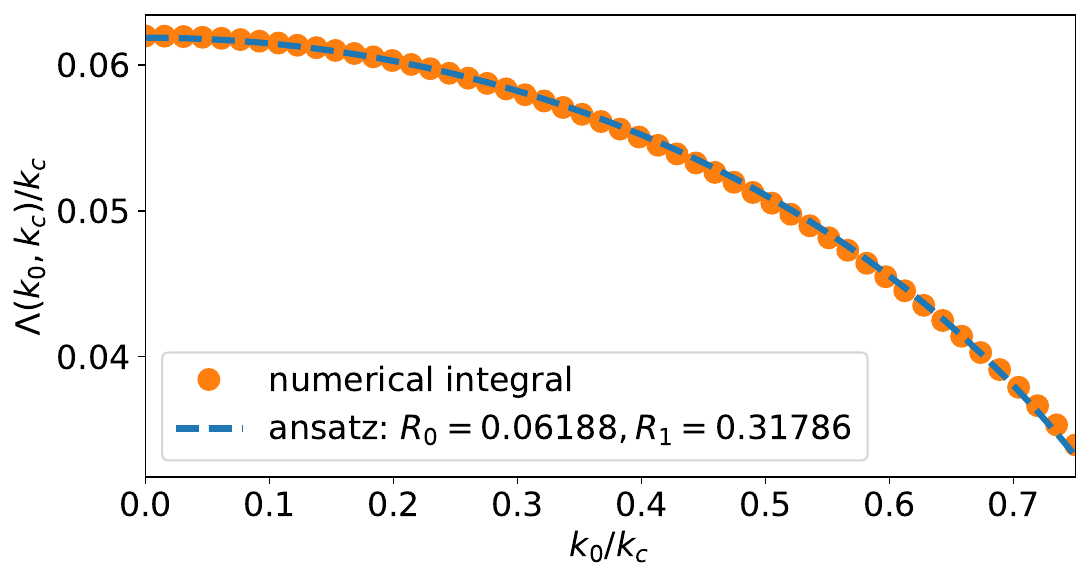}
	\caption{\respch{Numerically computed integral~(\ref{eq:Lambda_s_BZ}) together with the fitted ansatz~(\ref{eq:Lambdas-R}), for $A_{+}=1$. The agreement supports the validity of the proposed ansatz.}}
	\label{fig:Lambda}
\end{figure}

Following the spherical cutoff prescription, Eq.~(\ref{eq:spherical-Lambda-c}), the modified regularization function is 
\begin{equation}\label{eq:Lambda_s_BZ}
\Lambda_c(k_0,k_c) 
  = \frac{1}{A_{+}} \int_{-k_c}^{k_c}\frac{dk_x}{2\pi}
     \int_{-k_c}^{k_c}\frac{dk_y}{2\pi}
     \int_{-k_c}^{k_c}\frac{dk_z}{2\pi} 
     \frac{1}{k^2-k_0^2 + i 0^{+}}.
\end{equation}
Unfortunately, this integral does not have a compact analytical expression. The regularization function must therefore be computed numerically. The result of such a calculation is shown in Fig.~\ref{fig:Lambda}.
It turns out that the numerically computed integral is well approximated by the function 
\begin{equation}\label{eq:Lambdas-R}
    \Lambda_c(k_0,k_c) = \frac{R_0 k_c}{A_{+}}\left(1 - R_1\frac{k_0}{k_c}
    \ln\frac{k_c + k_0}{k_c - k_0}\right),
\end{equation}
where $R_0=0.06188$ and $R_1=0.31786$. This ansatz for the regularization function can be motivated on physical grounds. The first term is precisely the leading contribution obtained analytically in Eq.~(\ref{eq:box-Lambda}). The logarithmic contribution originates from the local neighborhood of the pole and is therefore expected to be largely insensitive to the global geometry of the cutoff, provided that the Fermi sphere remains well inside the Brillouin zone. This form separates the ultraviolet contribution, controlled by $R_0$, from the nonanalytic dependence encoded in the logarithmic term. In this approach, deviations due to the cubic geometry enter primarily through smooth corrections and can be effectively absorbed into the coefficients $R_0$ and $R_1$. Fig.~\ref{fig:Lambda} demonstrates good agreement between the numerically computed $\Lambda_c$ and the ansatz~(\ref{eq:Lambdas-R}), which allows us to avoid numerical integration in practical calculations.

In the box regularization scheme, the associated energy cutoff scale can be estimated as
\begin{equation}
    E_c \approx 3\frac{k_c^2}{2}=\frac{3\pi^2}{2\Delta x^2},
\end{equation}
which approximates the highest energy scale represented in the computation. Note, however, that the box regularization scheme does not use the scale $E_c$ explicitly. Instead, it uses $k_c=\pi/\Delta x$, which is fixed by the lattice spacing. Consequently, the number of evolved states is equal to the number of lattice points,
\begin{equation}
   N_{\textrm{wf}} = N_x N_y N_z. 
\end{equation}

\respch{
\subsubsection{Correction to the mean-field}

The regularization function $\Lambda_c$ depends on densities, via $k_0$ and $A_{+}$ coefficients. This dependence will induce a correction to the mean-field. Consider the contribution to the mean field arising from the pairing term:
\begin{equation}
    { V}_{\sigma}^{C}=\dfrac{\delta}{\delta \rho_{\sigma}}\left(\frac{\Cc}{\rho^{1/3}}|\nu|^2\right)
        =-\Bigl[\frac{1}{3\Cinf \rho^{2/3}}-\frac{\delta\Lambda_c}{\delta\rho_\sigma}\Bigr]|\Delta|^2,
\end{equation}
where we have used
\begin{equation}
    \Delta=-\frac{\Cc}{\rho^{1/3}}\nu,
\end{equation}
and Eq.~\ref{eq:Creg}. 

To evaluate $\frac{\delta\Lambda_c}{\delta\rho_\sigma}$, we introduce the dimensionless regularization function
\begin{equation}\label{eq:Lambda0-R}
    \Lambda_0(r) = \frac{bR_0}{r}\left[ 1 - R_1 \, r \ln\left(\frac{1+r}{1-r}\right) \right],
\end{equation}
where the dimensionless parameter $r$ is defined as
\begin{equation}
    r=\frac{b\kF}{k_c}.
\end{equation}
The coefficient $b$ is introduced through
\begin{equation}
    k_0=b\kF,
\end{equation}
and is typically close to unity. Its role is to account for the slight mismatch between the pole position $k_0$ and the local Fermi momentum $\kF$. For example, at unitarity, numerical calculations indicate that $b\approx 0.9$. 

With these definitions, the regularization function is given by
\begin{equation}
    \Lambda_c=\frac{\kF\Lambda_0}{A_{+}}.
\end{equation}
Following the spirit of SLDA, we take the local Fermi momentum to be density dependent
\begin{equation}
    \kF=[3\pi^2 \rho]^{1/3},
\end{equation}
while the other coefficient $b$ is assumed not to exhibit density dependence. Under these assumptions, we can derive the correction $\frac{\delta\Lambda_c}{\delta\rho_\sigma}$ explicitly. The user can control whether the correction is included via a suitable flag in \filename{predefines.h}. 

In Fig.~\ref{fig:reg-scheme-comparison}, we show the total energy, chemical potential, and pairing gap of the uniform unitary Fermi gas as functions of the ratio between the Fermi momentum and the cutoff scale, $k_F/k_c$. We observe that the energy is nearly independent of the cutoff scale $k_c$ over a wide range of densities. Moreover, the relation
\begin{equation}
    E/\Effg=\mu/\eF,
\end{equation}
is satisfied with high accuracy, as required by scale invariance at unitarity. The pairing gap exhibits a residual dependence on the cutoff scale. This dependence is weak, and the resulting variations remain smaller than the typical uncertainty associated with extracting the gap from quantum Monte Carlo calculations.
\begin{figure}[h!]
	\centering
	\includegraphics[width=0.6\linewidth]{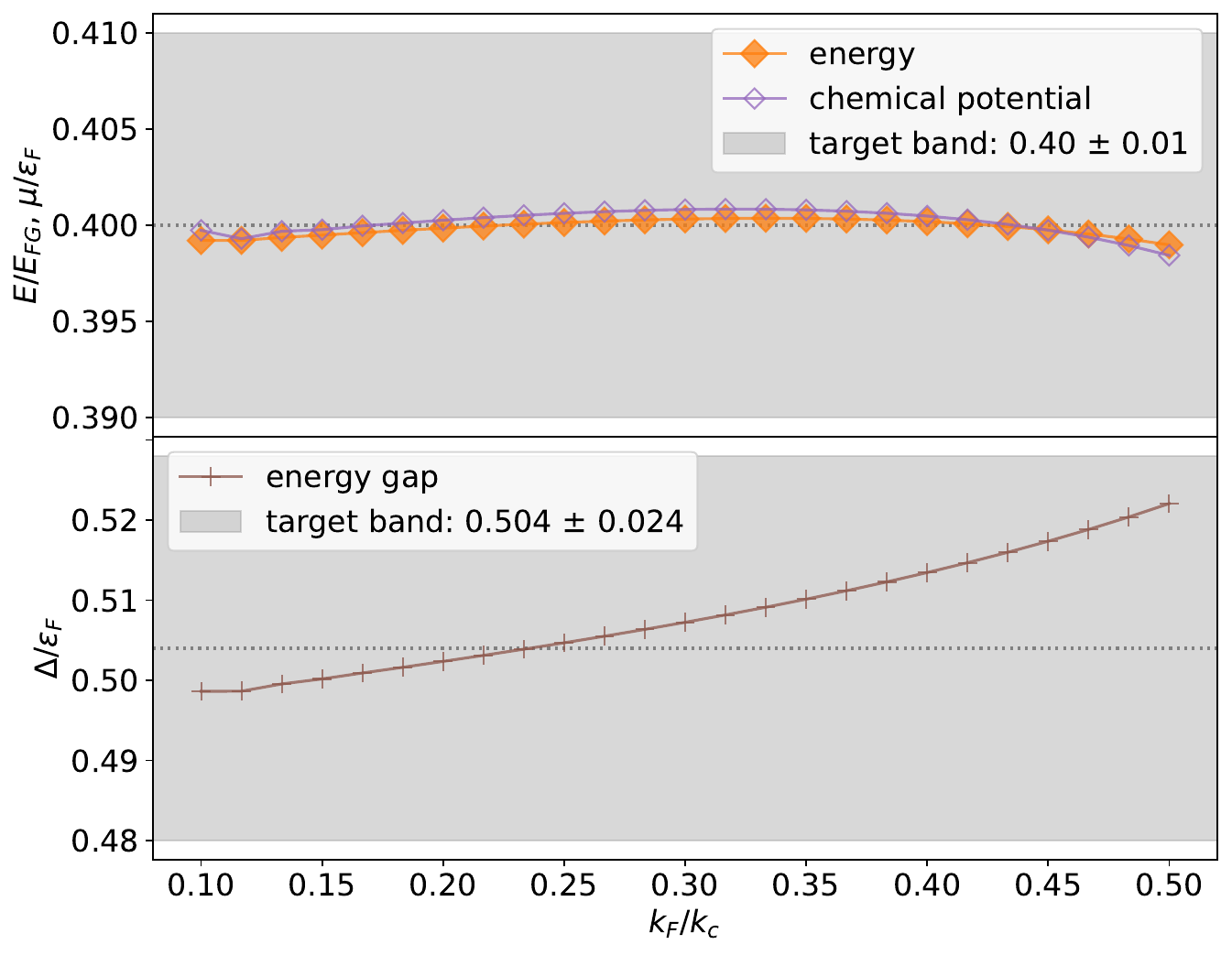}
	\caption{\respch{Total energy $E/\Effg$ and chemical potential $\mu/\eF$ (top), and pairing gap $\Delta/\eF$ (bottom), for the uniform unitary Fermi gas as functions of the ratio $k_F/k_c$, where $k_c=\pi/\Delta x$ is the cutoff scale. In this test, the ASLDA functional was used together with the box regularization scheme. The gray bands indicate $E/\Effg=0.40(1)$ and $\Delta/\eF=0.504(24)$~\cite{Bulgac2012}, corresponding to the quantum Monte Carlo results used as constraints in the construction of the ASLDA functional.}}
	\label{fig:reg-scheme-comparison}
\end{figure}
}

\respch{
\subsubsection{Regularization scheme and energy conservation in time-dependent problems}
\label{sec:reg_and_energy_conservation}

One may be tempted to use the energy cutoff scale $E_c$ as small as possible to reduce the number of considered quasiparticle states. This strategy is legitimate if one aims only for static problems. The perspective changes radically if one delves into time-dependent problems. 
}

\begin{figure}[h!] \centering \includegraphics[width=0.9\linewidth]{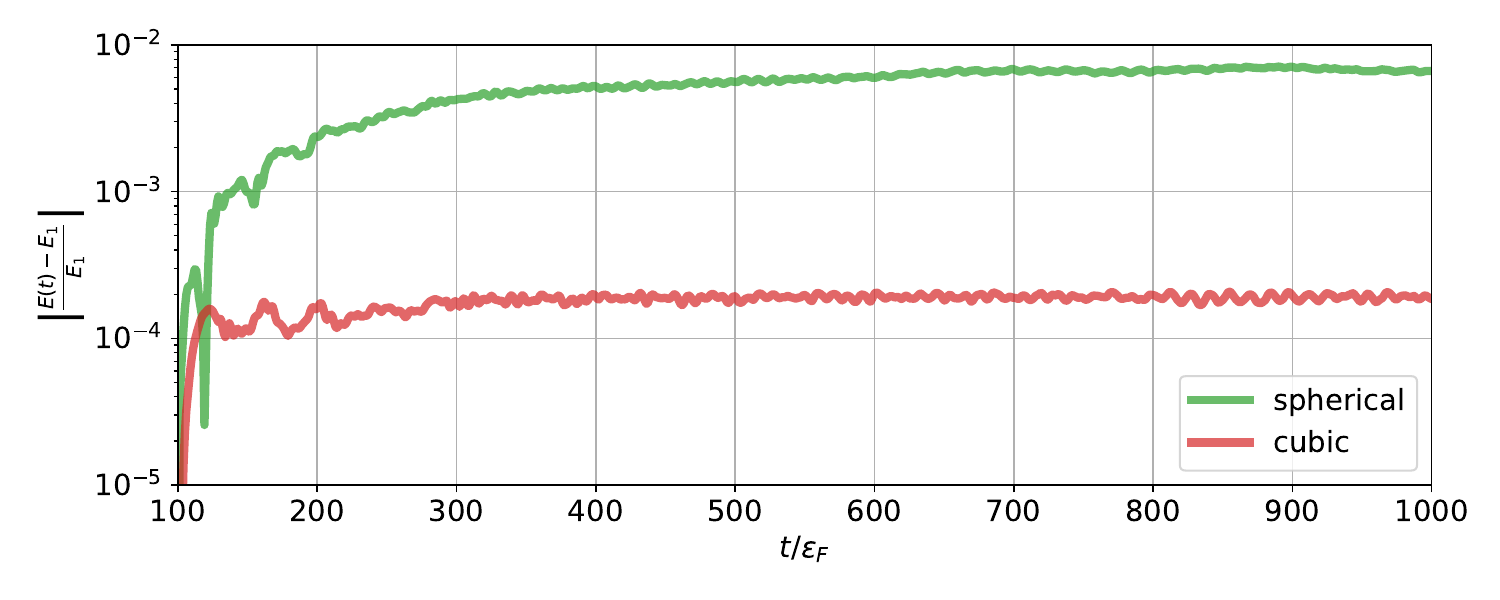} \caption{\respch{Deviation of the total energy from the reference value $E_1$, defined as the total energy at time $t\eF=100$. The simulation corresponds to a unitary Fermi gas excited by a time-dependent external perturbation of the form $\Vext{}(x,y,t)=A(t)\exp\left(-\frac{x^2}{2\sigma_x^2}-\frac{y^2}{2\sigma_y^2}\right)$, with $\sigma_x\kF=4$, $\sigma_y\kF=6$, and $\max[A(t)]/\eF=0.5$. This perturbation increases the total energy by approximately $(E_1-E_0)/E_0\approx 5\%$, where $E_0$ denotes the ground-state energy.}} 
\label{fig:cmp-reg} 
\end{figure}

It should be emphasized that the time-dependent formulation strictly conserves energy only in the limit where $E_c$
is sufficiently large to include all single-particle states. In the case of a cubic lattice, this would require
$E_c \ge 3\frac{k_c^2}{2}$. Otherwise, the Bogoliubov transformation defining the unitary time evolution violates the completeness relation (see discussion in Ref.~\cite{APPBMagierski2018}). As a result, energy ceases to be conserved during long-time evolution~\cite{APPBMagierski2018,Grineviciute2018,arXiv-Bjelcic}. This issue is independent of the specific regularization scheme employed, as it is solely related to the properties of the time-evolution generator. Consequently, the choice of the energy cutoff $E_c$ represents a trade-off between computational cost and the duration over which energy conservation can be maintained. As the cutoff is reduced, the accuracy of energy conservation gradually deteriorates, particularly at longer evolution times.
For this reason, we recommend using the largest cutoff energy permitted by the lattice in practical calculations. 
\respch{
In particular, the box cutoff scheme, sec.~\ref{sec:cubic-cutoff-scheme}, with $E_c$ being consistent with the lattice ultraviolet cutoff scale $k_c=\frac{\pi}{\Delta x}$, assures the highest computational precision in time-dependent runs. 

In Fig.~\ref{fig:cmp-reg}, we show the time evolution of the total energy of a unitary Fermi gas simulated with the SLDA functional. The evolution is initialized from a uniform state, which is then excited by an external Gaussian perturbation switched on and off during the interval $t\eF\in[0,100)$. For $t\eF\ge 100$, the external potential is time independent, and the total energy should therefore be conserved. The perturbation is strong (the amplitude of the Gaussian exceeds the chemical potential), and generates substantial fluctuations in the density and pairing field. The comparison shows that the box regularization scheme provides noticeably better energy conservation than the spherical cutoff scheme.
}

\section{Algorithms}
\subsection{Discretization of the problem and symmetries }
The computation is performed in a discretized Cartesian space. All functions are represented on a spatial mesh of size $N_x \times N_y \times N_z$, with lattice spacings $\Delta x$, $\Delta y$, and $\Delta z$ in the respective directions. A function $f(x,y,z)$ is discretized as
\begin{equation}
f_{lmn}=f(x_0+l\Delta x, y_0+m\Delta y,z_0+n\Delta z)
\end{equation}
with indices $l=0,1,\dots,N_x-1$ (and analogously for $m$ and $n$), and represented as a vector. Without loss of generality, we set $x_0 = y_0 = z_0 = 0$, so that the origin lies at the corner of the computational box. 

Periodic boundary conditions are imposed:
\begin{subequations}
  \begin{align}
    f(x,y,z) &= f(x+L_x, y, z),\\
    f(x,y,z) &= f(x, y+L_y, z),\\
    f(x,y,z) &= f(x, y, z+L_z),
  \end{align}
\end{subequations}
where $(L_x,L_y,L_z)=(N_x\Delta x, N_y\Delta y, N_z\Delta z)$ define the spatial extent of the simulation domain. These boundary conditions enable efficient use of Fourier transforms. The Fourier components are discretized as
\begin{equation}
  k_l^{(d)}=\begin{cases}
      \frac{2\pi}{L_d}l,&l=0,1,\dots,\frac{N_d}{2}-1\,,\\
      \frac{2\pi}{L_d}(l-N_d),&l=\frac{N_d}{2},\frac{N_d}{2}+1,\dots,N_d-1\,,
  \end{cases}
\end{equation}
where $d = x,y,z$, and we assume that the number of lattice points $N_d$ in each direction is even. 

This lattice representation allows for efficient application of spectral methods to compute derivatives~\cite{Shen2011}. The $r$-th derivative is obtained as
\begin{equation}\label{eq:derivative-spectral}
  \frac{\partial^rf}{\partial x^r} = \frac{1}{L_x L_y L_z}\sum_{lmn} \left(ik_l^{(x)}\right)^r \tilde{f}_{lmn}\exp\left[i(k_l^{(x)} x+k_m^{(y)}y+k_n^{(z)}z)\right],
\end{equation}
with analogous expressions for $\frac{\partial^r}{\partial y^r}$ and $\frac{\partial^r}{\partial z^r}$. Here, $\tilde{f}_{lmn}$ denotes the discrete Fourier transform of $f$, which can be computed efficiently using FFT algorithms. For odd derivatives ($r=1,3,\dots$), setting the contribution from the Nyquist mode ($l = N_x/2$) to zero improves numerical accuracy.\footnote{This ensures that, for real $f$, the odd derivative remains real.}

The derivative operator can also be expressed in matrix form:
\begin{equation}\label{eq:derivative-D}
\frac{\partial^rf}{\partial x^r} = \sum_{l^{\prime}m^{\prime}n^{\prime}}D^{(r)}_{lmn,l^{\prime}m^{\prime}n^{\prime}}f_{l^{\prime}m^{\prime}n^{\prime}}.
\end{equation}
For $r=1,2$, the derivative matrices are given by~\cite{Shen2011,PhysRevC.87.051301}:
\begin{subequations}
  \begin{align}
    \label{eq:D1}
    D^{(1)}_{lmn,l^{\prime}m^{\prime}n^{\prime}}
    &=\frac{\pi}{L_x}
      (-1)^{l-l^{\prime}}
      \cot \left[\frac{\pi(l-l^{\prime})}{N_x}\right]
      (1-\delta_{ll^{\prime}})\delta_{mm^{\prime}}\delta_{nn^{\prime}},\\
    \label{eq:D2}
    D^{(2)}_{lmn,l^{\prime}m^{\prime}n^{\prime}}
    &=\frac{2\pi^2}{L_x^2} (-1)^{l-l^{\prime}-1} \sin^{-2}\left[\frac{\pi(l-l^{\prime})}{N_x}\right]
      (1-\delta_{ll^{\prime}})\delta_{mm^{\prime}}\delta_{nn^{\prime}}- \frac{\pi^2}{3\Delta x^2}\left(1+\frac{2}{N_x^2}\right)\delta_{ll^{\prime}}\delta_{mm^{\prime}}\delta_{nn^{\prime}}.
  \end{align}
\end{subequations} 
By analogy, derivative matrices along the $y$ and $z$ directions can be constructed. Importantly, $D^{(1)}$ and $D^{(2)}$ provide an exact discrete representation of the spectral differentiation method. Namely, \cref{eq:derivative-spectral,eq:derivative-D} yield identical results (up to machine precision) for $r=1$ and $r=2$.

\begin{figure}[t]
	\centering
	\includegraphics[width=1.0\linewidth]{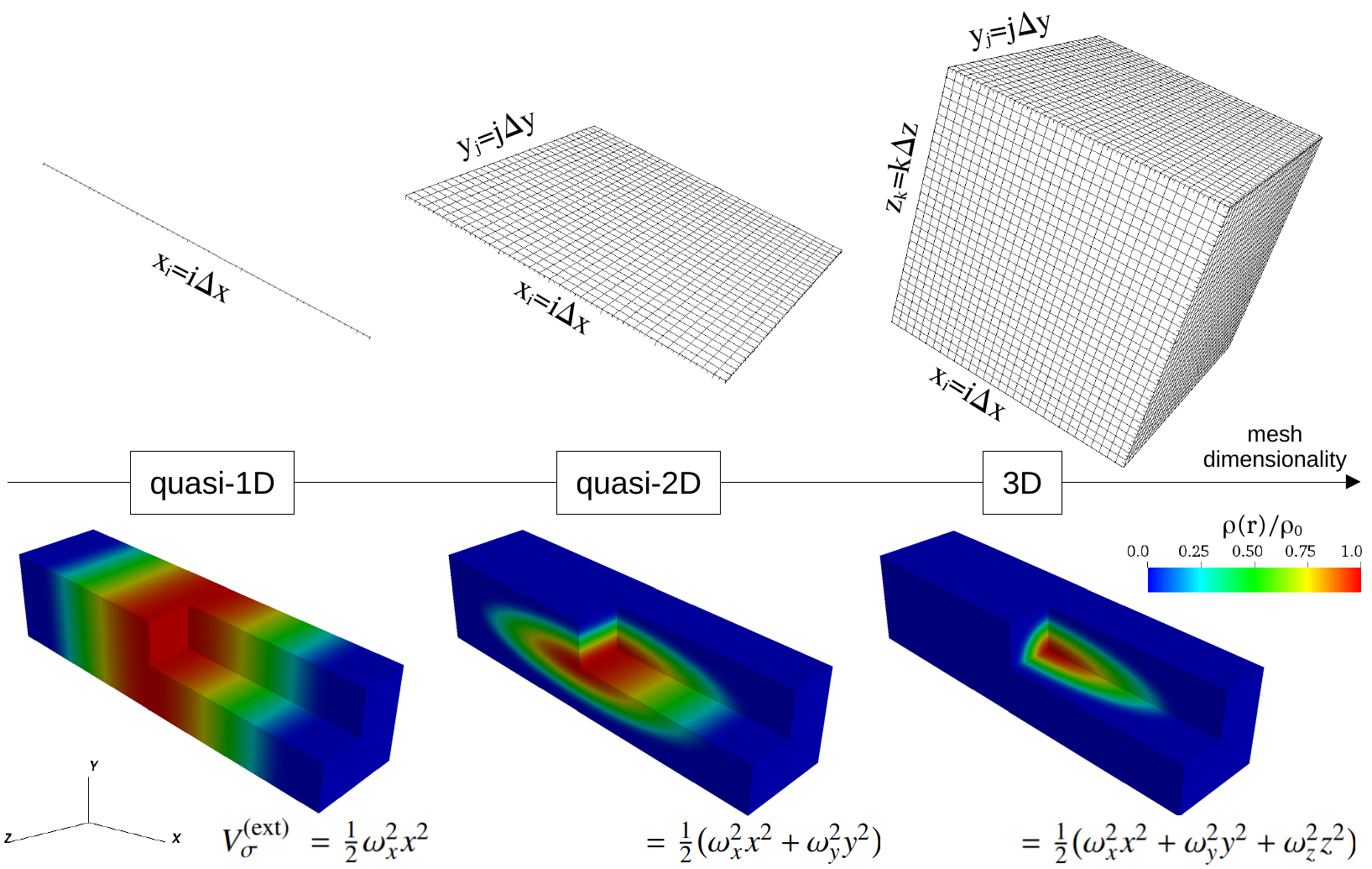}
	\caption{Types of meshes used in the W-SLDA Toolkit. If the system possesses translational symmetries along two directions ($y$ and $z$), a 1D mesh is employed. If the symmetry is present only along one direction ($z$), functions are represented on a 2D mesh. In the absence of such symmetries, a full 3D mesh is used. Examples are shown in the bottom row, where solutions obtained on the lattice $128\times 32\times 32$ are presented for the \respch{quasi-}1D harmonic oscillator potential $\Vext{\sigma}(x)=\frac{1}{2}\omega_x^2 x^2$, the 2D harmonic oscillator potential $\Vext{\sigma}(x,y)=\frac{1}{2}\bigl(\omega_x^2 x^2+\omega_y^2 y^2\bigr)$, and for 3D harmonic potential $\Vext{\sigma}(x,y,z)=\frac{1}{2}\bigl(\omega_x^2 x^2+\omega_y^2 y^2 +\omega_z^2 z^2 \bigr)$. The colormaps show the density distribution normalized to the value in the center of the box. }
	\label{fig:mesh-dim}
\end{figure}
Three-dimensional calculations are typically numerically demanding, as functions are represented by vectors of length $M^{(3)} = N_x N_y N_z$, and the Hamiltonian matrix has dimension $2M^{(3)} \times 2M^{(3)}$. Although the toolkit supports fully unrestricted 3D calculations, the computational cost can be substantially reduced by exploiting translational symmetries.
\respch{It is important to note that while the code can simulate strictly 1D and 2D systems, the built-in functionals are only calibrated and designed for 3D 
systems.  Thus, one should think here always of \respch{quasi-}1D and \respch{quasi-}2D representing 3D systems with translational invariance.}

The W-SLDA codes can take advantage of the following cases:
\begin{description}
\item[quasi-2D]
  \begin{subequations}
    Densities and potentials depend only on two spatial coordinates, e.g.\ $\rho_{\sigma}(\vbr) \rightarrow \rho_{\sigma}(x,y)$. This situation commonly arises when the external potentials are independent of the third ($z$) direction, although this condition alone is not sufficient, as solutions of the equations of motion may spontaneously break the symmetry. Imposing translational symmetry along $z$ amounts to assuming that the Bogoliubov quasiparticle wave functions have the generic form
    \begin{equation}\label{eq:wf-quasi2D}
      \varphi_{n}(\vbr)\rightarrow\phi_{ln}(x,y)\frac{1}{\sqrt{L_z}}e^{ik_n^{(z)}z},
    \end{equation}
    labeled by two quantum numbers $l$ and $n$. In practice, only the functions $\phi_{ln}(x,y)$ need to be represented numerically, on an $N_x \times N_y$ lattice, while all operations related to the $z$ direction can be carried out analytically. When $N_z = 1$, the $k_z$ vector can take only the single value $k_z = 0$, in which case the W-SLDA code performs strict 2D calculations. For $N_z > 1$, the code includes all $k_n^{(z)}$ modes with $n = 0,1,\dots,N_z-1$.
    
    The derivatives with respect to the $z$ coordinate can be evaluated analytically. The densities \crefrange{eq:normal-density}{eq:anomalous-density} are given by
    \begin{align}
      \rho_{\sigma}(x,y) &= \frac{1}{L_z}\sum_{\abs{E_{ln}}<E_c}\abs{v_{ln,\sigma}(x,y)}^2 f_{\beta}(-E_{ln}),\\
      \tau_{\sigma}(x,y) &= \frac{1}{L_z}\sum_{\abs{E_{ln}}<E_c}\left[ \abs{k_n^{(z)}v_{ln,\sigma}(x,y)}^2+ \abs{\vb{\nabla}_{\textrm{2d}}v_{ln,\sigma}(x,y)}^2\right] f_{\beta}(-E_{ln}),\\
      \vbj_\sigma(x,y) &= \frac{1}{L_z}\sum_{\abs{E_{ln}}<E_c} \Im[v_{ln,\sigma}(x,y)\vb{\nabla}_{\textrm{2d}} v_{ln,\sigma}^*(x,y)] f_{\beta}(-E_{ln}),\\
      \nu(x,y) &= \frac{1}{2L_z}\sum_{\abs{E_{ln}}<E_c} \left [u_{ln,a}(x,y)v_{ln,b}^{*}(x,y)
                 -u_{ln,b}(x,y)v_{ln,a}^{*}(x,y)\right ]f_{\beta}(-E_{ln}),
    \end{align}
    with the $z$-component of the current identically vanishing (the formulas can be further simplified if the spin-symmetric case is assumed). The quasiparticle wave functions are solutions of \cref{eqn:HpsiEpsi} under the substitutions:
    \begin{align}
      \vbr \rightarrow & (x,y),\nonumber\\
      \vb{\nabla} \rightarrow & \vb{\nabla}_{\textrm{2d}}=[\frac{\partial}{\partial x}, \frac{\partial}{\partial y}],\nonumber\\
      \mu_\sigma\rightarrow & \mu_\sigma - \frac{1}{2}\alpha_\sigma(x,y)\left(k_n^{(z)}\right)^2,\nonumber\\
      \textvisiblespace_n\rightarrow &\textvisiblespace_{ln}\nonumber,
    \end{align}
    In the quasi-2D case, one must handle $N_z$ Hamiltonian matrices (one for each $k_n^{(z)}$), each of dimension $2M^{(2)}\times 2M^{(2)}$, where $M^{(2)} = N_x N_y$. In practice, the computational cost can be further reduced by half, since the contributions to the densities from $+k^{(z)}$ and $-k^{(z)}$ are identical; thus, it suffices to include only modes with $k_n^{(z)} \geq 0$. 
  \end{subequations}
\item[quasi-1D]
  \begin{subequations}
    Analogously, observables are assumed to depend only on a single coordinate, e.g.\ $\rho_{\sigma}(\vbr) \rightarrow \rho_{\sigma}(x)$. The Bogoliubov quasiparticle wave functions are then taken as
    \begin{equation}\label{eq:wf-quasi1D}
      \varphi_{n}(\vbr)\rightarrow\phi_{lmn}(x)\frac{1}{\sqrt{L_y}}e^{ik_m^{(y)}y}\frac{1}{\sqrt{L_z}}e^{ik_n^{(z)}z},
    \end{equation} 
    and are labeled by the quantum numbers $l$, $m$, and $n$. When $N_y = N_z = 1$, the problem reduces to a strict 1D case. Otherwise, all modes with $m = 0,1,\dots,N_y-1$ and $n = 0,1,\dots,N_z-1$ are included in the simulations.
    
    Analogously to the quasi-2D case, the densities are given by
    \begin{align}
      \rho_{\sigma}(x)
      &= \frac{1}{L_y L_z}\sum_{\abs{E_{lmn}}<E_c}\abs{v_{lmn,\sigma}(x)}^2 f_{\beta}(-E_{lmn})\,,\\
      \tau_{\sigma}(x)
      &= \frac{1}{L_y L_z}\sum_{\abs{E_{lmn}}<E_c}\left[ \abs{k_m^{(y)}v_{lmn,\sigma}(x)}^2+ 
        \abs{k_n^{(z)}v_{lmn,\sigma}(x)}^2 +\abs{\vb{\nabla}_{\textrm{1d}}v_{lmn,\sigma}(x)}^2\right] f_{\beta}(-E_{lmn})\,,\\
      \vbj_\sigma(x)
      &= \frac{1}{L_y L_z}\sum_{\abs{E_{lmn}}<E_c} \Im[v_{lmn,\sigma}(x)\vb{\nabla}_{\textrm{1d}} v_{lmn,\sigma}^*(x)] f_{\beta}(-E_{lmn})\,,\\
      \nu(x)
      &= \frac{1}{2L_y L_z}\sum_{\abs{E_{lmn}}<E_c} \left [u_{lmn,a}(x)v_{lmn,b}^{*}(x)-u_{lmn,b}(x)v_{lmn,a}^{*}(x)\right ]f_{\beta}(-E_{lmn})\,,
    \end{align}
    and further simplification can be applied for a spin-symmetric system. The substitutions are:
    \begin{align*}
      \vbr \rightarrow & x\,,\\
      \vb{\nabla} \rightarrow & \vb{\nabla}_{\textrm{1d}}=\frac{d}{d x}\,,\\
      \mu_\sigma\rightarrow & \mu_\sigma 
                              - \frac{1}{2}\alpha_\sigma(x)\left(k_m^{(y)}\right)^2
                              - \frac{1}{2}\alpha_\sigma(x)\left(k_n^{(z)}\right)^2\,,\\
      \textvisiblespace_n\rightarrow &\textvisiblespace_{lmn}\,.
    \end{align*}
    In this case, one deals with $N_y N_z$ Hamiltonian matrices, each of size $2N_x\times 2N_x$.
    The computational cost can be further reduced by considering only unique values of $\big(k^{(y)}\big)^2 + \big(k^{(z)}\big)^2$.
  \end{subequations}
\end{description}

The supported computational modes (full 3D, quasi-2D, and quasi-1D) are illustrated in \cref{fig:mesh-dim}.
In the full 3D mode, functions are represented as vectors of length $M^{(3)} = N_x N_y N_z$, meaning that a value is assigned to every lattice point in the 3D grid.
In the quasi-2D and quasi-1D modes, the dimensionality of the representation is reduced according to the assumed translational symmetries of the system.
Specifically, in the quasi-2D case, functions are stored as vectors of length $M^{(2)} = N_x N_y$, corresponding to a single $x$–$y$ plane, while in the quasi-1D case, the representation is reduced further to vectors of length $M^{(1)} = N_x$, with values defined only along the $x$ axis.
As an example, \cref{fig:mesh-dim} (bottom row) presents the 3D density profiles obtained for an atomic gas in the unitary regime confined in one-, two-, and three-dimensional harmonic oscillator potentials, all computed using lattices of identical size.
This visualization highlights how the same physical system can be efficiently modeled using reduced-dimensional modes when translational symmetries are present.

It is important to note that the regularization procedure described in \cref{sec:regularization} is valid only in 3D. For strict 1D or 2D calculations, it must be modified accordingly, which lies outside the scope of this work. Nevertheless, the toolkit provides an interface that allows users to implement custom regularization schemes. 

\respch{
Note that the current implementation supports only periodic boundary conditions. This choice is tied to the use of FFT-based spectral derivatives, which provide high numerical accuracy and computational efficiency. The interparticle interaction is short-ranged for cold atoms, and thus the effect of interaction of the simulation domain with its images is usually strongly suppressed. 
In applications to trapped ultracold atomic gases, the simulation box should therefore be chosen large enough that the density, pairing field, and relevant currents are strongly suppressed near the boundaries. External potentials should also be smoothed near the edges of the computational domain to avoid artifacts arising from boundary discontinuities. Other types of boundary conditions are not implemented in the current version and are left for future extensions of the toolkit.
}

\subsection{Static problems}
The static problem requires finding solutions of \cref{eqn:HpsiEpsi}.
This is a nonlinear problem, since the potentials ($V_\sigma$, $\Delta$, $\vbA_{\sigma}$, and $\alpha_\sigma$) depend on the densities ($\rho_{\sigma}$, $\nu$, $\vbj_\sigma$, and $\tau_{\sigma}$), which in turn depend on the quasiparticle wave functions $\varphi_n = [u_{n,a}, v_{n,b}]^T$. 
It is not sufficient to simply construct and diagonalize the Hamiltonian matrix.
The problem \cref{eqn:HpsiEpsi} must be solved self-consistently.
In fact, it can be cast as a \textit{fixed-point problem}
\begin{equation}
  \label{eq:fixed-point-problem}
  K[\vb{p}]=\vb{p},
\end{equation}
where $\vb{p}=[V_\sigma, \alpha_\sigma, \Delta, \vbA_{\sigma}]$ denotes the set of potentials in shorthand notation, and the map $K$ represents the sequence of operations:

\begin{algorithm}[h!]
  \begin{algorithmic}[1]
  \item Construct the Hamiltonian matrix [\cref{eqn:HpsiEpsi}]
  \item Diagonalize the Hamiltonian matrix: $\rightarrow \{(u_{n,a},v_{n,b}), E_n\}$
  \item Compute densities [\crefrange{eq:normal-density}{eq:anomalous-density}]:
    $\{(u_{n,a},v_{n,b}), E_n\} \rightarrow
     [\rho_{\sigma}^{}, \tau_{\sigma}^{}, \nu^{}, \vbj_\sigma^{}]$
  \item Compute potentials: $[\rho_{\sigma}, \tau_{\sigma}, \nu, \vbj_\sigma] \rightarrow [V_\sigma^{\mathrm{out}}, \alpha_\sigma^{\mathrm{out}}, \Delta^{\mathrm{out}}, \vbA_{\sigma}^{\mathrm{out}}]$
  \item {\bf return} $[V_\sigma^{\mathrm{out}}, \alpha_\sigma^{\mathrm{out}}, \Delta^{\mathrm{out}}, \vbA_{\sigma}^{\mathrm{out}}]$
  \end{algorithmic}
  \caption{Pseudo-code for the fixed-point map $K[\vbp]$: $[V_\sigma^{\mathrm{in}}, \alpha_\sigma^{\mathrm{in}}, \Delta^{\mathrm{in}}, \vbA_{\sigma}^{\mathrm{in}}]
  \mapsto
  [V_\sigma^{\mathrm{out}}, \alpha_\sigma^{\mathrm{out}}, \Delta^{\mathrm{out}}, \vbA_{\sigma}^{\mathrm{out}}]$.}
\label{alg:MapK}  
\end{algorithm}
\noindent
The self-consistent solution is defined by the condition
\begin{equation}
  \abs{\vb{p}^{\textrm{in}}-\vb{p}^{\textrm{out}}}<\varepsilon
\end{equation}
where $\varepsilon$ specifies the computational accuracy (an algorithmic parameter).
In practice, however, it is more convenient to monitor the convergence of quantities derived from the self-consistent fields rather than the fields themselves.
Among such quantities, energy-based indicators are particularly useful, as their physical meaning allows for an intuitive estimation of acceptable tolerances. 

\begin{subequations}
  The following contributions to the intrinsic energy are defined:
  \begin{description}
  \item[(Intrinsic) Kinetic Energy]
    \begin{equation}
      E_{\textrm{kin}} = \int\sum_\sigma  \frac{A_{\sigma}(\vbr)}{2}\left(\tau_{\sigma}(\vbr)-\frac{\vbj_\sigma^2(\vbr)}{\rho_{\sigma}(\vbr)}\right)\,d\vbr,
    \end{equation} 
  \item[Potential Energy]
    \begin{equation}
      E_{\textrm{pot}} = \int \frac{3}{5}B(\vbr) \rhoP(\vbr) \eF(\vbr)\,d\vbr,
    \end{equation}
  \item[Pairing Energy]
    \begin{equation}
      E_{\textrm{pair}} = 
      -\int \Delta^*(\vbr) \nu(\vbr)\, d\vbr,
    \end{equation}
  \item[Current (Flow) Energy]
    \begin{equation}
      E_{\textrm{curr}} = \int \sum_\sigma  \frac{\vbj_\sigma^2(\vbr)}{2\rho_{\sigma}(\vbr)}\,d\vbr.
    \end{equation}
  \end{description}
  Note that the kinetic energy $E_{\textrm{kin}}$ and pairing energy $E_{\textrm{pair}}$ are formally divergent as one takes the cutoff $E_c \rightarrow \infty$, but the combination $E_{\textrm{kin}} + E_{\text{pair}}$ remains finite. Thus, for physics interpretation, one should always consider their sum, whereas for checking the convergence of the algorithm, we check them separately, as each is sensitive to variations across different potentials and densities. 
  
  In addition, external potentials contribute as follows:
  \begin{description}
  \item[External Potential Energy]
    \begin{equation}
      E_{\textrm{pot.ext}} = \int \sum_{\sigma}\Vext{\sigma}(\vbr)\rho_{\sigma}(\vbr)\,d\vbr,
    \end{equation}
  \item[External Pairing Energy]
    \begin{equation}
      E_{\textrm{pair.ext}} = -\int \left(\Dext(\vbr)\nu^*(\vbr)+\textrm{h.c.}\right)\,d\vbr,
    \end{equation}
  \item[External Flow Energy]
    \begin{equation}
      E_{\textrm{vel.ext}} = -\int \sum_{\sigma} \velext{\sigma} (\vbr)\cdot\vbj_{\sigma}(\vbr)\,d\vbr,
    \end{equation}
  \end{description}
\end{subequations}
The total energy of the system is obtained as the sum of all these contributions.
Within the W-SLDA Toolkit, the iterative process is considered converged when the relative change of each energy term (as well as of the total energy) between two consecutive iterations satisfies
\begin{equation}\label{eq:E-conv}
    \frac{\abs{E_{\textrm{label}}^{(i)}-E_{\textrm{label}}^{(i-1)}}}{\Effg}<\varepsilon_{\textrm{energy}},
\end{equation}
where $\varepsilon_{\textrm{energy}}$ denotes the convergence threshold (typically set to $\varepsilon_{\textrm{energy}} = 10^{-6}$), and $\Effg$ is the Fermi energy of the reference uniform system.
This criterion ensures rejection of cases where individual terms change from iteration to iteration (meaning that densities and potentials also change), but the changes cancel out such that the total energy remains unchanged.

We note that, from a formal perspective, it is equivalent to search for densities ($\rho_{\sigma}$, $\tau_{\sigma}$, $\nu$, $\vbj_\sigma$) or for potentials ($V_\sigma$, $\alpha_\sigma$, $\Delta$, $\vbA_{\sigma}$) that remain unchanged under the map. In practice, however, the choice affects the convergence properties of the iterative algorithm. Our experience shows that the definition of the solution vector in terms of potentials more often yields superior convergence compared to the density-based definition. For this reason, the potential-based formulation is set as the default option in W-SLDA. 

A variety of methods exist for solving the fixed-point problems; see Ref.~\cite{Woods2019} for a review. The W-SLDA Toolkit supports the following approaches:
\begin{description}
 \item[Linear mixing] the most standard method. The $(n+1)$-th approximation of the solution is constructed as
 \begin{equation}
 \vb{p}_{n+1}=m_p K[\vb{p}_{n}] + (1-m_p)\vb{p}_{n},
 \end{equation}
with $m_p \in [0,1]$ the mixing parameter. For any problem, there exists a value of $m_p$ that ensures convergence, although this value is not known \textit{a priori}~\cite{Ryu2016}. As a rule of thumb, $m_p \approx 0.5$ is chosen; if the algorithm fails to converge, $m_p$ should be decreased.

\item[Broyden algorithm] reformulates the problem as finding the root of
  \begin{equation}
    R[\vb{p}]\equiv K[\vb{p}]-\vb{p}=0.
  \end{equation}
  The classical Broyden method requires the Jacobian matrix of the nonlinear equations, the computation of which is numerically costly.
  However, it has been shown that an approximate Jacobian, constructed from the last $k$ iterations, is sufficient to accelerate convergence.
  This approach was applied in Refs.~\cite{PhysRevB.30.6118,PhysRevB.38.12807,Eyert1996}.
 The W-SLDA Toolkit implements a variant of the method tailored for strongly interacting systems, described in Ref.~\cite{PhysRevC.78.014318}.
\end{description}

Depending on the problem, the solution must be obtained either for fixed values of the chemical potentials $\mu_\sigma$ or for fixed particle numbers $N_\sigma$. In the latter case, the chemical potentials are updated at each iteration according to the prescription (an additional step in the map $K[\vb{p}]$):
\begin{equation}\label{eq:update-mu}
\mu_\sigma^{\textrm{out}} = \mu_\sigma^{\textrm{in}} + \frac{a_\sigma^{(\mu)}}{N_\sigma^{\textrm{req.}}}\left[N_\sigma^{\textrm{req.}}-\int \rho_\sigma(\vbr)d\vbr\right],
\end{equation}
where $N_\sigma^{\textrm{req.}}$ denotes the target number of particles for species $\sigma$, and $a_\sigma^{(\mu)}$ is an algorithmic parameter controlling the step size of the adjustment.
In practice, values of $a_\sigma^{(\mu)}$ in the range $0.1$-$1$ are typically effective. When solving for a fixed particle number, an additional convergence criterion is imposed:
\begin{equation}\label{eq:N-conv}
\abs{N_\sigma^{\textrm{req.}} - N_{\sigma}^{(i)}} < \varepsilon_{\textrm{npart}},
\end{equation}
where $\varepsilon_{\textrm{npart}}$ is the tolerance parameter (typically set to $10^{-6}$), and $N_{\sigma}^{(i)}$ denotes the particle number obtained at the $i$-th iteration.
Setting $a_\sigma^{(\mu)} = 0$ and $\varepsilon_{\textrm{npart}} \to \infty$ corresponds to performing calculations at fixed chemical potential.

\respch{
We note that convergence criteria based on changes in the individual energy components between successive iterations do not, by themselves, guarantee that a true fixed point has been reached. For example, when the mixing parameter $m_p$ is chosen to be very small, the algorithm may advance only slowly through configuration space, leading to correspondingly small changes in the energy components even though the residuals $R[\vb{p}]$ remain appreciable. Users should therefore verify that the residuals, reported in the \texttt{stdout} file, have decreased to a level appropriate for the problem under consideration before accepting the solution as a genuine fixed point.
}

\begin{subequations}
  To construct the matrix representation of the Hamiltonian, which is diagonalized during execution of the map $K$, note that local fields are represented by diagonal matrices:
  \begin{equation}
    \bigl(A(\vbr)\bigr)_{lmn,l'm'n'}=A(l\Delta x,m\Delta y,n\Delta z)\delta_{ll'}\delta_{mm'}\delta_{nn'}.
  \end{equation}
  The only non-diagonal contributions arise from the derivative operators, whose representations have already been given in \cref{eq:D1,eq:D2}. The matrix representation of the product of a local field with a derivative operator is therefore
  \begin{align}
    \label{eq:Ad_dx}
    \left(A(\vbr)\frac{\partial}{\partial x}\right)_{lmn,l'm'n'} =
    &\sum_{l_1 m_1 n_1}\left(A(\vbr)\right)_{lmn,l_1 m_1 n_1}\left(\frac{\partial}{\partial x}\right)_{l_1 m_1 n_1,l'm'n'}
      =A(l\Delta x,m\Delta y,n\Delta z)D^{(1)}_{lmn,l'm'n'},
  \end{align}
  and analogously,
  \begin{align}
    \label{eq:d_dxA}
    \left(\frac{\partial}{\partial x}A(\vbr)\right)_{lmn,l'm'n'}
    =&D^{(1)}_{lmn,l'm'n'}A(l'\Delta x,m'\Delta y,n'\Delta z).
  \end{align}
\end{subequations}

Such terms arise in the single-particle Hamiltonian due to the explicit dependence of the energy functional on the currents $\vbj_{\sigma}$. From the standpoint of both numerical accuracy and computational efficiency, it is advantageous to rewrite products of first derivatives in the kinetic operator as
  \begin{equation}
    \frac{\partial}{\partial x}\alpha(\vbr)\frac{\partial}{\partial x}\varphi(\vbr)=\frac{1}{2}\left(
\frac{\partial^2[\alpha(\vbr)\varphi(\vbr)]}{\partial x^2}
+\alpha(\vbr)\frac{\partial^2\varphi(\vbr)}{\partial x^2}
-\frac{\partial^2\alpha(\vbr)}{\partial x^2}\varphi(\vbr)
\right).
\label{eq:kinop-alpha}
\end{equation}
This prescription yields higher numerical accuracy compared to the standard approach of computing successive first derivatives. The matrix representations of the terms contributing to the kinetic energy are analogous to \cref{eq:Ad_dx,eq:d_dxA}, with the difference that the $D^{(2)}$ matrix is used instead of $D^{(1)}$. 

\subsection{Integration of time-dependent problems}
\label{subsec:integation-td}
The time-dependent problem \cref{eqn:dpsidt-Hpsi} can be written in shorthand notation as
\begin{equation}\label{eq:shorttdproblem}
\I\frac{\partial \varphi_{n}(\vbr,t)}{\partial t} = \calH(\vb{\varphi},t)\varphi_{n}(\vbr,t),
\end{equation}
where $\calH(\vb{\varphi},t)$ is the Hamiltonian matrix, which depends on all quasiparticle states $\vb{\varphi}\equiv [\varphi_1, \varphi_2, ...]$ and on time (through the external potentials).
The formal solution is 
\begin{equation}
  \varphi_{n}(\vbr,t) = \calT \exp\left[-\I\int_{0}^{t'}\calH(\vb{\varphi},t')dt' \right]\varphi_{n}(\vbr,0),
\end{equation}
where $\calT$ denotes the time-ordering operator.

In the simplest case, where the Hamiltonian is time-independent, i.e., no time-dependent external potentials, $\calH(\vb{\varphi},t)\rightarrow \calH(\vb{\varphi})$, and the initial state $\varphi_{n}(\vbr,0)$ is the self-consistent solution of the associated static problem, the quasiparticle states evolve as
\begin{equation}
  \varphi_{n}(\vbr,t)=\exp(-\I E_n t)\varphi_{n}(\vbr,0).
\end{equation}
This demonstrates that the evolution involves both slowly and rapidly oscillating in time wave functions, since the quasiparticle energies span the interval $[-E_c,E_c]$ with $E_c \gg \eF$.
As a result, the integration becomes numerically demanding.
The toolkit employs algorithms with a fixed integration step size $\Delta t$.
To avoid the need for prohibitively small time steps, required to resolve high-frequency oscillations, we modify the equations of motion to the following form:
\begin{subequations}
  \begin{equation}
    \label{eq:shorttdproblem-mod}
    \I\frac{\partial \varphi_{n}(\vbr,t)}{\partial t} =\left[ \calH(\vb{\varphi},t)-\avg{\calH(\vb{\varphi},t)}_n\right]\varphi_{n}(\vbr,t),
  \end{equation}
  where 
  \begin{equation}
    \avg{\calH(\vb{\varphi},t)}_n = \int \varphi_n^{\dagger}(\vbr,t)\calH(\vb{\varphi},t)\varphi_n(\vbr,t)d\vbr,
  \end{equation}
\end{subequations}
is the \textit{instantaneous quasiparticle energy}.
The energy shift $\avg{\calH}_n$ suppresses the phase factor.
For example, in the stationary case one recovers simply  $\varphi_{n}(\vbr,t)=\varphi_{n}(\vbr,0)$. The solutions of \cref{eq:shorttdproblem,eq:shorttdproblem-mod} differ only by an overall phase factor, $\exp\left[-\I\int_{0}^{t'}\avg{\calH(\vb{\varphi},t')}_n dt'\right]$ which cancels out when computing the densities [\crefrange{eq:normal-density}{eq:anomalous-density}].
Although computing the instantaneous energy shift introduces additional numerical cost, it substantially improves the stability of the time integration and enables the use of significantly larger time steps compared to the unmodified formulation.

For time integration, we employ the multistep Adams-Bashforth-Moulton (ABM) method.
Multistep schemes typically offer higher stability than single-step methods (such as Runge-Kutta), at the expense of increased memory consumption, since several past states of the system must be stored.
To formulate the ABM algorithm, we rewrite the equation of motion~\cref{eq:shorttdproblem-mod} as
\begin{subequations}
  \begin{equation}
    \dot{\vb{\varphi}} = \vbf(\vb{\varphi},t),
  \end{equation}
  with
  \begin{equation}\label{eq:ABM-f}
    f(\vb{\varphi},t)=-\I\left[ \calH(\vb{\varphi},t)-\avg{\calH(\vb{\varphi},t)}_n\right]\vb{\varphi}(\vbr,t).
  \end{equation}
\end{subequations}
We define $\vb{\varphi}_k \equiv \vb{\varphi}(k\Delta t)$ and $\vbf_k \equiv \vbf(\vb{\varphi}_k,k\Delta t)$, where $\Delta t$ is the integration time step. 
The ABM scheme consists of a predictor step (Adams-Bashforth) and a corrector step (Adams-Moulton). The predictors are:
\begin{description}
\item[AB3] (3rd order)
  \begin{equation}
    \label{eq:AB3}
    \vb{\varphi}_k^{(p)} = \vb{\varphi}_{k-1} + \dfrac{23}{12}\Delta t \vbf_{k-1} - \dfrac{16}{12}\Delta t \vbf_{k-2} + \dfrac{5}{12}\Delta t \vbf_{k-3},
  \end{equation}
\item[AB4] (4-th order)
  \begin{equation}
    \label{eq:AB4}
    \vb{\varphi}_k^{(p)} = \vb{\varphi}_{k-1} + \dfrac{55}{24}\Delta t \vbf_{k-1} - \dfrac{59}{24}\Delta t \vbf_{k-2} + \dfrac{37}{24}\Delta t \vbf_{k-3} - \dfrac{9}{24}\Delta t \vbf_{k-4},
  \end{equation}
\item[AB5] (5-th order)
  \begin{align}
    \label{eq:AB5}
    \vb{\varphi}_k^{(p)} = \vb{\varphi}_{k-1}  + \dfrac{1901}{720}\Delta t \vbf_{k-1} - \dfrac{2774}{720}\Delta t \vbf_{k-2} 
    + \dfrac{2616}{720}\Delta t \vbf_{k-3} - \dfrac{1274}{720}\Delta t \vbf_{k-4} + \dfrac{251}{720}\Delta t \vbf_{k-5}.
  \end{align}
\end{description}
The considered correctors are:
\begin{description}
\item[AM4] (4-th order)
  \begin{align}
    \label{eq:AM4}
    \vb{\varphi}_k = \vb{\varphi}_{k-1}  + \dfrac{9}{24}\Delta t \vbf(\vb{\varphi}_k^{(p)},k\Delta t)  
    + \dfrac{19}{24}\Delta t \vbf_{k-1}
    -\dfrac{5}{24}\Delta t \vbf_{k-2}+\dfrac{1}{24}\Delta t \vbf_{k-3},
  \end{align}
\item[AM5] (5-th order)
  \begin{align}
    \label{eq:AM5}
    \vb{\varphi}_k = \vb{\varphi}_{k-1} + \dfrac{251}{720}\Delta t \vbf(\vb{\varphi}_k^{(p)},k\Delta t) + \dfrac{646}{720}\Delta t \vbf_{k-1} 
    -\dfrac{264}{720}\Delta t \vbf_{k-2}+\dfrac{106}{720}\Delta t \vbf_{k-3} - \dfrac{19}{720}\Delta t \vbf_{k-4}.
  \end{align}
\end{description}
The integrator combines predictor and corrector as AB$x$AM$y$, where $x$ and $y$ denote their respective orders. The toolkit allows the user to choose between AB3AM4, AB4AM5, and AB5AM5. These integrators have been tested in applications to DFT equations~\cite{Rehn2019}, and their stability was analyzed in Ref.~\cite{Ghrist2014}.

\respch{
According to our experience, $\Delta t\eF=0.0035$ is a reasonable choice for the integration time step and is set as the default value.
The default value of the integration time step, should be understood as a practical recommendation based on experience accumulated in applications of the toolkit, summarized in Sec.~\ref{sec:applications}. It is not a universal stability bound. The optimal time step depends on the interaction regime, the regularization scheme, the time dependence of external fields, and the characteristic excitation energies of the system. Therefore, users should verify the stability and convergence of their results by repeating representative calculations with smaller values of $\Delta t$.
}

Among these, AB5AM5 provides the highest accuracy but is also the most memory demanding.
It requires six buffers to operate: $\vbf_{k-1}$, $\vbf_{k-2}$, $\vbf_{k-3}$, $\vbf_{k-4}$, $\vbf_{k-5}$ (historical states), and one additional buffer for storing the solution $\vb{\varphi}_{k-1}$.
Since each buffer may be large, the memory cost can be significant.
Thus, in addition to integration accuracy, efficient memory management is a critical factor.
The memory demand can be reduced to five buffers (AB4AM5) or four buffers (AB3AM4), at the cost of reduced integration accuracy.
In practice, lowering the order of the predictor-corrector scheme also requires decreasing the integration time step $\Delta t$ to keep numerical errors under control.
Therefore, users can trade memory consumption against the integration step size, depending on the specifications of the computing system.
\Cref{alg:AB4AM5} illustrates the memory management for the AB4AM5 integrator; the other schemes are implemented in an analogous manner.

\begin{algorithm}[h!]
  \verb|Integrator_AB4AM5|[$\vb{\varphi}_{k-1}$, $\vbf_{k-1}$, $\vbf_{k-2}$, $\vbf_{k-3}$, $\vbf_{k-4}$]
  \begin{algorithmic}[1]
  \item Update states
    \Comment{Operations in brackets must be executed simultaneously.}
    \begin{flalign*}
      &\begin{bmatrix}
        \vb{\varphi}_{k-1} + \frac{55}{24}\Delta t \vbf_{k-1} - \frac{59}{24}\Delta t \vbf_{k-2} + \frac{37}{24}\Delta t \vbf_{k-3} - \frac{9}{24}\Delta t \vbf_{k-4}\\
        \vb{\varphi}_{k-1} +\frac{646}{720}\Delta t \vbf_{k-1}-\frac{264}{720}\Delta t \vbf_{k-2}+\frac{106}{720}\Delta t \vbf_{k-3} - \frac{19}{720}\Delta t \vbf_{k-4}
      \end{bmatrix}
        \rightarrow
        \begin{bmatrix}
          \vb{\varphi}_{k-1}\\
          \vbf_{k-4}
        \end{bmatrix}&&
    \end{flalign*}
    \Comment{This can be done with two temporary registers in the loop.}
  \item Normalize wave-function stored in $\vb{\varphi}_{k-1}$
  \item $\vbf(\vb{\varphi}_{k-1},k\Delta t) \rightarrow \vb{\varphi}_{k-1}$
    \Comment{This is the computationally intensive step}
  \item $\frac{251}{720}\Delta t \vb{\varphi}_{k-1} + \vbf_{k-4}\rightarrow \vb{\varphi}_{k-1}$
  \item $\vbf_{k-3} \rightarrow \vbf_{k-4}$
  \item $\vbf_{k-2} \rightarrow \vbf_{k-3}$
  \item $\vbf_{k-1} \rightarrow \vbf_{k-2}$
  \item Normalize wave-function stored in $\vb{\varphi}_{k-1}$
  \item $\vbf(\vb{\varphi}_{k-1},k\Delta t) \rightarrow \vbf_{k-1}$
    \Comment{computationally intensive step}
  \item $k+1\rightarrow k$
    \Comment{Now buffer $\vb{\varphi}_{k-1}$ has state propagated forward by $\Delta t$}
  \end{algorithmic}
  \caption{Pseudo-code that evolves state $\vb{\varphi}$ by time interval $\Delta t$.}
  \label{alg:AB4AM5}
\end{algorithm}

\begin{figure}[t]
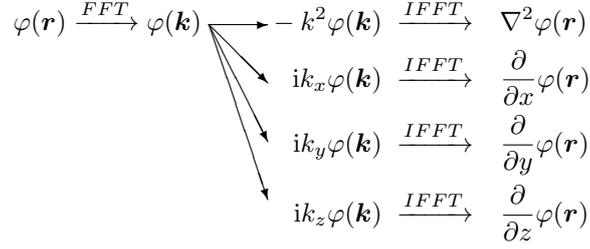

	\centering
  \begin{alignat*}{4}
    \varphi(\vbr) \xrightarrow{FFT} \varphi(\vbk)\;
    \smash{\raisebox{2pt}{\vector(1,0){22}%
                          \llap{\vector(1,-1){22}}%
                          \llap{\vector(1,-2){22}}%
                          \llap{\vector(1,-3){22}}}}
    -k^2&\varphi(\vbk) &\;&\xrightarrow{IFFT} &\quad &\nabla^2\varphi(\vbr)\\
    \I k_x&\varphi(\vbk) &&\xrightarrow{IFFT} &&\frac{\partial}{\partial x}\varphi(\vbr)\\
    \I k_y&\varphi(\vbk) &&\xrightarrow{IFFT} &&\frac{\partial}{\partial y}\varphi(\vbr)\\
    \I k_z&\varphi(\vbk) &&\xrightarrow{IFFT} &&\frac{\partial}{\partial z}\varphi(\vbr)
  \end{alignat*}
	\caption{Schematic illustration of the computation of gradients and the Laplacian of quasiparticle wave functions.
    The procedure involves one forward Fourier transform and four inverse Fourier transforms.
    This approach increases memory requirements, as intermediate results must be stored in auxiliary buffers.
  }
  \label{fig:td-derivative}
\end{figure}

The most time-consuming step is the evaluation of $f(\vb{\varphi},t)$, since it requires applying the Hamiltonian to the wave functions. This, in turn, involves the computation of spatial derivatives (gradients and Laplacians), which constitute the main computational cost.
Spatial derivatives are computed using spectral methods: Fast Fourier Transforms (FFT) are employed to switch between coordinate and momentum spaces. The procedure is illustrated schematically in \cref{fig:td-derivative}. Computing derivatives requires a total of five Fourier transforms (one forward and four inverse) and four additional buffers for intermediate storage (one for the Laplacian and three for gradients).
Moreover, evaluating the kinetic operator [\cref{eq:kinop-alpha}] requires the Laplacian of $\alpha \vb{\varphi}$, which adds two more Fourier transforms and one additional storage buffer. Altogether, a single application of the Hamiltonian to the quasiparticle wave functions requires seven Fourier transforms and five buffers for temporary results. This computational burden can be significantly reduced for the functional variant with $A_{\sigma} = 1$, where only the Laplacian needs to be computed (see \cref{sec:EDFs}).
In this case, the expensive calculation of gradients and of the Laplacian of the product $\alpha \vb{\varphi}$ can be omitted entirely, resulting in a substantial speed-up of the algorithm.

To initialize the ABM algorithm, one must provide historical values of
\begin{equation}
    \vbf_k = -\I\left[ \calH(\vb{\varphi},t_k)-\avg{\calH(\vb{\varphi},t_k)}\right]\vb{\varphi}(\vbr,t_k).
\end{equation}
If the algorithm starts from the stationary solution of \cref{eqn:HpsiEpsi}, then by definition $\vbf_k=0$, making the start-up straightforward.
If instead the evolution begins from wave functions $\vb{\varphi}(\vbr,0)$ that are not eigenstates of the Hamiltonian, the first $n$ steps must be performed with a single-step integrator.
In the W-SLDA package, this role is fulfilled by a method based on the Taylor expansion of the evolution operator:
\begin{align}\label{eq:Taylor-exp}
\varphi_{n}(\vbr,t+\Delta t) = \calT \exp\left[-\I\int_{t}^{t+\Delta t}\calH(\vb{\varphi},t')dt' \right]\varphi_{n}(\vbr,t)
 &\approx \exp\left[-\I\calH\left(\vb{\varphi},t+\frac{\Delta t}{2}\right)\Delta t \right]\varphi_{n}(\vbr,t)\nonumber\\
&\approx \sum_{k=0}^{r}\frac{(-\I\Delta t)^k}{k!}\left[\calH\left(\vb{\varphi},t+\frac{\Delta t}{2}\right)\right]^k \varphi_{n}(\vbr,t),
\end{align}
where the truncation order $r$ is chosen consistently with the order of the ABM corrector.
To achieve high accuracy, the Hamiltonian should be evaluated at the midpoint $t+\tfrac{\Delta t}{2}$.
Following the strategy adopted in nuclear DFT solvers such as Sky3D~\cite{MARUHN20142195} or LISE~\cite{Jin2021}, we employ a predictor-corrector approach:
\begin{enumerate}
\item Estimate state for the midtime (predictor) 
  \begin{equation}
    \varphi_{n}^{(p)}\left(\vbr,t+\frac{\Delta t}{2}\right)\approx\sum_{k=0}^{r}\frac{(-\I\frac{\Delta t}{2})^k}{ k!}\left[\calH(\vb{\varphi},t)\right]^k \varphi_{n}(\vbr,t)\,.
  \end{equation}
\item Perform the integration step, (corrector)
  \begin{equation}
    \varphi_{n}(\vbr,t+\Delta t)\approx\sum_{k=0}^{r}\frac{(-\I\Delta t)^k}{ k!}\left[\calH(\vb{\varphi}^{(p)},t+\frac{\Delta t}{2})\right]^k \varphi_{n}(\vbr,t)\,.
  \end{equation}
\end{enumerate}
The Taylor expansion method requires $2r$ Hamiltonian applications per integration step, making it substantially slower than the ABM integrator, which applies the Hamiltonian only twice per step.
On the other hand, ABM requires storing historical states, which makes it memory-intensive compared to the Taylor approach.
The W-SLDA Toolkit is optimized for performance; therefore, the ABM algorithm is generally preferred.
The Taylor expansion method is only used for starting the ABM integrator.

\section{Implementation}
\subsection{Code types}
The toolkit consists of two main branches of codes:

\begin{description}
\item[Static codes] used to solve the static DFT equations \cref{eqn:HpsiEpsi} self-consistently. Their names follow the convention \filename{st-wslda-?d}, where \filename{?} denotes the dimensionality of the code: 1, 2, or 3 for quasi-1D, quasi-2D, and full 3D cases, respectively. The codes are written in C (C99 standard) and can be executed either on standard CPU machines or on hybrid CPU+GPU systems. For hybrid execution, the ELPA library with GPU acceleration must be installed first~\cite{Yu2021,Wlazlowski2024}. The ELPA library supports both NVIDIA and AMD GPUs.

\item[Time-dependent codes] used to solve the time-dependent DFT equations \cref{eqn:dpsidt-Hpsi} self-consistently. Their respective names are \filename{td-wslda-?d}, with \filename{?} again indicating the dimensionality. These codes are written in a mixed C+CUDA language, with the C part conforming to the C99 standard. They can only be run on hybrid CPU+GPU systems. The toolkit also provides a script (\filename{hipify.sh}, located in the main directory) that automatically converts the CUDA code to the HIP framework, thereby enabling execution on both NVIDIA and AMD GPUs.
\end{description}
All code variants rely on the Message Passing Interface (MPI)~\cite{MPI-3.1} to enable efficient parallel computation. The implementation assumes that the number of lattice points is even in each direction. 

In addition, the package includes a set of auxiliary codes and scripts designed to facilitate the computational workflow and the analysis of results. These are written in C or Python and are located in the \filename{extensions} folder of the main directory.

\subsection{Requirements}

To use the W-SLDA Toolkit, the following software stack must be installed:
\begin{itemize}
    \item A C compiler, e.g., \texttt{gcc} or the Intel compiler \texttt{icc}.
    \item An MPI C compiler, e.g., \texttt{mpicc} or the Intel MPI compiler \texttt{mpiicc}.
    \item Python (basic distribution).
    \item FFTW library~\cite{FFTW}.
    \item LAPACK library~\cite{LAPACK}.
    \item ScaLAPACK library~\cite{ScaLAPACK}.
    \item ELPA library~\cite{BungartzHansJoachim2020} (\textit{optional}).
    \item CUDA Toolkit~\cite{CUDAToolkit} for systems equipped with NVIDIA GPUs, or ROCm~\cite{ROCm} for systems equipped with AMD GPUs (\textit{required only for time-dependent codes}).
\end{itemize}

Most high-performance computing (HPC) clusters provide compilers and numerical libraries such as FFTW, LAPACK, and ScaLAPACK as part of their standard software environment.  
If the computing system is equipped with GPUs, the corresponding GPU software stack (CUDA or ROCm) is typically preinstalled as well. Many HPC centers also provide precompiled versions of the ELPA library. If not, it can be compiled manually following the instructions provided on the official web page~\cite{ELPAwww}.  
Note that the LAPACK and ScaLAPACK libraries are also included in the Intel Math Kernel Library (MKL).  

It is recommended to use the ELPA library whenever possible for static calculations, as it provides a significant speed-up of the diagonalization routines, even when running on CPU-only systems.  
Time-dependent codes, in contrast, require GPU acceleration and cannot be executed on CPU-only architectures, whereas static codes can operate entirely on CPUs.

\subsection{Usage model}
\begin{figure}[t]
	\centering
	\includegraphics[width=0.7\linewidth]{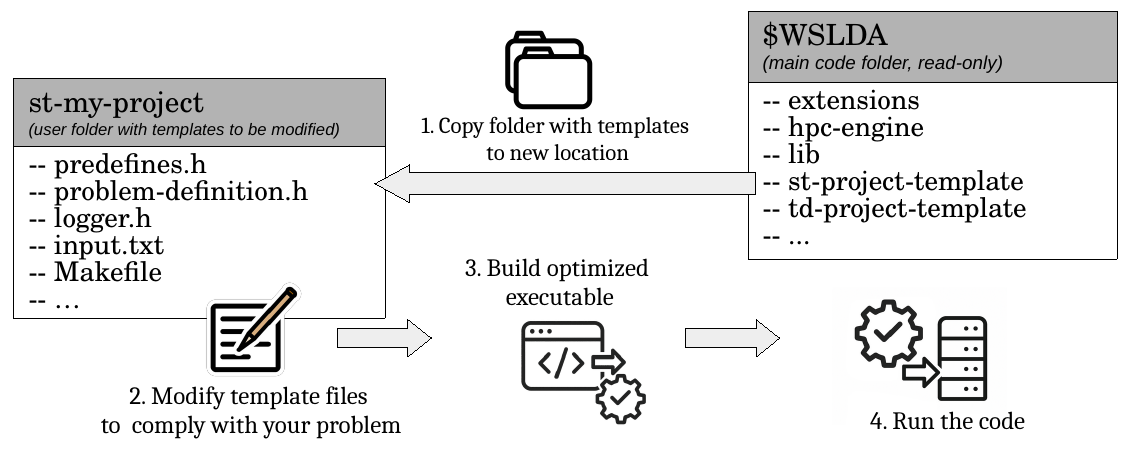}
	\caption{Template-based workflow in W-SLDA: the user copies a template directory to a new location (1), modifies the template files (2), compiles (3), and runs the executable (4).}
	\label{fig:usage-model}
\end{figure}

From the user’s perspective, W-SLDA follows a template-based programming model. A project starts from a template where the user specifies items such as the external potential, external pairing potential, and external velocity field in C. The toolkit separates the computational engine from user-modifiable files. In practice, direct modification of the engine is rarely required. The engine resides in \filename{$WSLDA/hpc-engine}, while templates for static and time-dependent calculations are located in \filename{$WSLDA/st-project-template} and \filename{$WSLDA/td-project-template}, respectively, where \filename{$WSLDA} denotes the path to the main toolkit directory.
A typical workflow begins with copying one of these template folders into a new working directory. For example, for static calculations:
\begin{lstlisting}[style=Commands]
cp -r $WSLDA/st-project-template ./st-my-project
\end{lstlisting}
This procedure is illustrated in \cref{fig:usage-model}.
The copied directory (e.g., \filename{st-my-project}) becomes the working folder where the user modifies templates, compiles, and executes the code. The template files are thoroughly documented and contain links to online resources for further reference.

The design priorities of the W-SLDA Toolkit are, in order:
\begin{enumerate}\setlength{\itemsep}{0pt}
\item Correctness of computation.
\item Performance.
\item User-friendly interface.
\end{enumerate}
The highest priority is correctness, which is essential in computational physics, and will be discussed later in \cref{sec:CodeResultsQuality}. Performance is the next priority and has direct implications for the usage model. Specifically, for each type of problem, the user compiles a dedicated executable optimized for that problem. For this reason, the toolkit does not provide universal precompiled binaries controlled solely through input files. Instead, the user must supply problem-defining information at compile time.
For example, the file \filename{predefines.h} specifies the computational mesh size and spacing:
\begin{lstlisting}
// in predefines.h
/**
 * Define lattice size and lattice spacing.
 * */
#define NX (8)
#define NY (10)
#define NZ (12)

#define DX (1.0)
#define DY (1.0)
#define DZ (1.0)
// ...
\end{lstlisting}
Other configuration tags must also be set in this file. Defining the mesh size and dimensionality at compilation ensures that the code is optimized for the specific case. For instance, preparing code for \lstinline{NX=NY=NZ=100} in a full 3D simulation is very different from preparing code for \lstinline{NX=NY=NZ=32} in a quasi-2D case. The former may require large-scale HPC resources, while the latter may run efficiently on a modest computing cluster.

\noindent A typical workflow proceeds as follows:
\begin{enumerate}\setlength{\itemsep}{0pt}
\item Modify template files (e.g., \filename{predefines.h}, \filename{problem-definition.h}, …) in the working directory.
\item Compile an executable optimized for the problem, e.g.,
\begin{lstlisting}[style=Commands]
make 3d
\end{lstlisting}
to build code for unrestricted 3D simulations.
\item Run the calculations.
\end{enumerate}
In this workflow, the user does not merely execute a pre-built program, but instead participates in the creation of an optimized binary tailored to the chosen physical setup.
Although this model requires compiling a new executable for each problem type, which may seem inconvenient, experience shows that it pays off in the long run. It ensures efficient utilization of computational resources, which are often scarce in research settings. Importantly, compilation is not required for every run. Many parameters, such as particle number or external potential characteristics, can be set at runtime via the input file. Compilation is only required when changing generic attributes of the problem, such as dimensionality, mesh size, or static vs. time-dependent formulation. 

User-friendliness in W-SLDA is realized through the template-based approach. By separating the computational engine from user-modifiable files, the toolkit enables users to focus on defining the physical problem while relying on a robust and optimized backend. This design minimizes the need for direct modifications of the core code, lowers the entry barrier for new users, and ensures both flexibility and efficiency in practical applications.

\subsection{Static codes}
\subsubsection{Overview of computation workflow}
\begin{figure*}[htbp]
	\centering
	\includegraphics[width=1\linewidth]{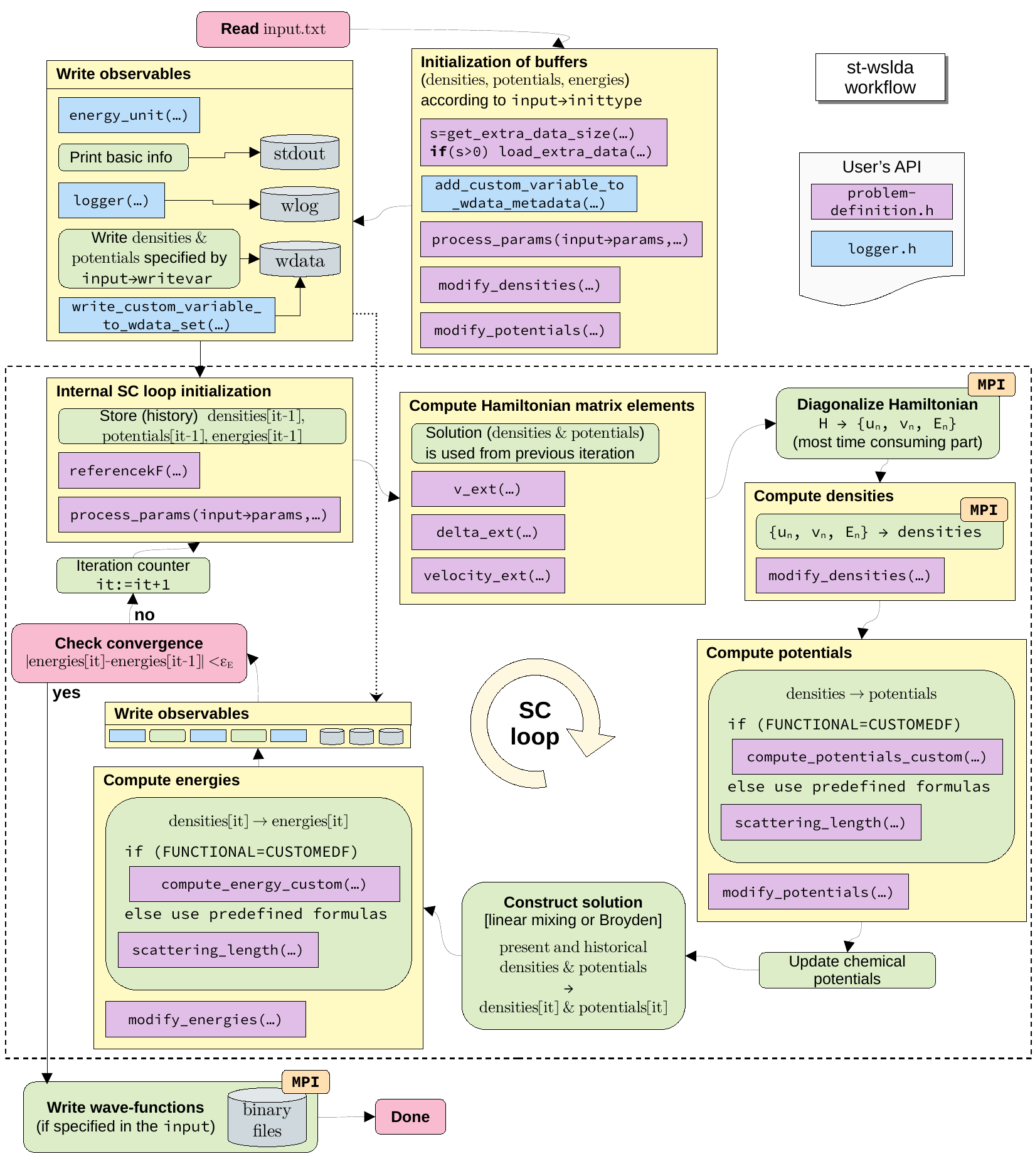}
	\caption{Workflow for the family of static codes. Yellow boxes indicate logical blocks of the code, with arrows denoting the order of execution. The process begins with reading \filename{input.txt}. The section enclosed by the dashed rectangle corresponds to the self-consistent loop, which is repeated until convergence is reached (or the maximum number of iterations is exceeded). The computation terminates at the block labeled \textsf{Done}. Functions shown in purple and blue boxes are located in the files \filename{problem-definition.h} and \filename{logger.h}, respectively, and can be customized by the user. The \textsf{MPI} icon marks blocks that involve communication between processes.}
	\label{fig:st-wslda-diagram}
\end{figure*}

The computational process implemented by the family of static codes \filename{st-wslda-?d} is illustrated in \cref{fig:st-wslda-diagram}.
The workflow is divided into logical blocks, each of which can be customized through template files:
\begin{description}
\item[\filename{problem-definition.h}] Contains user-defined functions that specify the physical problem. These functions are shown as purple boxes in the diagram. For instance, this is where the form of the external potential is implemented. 
\item[\filename{logger.h}] Contains functions responsible for writing results to output files. These functions are represented in blue boxes in the diagram. By modifying this file, the user can customize the format and content of the generated output.  
\end{description}

The computation starts with reading the input file. It is text, where each line has a simple structure: \lstinline{tag value}. For example
\begin{lstlisting}[style=Input, numbers=none]
# ------------------- INPUT/OUTPUT ------------------
outprefix               test    # all output file with start with this prefix
inprefix                none    # prefix of calculation you need to use as input
overwrite               1       # allow for overwriting results
checkpoint              1       # do checkpoints after each iteration
resetit                 1       # reset iterations counter (it): yes=1, no=0
writewf                 0       # write wave-functions at the end: yes=1, no=0
...
\end{lstlisting}
Next, the initialization of buffers is performed and writing to files information related to the initial state that will be used as input for the self-consistent loop. 

The core computation is executed within a self-consistent (SC) loop.
\begin{description}
    \item[Internal SC loop initialization] 
    Each iteration begins with reorganizing the history buffers and setting the reference scales. Throughout the computation, information about typical physical scales is exploited to guide convergence and stability.
    The most important reference scale is the Fermi momentum. For a uniform system it is defined, depending on dimensionality, as  
    \begin{align}
    k_F^{(1D)}=\frac{\pi \rhoP}{2}, \quad 
    k_F^{(2D)}=\sqrt{2\pi \rhoP}, \quad 
    k_F^{(3D)}=(3\pi^2 \rhoP)^{1/3},
    \end{align}  
    where $\rhoP$ denotes the total particle density. From $\kF$, other reference scales can be derived, such as the Fermi energy $\eF = \kF^2/2$.  

    These scales are essential for the algorithm, as they provide a sense of what constitutes a “small” or “large” variation in computed quantities. However, in non-uniform systems there is no unique definition of $\kF$, since the density varies in space (except for the special case of a uniform system). To handle this, the user must either (i) provide a prescription for evaluating $\kF$ given a density distribution (via function \lstinline{referencekF(...)}), or (ii) specify an estimated value directly via the input file (tag \lstinline{referencekF}).  

    The reference scale $\kF$ is also used when defining physical parameters in the input. For example, the user specifies the interaction strength through the dimensionless parameter $a_s \kF$.  

    \item[Compute Hamiltonian matrix elements] The matrix representation of the Hamiltonian~\cref{eqn:HpsiEpsi} is constructed. Since the Hamiltonian terms depend on densities and chemical potentials, the values generated in the previous iteration are used. This step also requires specifying the external potentials, which are provided through user-defined functions in \filename{problem-definition.h} (by default, these potentials are set to zero). The matrix is stored using a two-dimensional block cyclic distribution scheme, a standard data layout for dense matrix computations in HPC libraries such as ScaLAPACK~\cite{ScaLAPACK} or ELPA~\cite{Ks2019,BungartzHansJoachim2020}.

    \item[Diagonalize Hamiltonian] Eigenfunctions $[u_n, v_n]$ and corresponding eigenvalues $E_n$ are obtained by diagonalizing the Hamiltonian matrix. This is the most computationally expensive step in the static codes, typically consuming more than $95\%$ of the total execution time. It also involves intensive MPI communication. Because this block requires expertise in parallel computing, it is not intended for user modification. \respch{We emphasize that the superfluid variant of DFT considered here usually requires the extraction of a large fraction of the eigenvalues and eigenvectors of the Hamiltonian matrix, as discussed in Sec.~\ref{sec:regularization}. In the case of the so-called box cutoff scheme, which provides the highest accuracy for subsequent time-dependent simulations, one essentially has to extract the full spectrum of the Hamiltonian. For this reason, we rely on direct diagonalization rather than on iterative or subspace-projection eigensolvers, which are typically advantageous when only a small fraction of eigenstates is required.}

    \item[Compute densities] Densities are computed from the quasiparticle wave functions extracted in the previous step. The specific formulas depend on the code variant, for instance, \crefrange{eq:na3D}{eq:jb3D} for general 3D spin-imbalanced systems, or \crefrange{eq:nab3D}{eq:jab3D} for the spin-symmetric case, with further adjustments applied in quasi-2D or quasi-1D setups. If necessary, the user can introduce corrections or modifications to the computed densities through a dedicated function at the end of this block. Density computations are parallelized and executed with MPI communication. 

    \item[Compute potentials] Potentials are evaluated from the extracted densities, according to the formulas defined by the chosen energy functional. As with densities, the user can modify the computed potentials through a dedicated function at the end of this block. Alternatively, when the \lstinline{CUSTOMEDF} mode is enabled in \filename{predefines.h}, the user may implement entirely custom formulas for deriving potentials from densities. For reference, the implementations of the predefined functionals can be found in \filename{$WSLDA/hpc-engine/wslda_functionals.c}.

    \item[Update chemical potentials] The chemical potentials are updated according to \cref{eq:update-mu}.

    \item[Construct solution] The new solution is formed using either linear mixing or Broyden’s algorithm, as specified in the input file.

    \item[Compute energies] Once the solution is updated, the total energy of the system is computed. If needed, the user can introduce corrections or modifications via a dedicated function at the end of this block, or provide a fully custom implementation when the \lstinline{CUSTOMEDF} mode is enabled.

    \item[Write observables] The observables computed within the self-consistent cycle are written to output files. By default, energies and fields (densities and potentials) are stored. The output format can be customized by modifying the relevant functions in \filename{logger.h}. 

    \item[Check convergence] Convergence is assessed by monitoring changes in the individual energy contributions between successive iterations. The solution is considered converged when all energy components, normalized to the free Fermi gas energy $\Effg$, vary by less than a specified tolerance $\varepsilon_{\textrm{energy}}$, as defined in \cref{eq:E-conv}. For simulations performed at fixed particle number, the additional condition given by \cref{eq:N-conv} must also be fulfilled. Both the convergence thresholds and the target particle numbers are user-defined parameters specified in the input file.

    \item[Iteration counter] If convergence is not reached, the iteration counter is increased by one, and the self-consistent loop is repeated. In the next cycle, the potentials from the previous iteration are used to construct the Hamiltonian matrix. To prevent infinite looping in cases where the algorithm fails to converge, the maximum number of iterations is limited by the input parameter \lstinline{maxiters}.   
\end{description}
The computation process concludes with writing the quasiparticle wave functions and corresponding eigen-energies to binary files. These files can then be used as initial states for subsequent time-dependent simulations. The wave functions are stored in the standardized W-data format, which ensures portability across different code variants. Details of this storage format are provided in \cref{sec:storage-format}.

\subsubsection{Parallelization scheme}
The most time-consuming part of the static codes is the diagonalization of the Hamiltonian matrix. This part of the computation is realized in a parallel fashion. To use the \texttt{st-wslda} codes efficiently, it is crucial to understand how the Hamiltonian matrix is decomposed among MPI processes. The decomposition strategy depends on the problem's dimensionality.

\paragraph{Parallelization scheme for 3D code}
\begin{figure}[tb]
	\centering
	\includegraphics[width=1\linewidth,trim={0cm 7.4cm 0cm 0cm},clip]{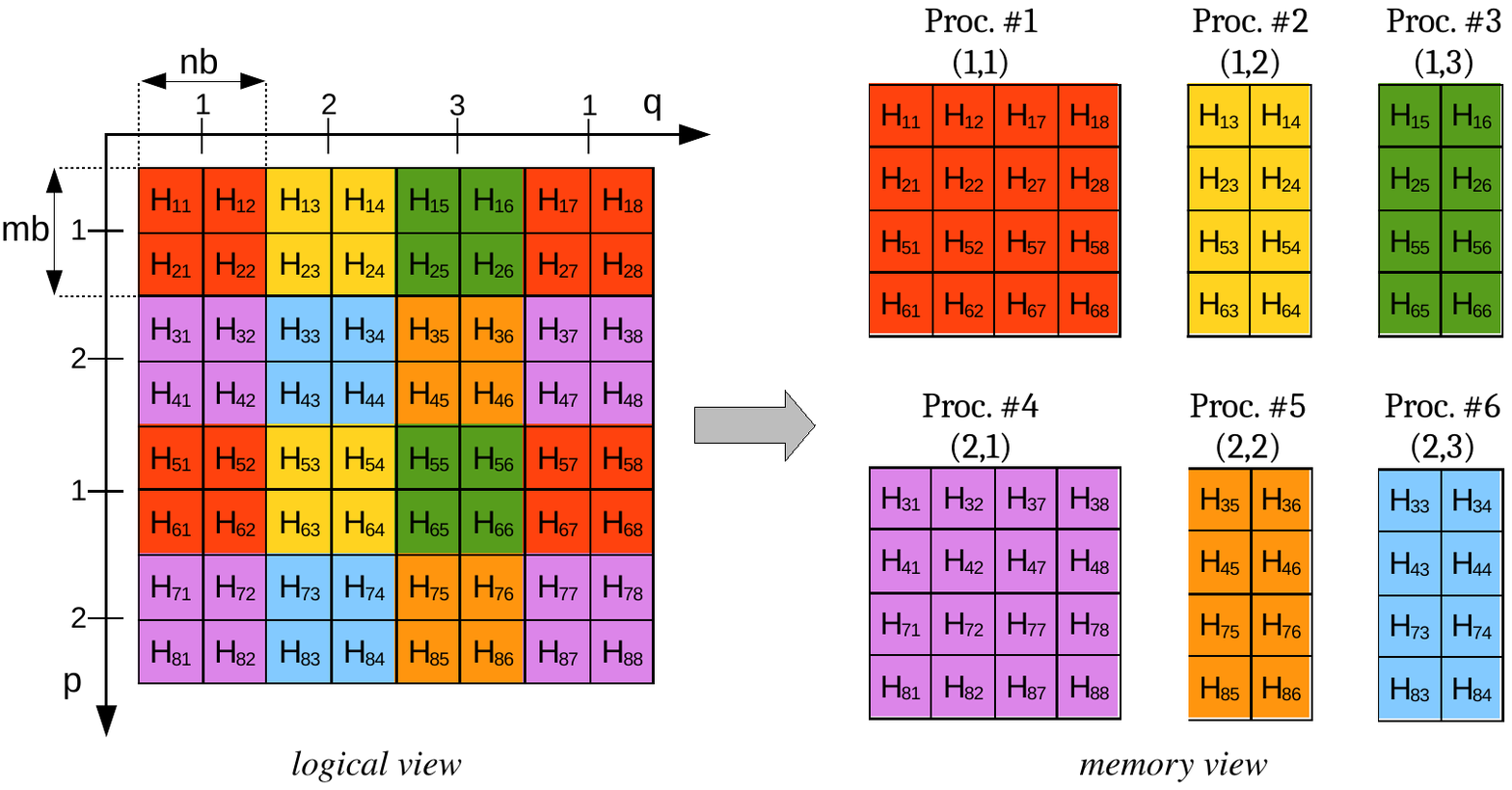}
	\caption{Example of block-cyclic decomposition of a matrix of size $8\times 8$.
    The matrix is distributed over \lstinline{np=6} processes arranged on a BLACS grid \lstinline{(p,q)=(2,3)}.
    The block size is \lstinline{(mb,nb)=(2,2)}.
    Left: logical view of the global matrix.
    Right: distribution of blocks across processes (memory view), with BLACS coordinates shown in parentheses.
    Note that the figure numbers the ranks starting from 1 for clarity, while the actual implementation (in C) follows the convention of starting from rank 0.}
	\label{fig:bc-decomp-1}
\end{figure}
In the 3D case, each iteration of the self-consistent loop requires diagonalizing a matrix of size $\texttt{M}=2N_x N_y N_z$. The matrix is distributed across MPI processes using a block-cyclic (BC) decomposition. To illustrate the idea, let us consider a toy Hamiltonian matrix of size $8 \times 8$ (in practice, such a small case does not occur; already for the smallest lattice $2^3$, the Hamiltonian matrix has size $16 \times 16$). Suppose we want to run the code on \texttt{np=6} processes. The MPI processes are arranged on a BLACS grid, which is a logical two-dimensional arrangement of processes used by distributed dense linear algebra libraries such as ScaLAPACK~\cite{ScaLAPACK} and ELPA. The grid size is denoted as \texttt{(p,q)}, with the constraint $\texttt{p}\cdot \texttt{q}=\texttt{np}$.
The Hamiltonian matrix is divided into submatrices (blocks) of fixed size \texttt{(mb,nb)}, which are then assigned to processes in the BLACS grid in a cyclic (round-robin) manner. Once each process has received a block, the assignment cycle repeats. These parameters are user-controllable through the input file. For example:
\begin{lstlisting}[style=Input, numbers=none]
# BLACS grid 
p                       2
q                       3
# data block size 
mb                      2
nb                      2
\end{lstlisting}
With these settings, the matrix distribution is shown in \cref{fig:bc-decomp-1}.
Different colors correspond to matrix elements stored on different MPI processes (six in total). Each process stores only its assigned subset of elements.

For performance, the choice of \texttt{(p,q)} is important. In general, the best results are obtained for square grids, i.e., \texttt{p=q}. If \texttt{p} and \texttt{q} are not specified, the code automatically chooses values satisfying this criterion. When a square grid cannot be achieved, we find empirically that settings with \texttt{p<q} often outperform \texttt{p>q}. For example, with \texttt{np=6}, the options are \texttt{(p,q)=(2,3)} or \texttt{(p,q)=(3,2)}, and \texttt{(2,3)} tends to yield better performance.

The block sizes \texttt{mb} and \texttt{nb} must satisfy the constraints $\texttt{mb} \leq \texttt{M}/\texttt{p}$ and 
$\texttt{nb} \leq \texttt{M}/\texttt{q}$. 
However, using the largest possible block sizes generally gives poor performance. Good results are typically obtained when $\texttt{mb}$ and $\texttt{nb}$ are much smaller than their upper limits, while still large enough so that the block contains at least a few hundred elements. In practice, powers of two such as 16, 32, or 64 often work well. We recommend experimenting with these values to optimize performance for a given system size and computing architecture. For the use of the ELPA library, a comprehensive overview of computational performance with respect to matrix decomposition parameters is presented in~\cite{Yu2021}. 

\begin{figure}[tb]
	\centering
	\includegraphics[width=0.9\linewidth,trim={0cm 0cm 0cm 0cm},clip]{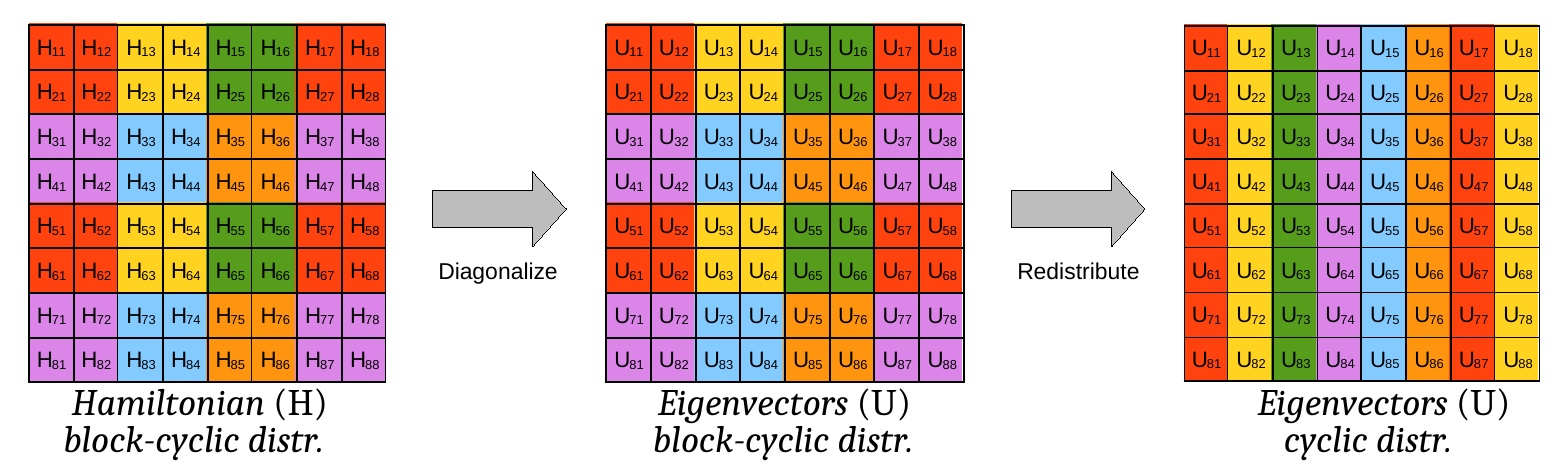}
	\caption{Computational workflow for extracting eigenvectors. The Hamiltonian matrix $H$, stored in block-cyclic form, is diagonalized, and the resulting eigenvector matrix $U$ is obtained. The matrix $U$ is then redistributed into cyclic form, so that each MPI process stores complete eigenvectors (distributed column by column). Colors indicate matrix elements handled by different MPI processes.}
	\label{fig:bc-decomp-2}
\end{figure}
The procedure for extracting eigenvectors (and the corresponding eigenvalues) is divided into three steps. First, the Hamiltonian matrix $H$ is constructed in block-cyclic form. Next, the eigenvectors are computed and stored in the matrix $U$. The matrix $U$ is initially produced in block-cyclic form with the same distribution parameters as $H$. However, this distribution is inconvenient for subsequent operations. Therefore, $U$ is redistributed among MPI processes into cyclic form, where each MPI process stores complete eigenvectors, i.e., the matrix $U$ is distributed column by column. These operations are illustrated in \cref{fig:bc-decomp-2}. Finally, each process performs computations on its subset of eigenvectors, such as evaluating the contribution to the densities. The partial results are then combined through a global reduction using the \lstinline{MPI_Allreduce(...)} routine.

Finally, if \texttt{p=q=1}, the code reduces to a serial execution (no MPI parallelization).
While this can be used for debugging or very small systems, it is not suitable for realistic 3D simulations.

\paragraph{Parallelization scheme for 2D code}
In this variant, the code assumes that the quasiparticle wave functions have the form~\cref{eq:wf-quasi2D}, meaning that along the $z$ direction plane waves are used, and 
\begin{equation}
    k^{(z)}_n = 0, \pm 1 \frac{2\pi}{L_z}, \pm 2 \frac{2\pi}{L_z}, \dots , +(N_z-1) \frac{2\pi}{L_z}
\end{equation}
becomes a good quantum number. Under this assumption, the Hamiltonian matrix acquires a block-diagonal form:
\begin{equation}
H = \begin{pmatrix} 
\begin{pmatrix} H_{\textrm{2d}}[k^{(z)}_0]\end{pmatrix} & & \\
& \begin{pmatrix} H_{\textrm{2d}}[k^{(z)}_1]\end{pmatrix} & \\
& & \ddots \\
\end{pmatrix}
\end{equation}
where each submatrix $H_{\textrm{2d}}$ has dimension $2N_x N_y$, and there are $N_z$ such submatrices in total. The diagonalization of the full Hamiltonian matrix $H$ can thus be performed more efficiently by diagonalizing the submatrices $H_{\textrm{2d}}$ separately. Moreover, due to translational symmetry one has $H_{\textrm{2d}}[k^{(z)}]=H_{\textrm{2d}}[-k^{(z)}]$, and in practice it is sufficient to diagonalize only the submatrices corresponding to non-negative $k^{(z)}$, which amounts to $N_z/2$ distinct values.  

\begin{figure}[tb]
	\centering
	\includegraphics[width=0.9\linewidth,trim={0cm 13cm 8cm 0cm},clip]{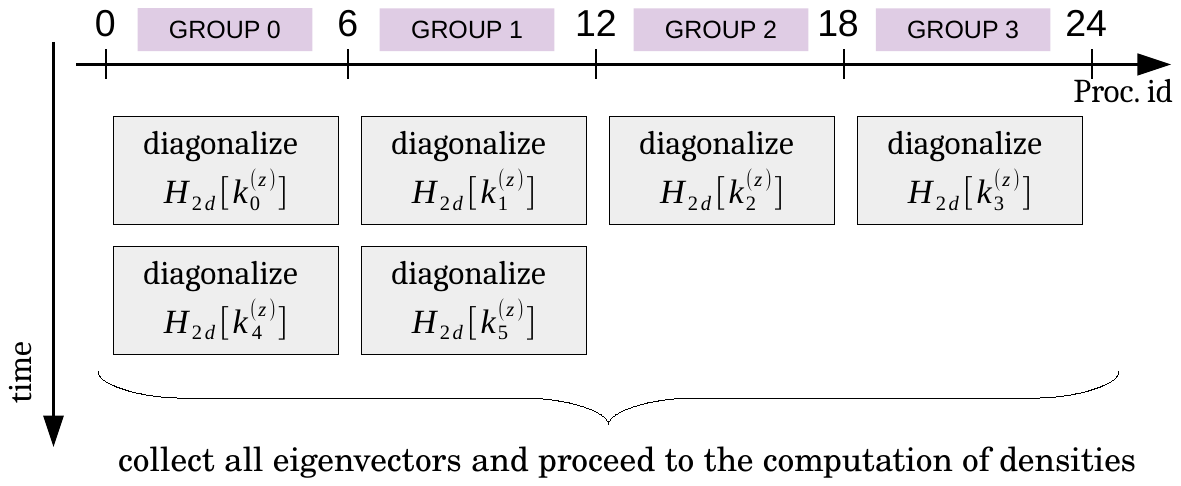}
	\caption{Parallelization scheme for the 2D static code. The total set of MPI processes is divided into independent groups. Each group diagonalizes a submatrix $H_{\textrm{2d}}$ using block-cyclic decomposition. After diagonalization, the extracted eigenvectors and eigenvalues are collected from all groups, and the code proceeds to compute the densities.}
	\label{fig:bc-decomp-3}
\end{figure}

The submatrices can be diagonalized simultaneously.  
To illustrate the parallelization scheme in the 2D case, consider the lattice $N_x \times N_y \times N_z = 8 \times 10 \times 12$. We employ the same \texttt{(p,q)} and \texttt{(nb,mb)} parameters as before, and assume execution with \texttt{np=24} processes. For these settings, a single iteration requires $N_z/2=6$ diagonalizations. The total set of processes is divided into computational groups of size $\texttt{p}\cdot\texttt{q}=6$, resulting in $24/6=4$ groups. Each submatrix $H_{\textrm{2d}}$ is distributed in block-cyclic form among the $\texttt{p}\cdot\texttt{q}$ processes within a group, as in the 3D case.  
The code output reports this decomposition:
\begin{lstlisting}[style=Input, numbers=none]
# CODE: ST-WSLDA-2D
# LATTICE: 8 x 10 x 12
 ...
# NUMBER OF k-MODES TO CONSIDER: 6
# SETTINGS 4 KZGROUPS, EACH WITH GRID PROCESSES OF SIZE [2 x 3]
# GROUP 0 WITH 6 PROCESSES HAS BEEN SUCCESSFULLY CREATED.
# GROUP 1 WITH 6 PROCESSES HAS BEEN SUCCESSFULLY CREATED.
# GROUP 2 WITH 6 PROCESSES HAS BEEN SUCCESSFULLY CREATED.
# GROUP 3 WITH 6 PROCESSES HAS BEEN SUCCESSFULLY CREATED.
# GROUP 0 COMPUTES FOR 2 k-values [0,2)
# GROUP 1 COMPUTES FOR 2 k-values [2,4)
# GROUP 2 COMPUTES FOR 1 k-values [4,5)
# GROUP 3 COMPUTES FOR 1 k-values [5,6)
 ...
\end{lstlisting}
The computational workflow for a single iteration is shown schematically in \cref{fig:bc-decomp-3}.  

By construction, the following constraint must be satisfied:  
\begin{gather}
  \frac{\texttt{np}}{\texttt{p}\cdot \texttt{q}} = \texttt{i},
\end{gather}
where \texttt{i} is an integer (the number of groups) and $\texttt{i}\leq N_z/2$.
If \texttt{p} and \texttt{q} are not specified in the input file, the code automatically selects them so as to maximize the number of simultaneous diagonalizations.
With respect to matrix distribution, all requirements described for the 3D case apply here as well.
The optimal configuration is obtained when the total number of processes is chosen as  
\begin{gather}
  \texttt{np}=\texttt{p}\cdot\texttt{q}\cdot \frac{N_z}{2}.
\end{gather}

\paragraph{\respch{Paralellization} scheme for 1D code}
The diagonalization scheme for the \respch{quasi-}1D code follows the same principle as in the \respch{quasi-}2D case, with the modification that the submatrices now depend on two wave vectors, $H_{\textrm{1d}}[k^{(y)},k^{(z)}]$. Each submatrix has dimension $2N_x$, and there are $N_y N_z$ such matrices in total. In practice, it is sufficient to diagonalize only those corresponding to unique values of $\bigl(k_m^{(y)}\bigr)^2+\bigl(k_n^{(z)}\bigr)^2$, which significantly reduces the computational cost.  

In the \respch{quasi-}1D case, one typically needs to diagonalize a large number of small matrices of size $2N_x$. For modest values of $N_x$ (not much larger than a thousand), the best performance is usually obtained with \texttt{p=q=1}, i.e., without block-cyclic distribution of the submatrices.

\subsubsection{Memory consumption}
There is no simple formula for estimating the memory consumption of the static codes, as it depends strongly on the code settings, in particular on the choice of the diagonalization routine. The code supports routines from both the ScaLAPACK and ELPA libraries, each with different memory requirements. The diagonalization routine is specified in the \filename{machine.h} file. The dominant contribution to memory consumption arises from storing large matrices, while additional buffers allocated internally by the code are comparatively small and can be neglected in memory estimates.  

During computation, memory must be allocated for the Hamiltonian matrix $H$, for the eigenvector matrix $U$, and for the working space required by the diagonalization routine. As a rule of thumb, the working space is of the same order as the storage required for $H$, although the precise factor depends on the selected routine. For example, the ScaLAPACK routine \lstinline{PZHEEVR} requires significantly less working memory than \lstinline{PZHEEVRD}.  
The order of the total memory consumption can be summarized as follows:
\begin{description}
\item[full 3D] $\texttt{mem}_{\textrm{st}}\sim k (2N_x N_y N_z)^2 \cdot 16\,\text{[Bytes]}$,
\item[quasi-2D] $\texttt{mem}_{\textrm{st}}\sim k (2N_x N_y)^2 \frac{N_z}{2} \cdot 16\,\text{[Bytes]}$,
\item[quasi-1D] $\texttt{mem}_{\textrm{st}}\sim k (2N_x)^2 \frac{N_y N_z}{4} \cdot 16\,\text{[Bytes]}$,
\end{description}
where $k$ is a parameter depending on the selected routine, typically in the range $3 \leq k \leq 10$. These expressions provide estimates for the total memory required. To obtain the memory consumption per computational process, the total estimate should be divided by the number of processes. 

\subsubsection{Scaling}\label{sec:Scaling-st}

\begin{figure}[tb]
	\centering
	\includegraphics[width=0.9\linewidth,trim={0cm 0cm 0cm 0cm},clip]{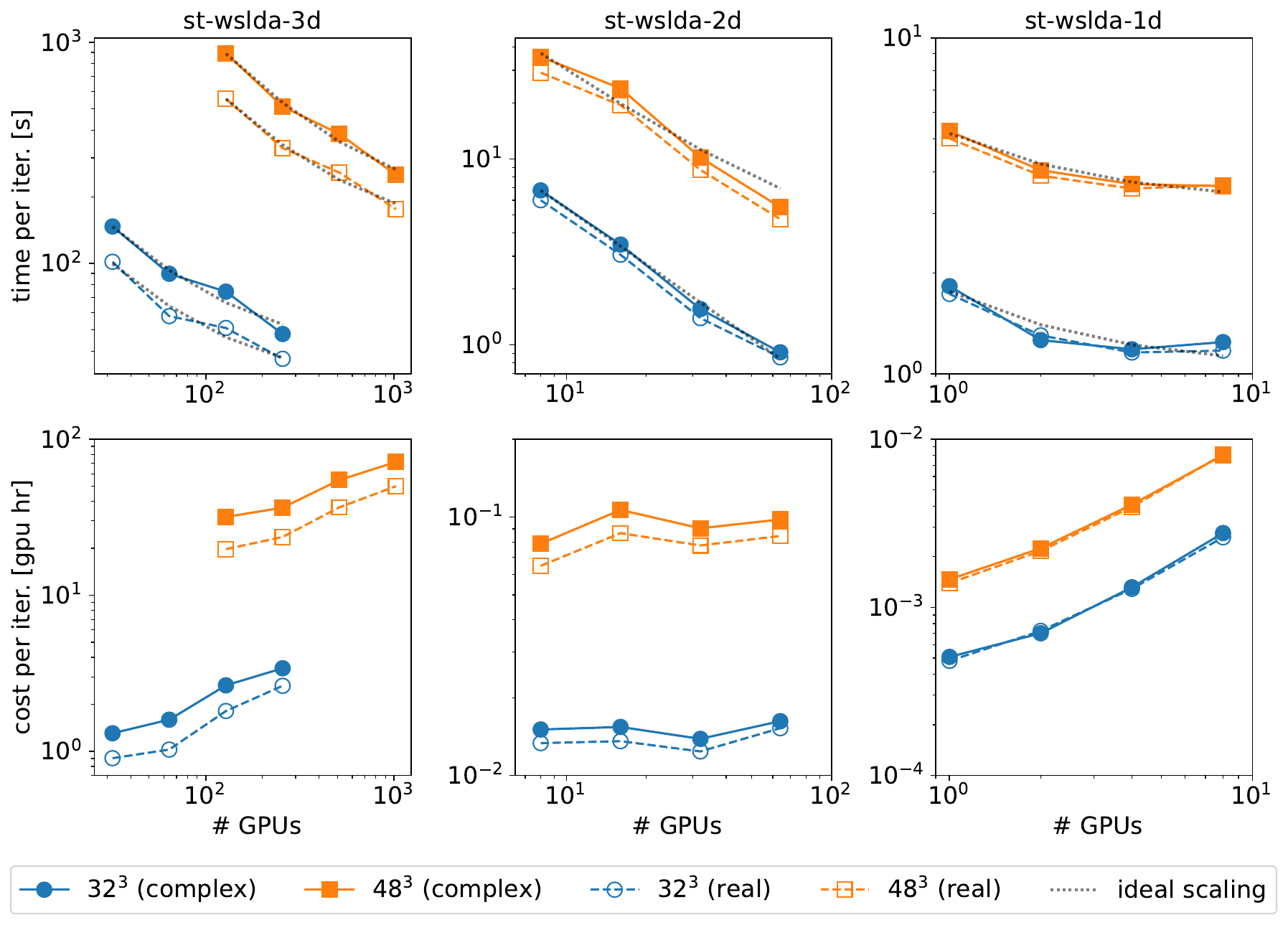}
	\caption{Scaling plots for the family of static codes. The top row shows the time per iteration (in seconds) versus the number of utilized GPUs. Two lattices were considered, $32^3$ and $48^3$, and tests were performed for two arithmetic modes: with the Hamiltonian matrix elements treated as complex or real. The gray dotted line indicates the ideal scaling, i.e., the computation time decreases by a factor of two when the number of GPUs is doubled. The bottom row shows the computation cost per iteration, measured in GPU hours, defined as the product of run time and the number of GPUs. For ideal scaling, this quantity should remain constant with respect to the number of GPUs. The diagonalization was performed using the ELPA library. Tests were executed on the LUMI supercomputer (EuroHPC/CSC, Finland), equipped with AMD MI250X GPUs.}
	\label{fig:st-scaling}
\end{figure}

\Cref{fig:st-scaling} presents scaling plots for the family of static codes. Two representative lattices, $32^3$ and $48^3$, were considered. The top row shows the time per iteration in seconds, as measured on LUMI (EuroHPC/CSC, Finland) supercomputer. To determine the total run time, this value must be multiplied by the number of iterations, which depends on the specific physical problem. Diagonalizations were carried out using the ELPA library with GPU acceleration. The bottom row shows the computation cost per iteration, measured in GPU hours, i.e., the product of run time and the number of GPUs.
\respch{
Since the computation cost is determined by the matrix diagonalizations, the measured scalings reflect the scaling of the diagonalization library; in this case, the ELPA library~\cite{Yu2021}. 
}

Several properties of the codes can be immediately observed. First, exploiting translational symmetries leads to substantial performance gains. Imposing translational symmetry in one direction reduces computation costs by approximately two orders of magnitude, and by about four orders of magnitude when symmetries in two directions are applied. Second, the codes exhibit good strong-scaling behavior. The observed increase in computation cost with larger GPU counts is due to overheads associated with MPI communication. Third, the toolkit supports both complex and real arithmetic. The latter can be used when it is known in advance that all Hamiltonian matrix elements are real. For 3D simulations, this mode yields roughly a twofold speed-up. In lower-dimensional cases, the gain is smaller, and in \respch{quasi-}1D simulations, where run times are already very short, it is typically negligible.  

The scaling with respect to the Hamiltonian matrix size follows the expected $\texttt{M}^3$ behavior, where $\texttt{M}$ denotes the matrix dimension. It depends on the dimensionality of the problem: $\texttt{M} = 2N_x N_y N_z$ for the 3D case, $\texttt{M} = 2N_x N_y$ for the \respch{quasi-}2D case, and $\texttt{M} = 2N_x$ for the \respch{quasi-}1D case.  
For example, in the 3D case, increasing the lattice size from $32^3$ to $48^3$ enlarges the matrix dimension by a factor of approximately 3.4, resulting in a computational cost increase of about 40, which is consistent with the measured data. More detailed scaling plots for the static codes are reported in Ref.~\cite{Wlazlowski2024}.  

The overall computational cost per iteration scales as
\[
\texttt{cost}_{\textrm{st}} \propto N_{\textrm{mat}} \cdot \texttt{M}^3,
\]
where $N_{\textrm{mat}}$ is the number of matrices diagonalized per iteration. Depending on the code dimensionality, this quantity is given by $N_{\textrm{mat}} = 1$ for the 3D variant, $N_{\textrm{mat}} = N_z/2$ for the \respch{quasi-}2D variant, and $N_{\textrm{mat}} \lesssim N_y N_z / 4$ for the \respch{quasi-}1D variant.  

Finally, we note that with access to leadership-class supercomputers such as LUMI, it is possible to carry out simulations where the Hamiltonian matrix dimension reaches up to 3.2 million. This enables full 3D calculations in boxes exceeding $100^3$ in size. An example of such a large-scale computation is presented in Ref.~\cite{Wlazlowski2024}.

\subsection{Time-dependent codes}

\subsubsection{Overview of computation workflow}
 
\begin{figure}[htbp]
	\centering
	\includegraphics[width=0.9\linewidth]{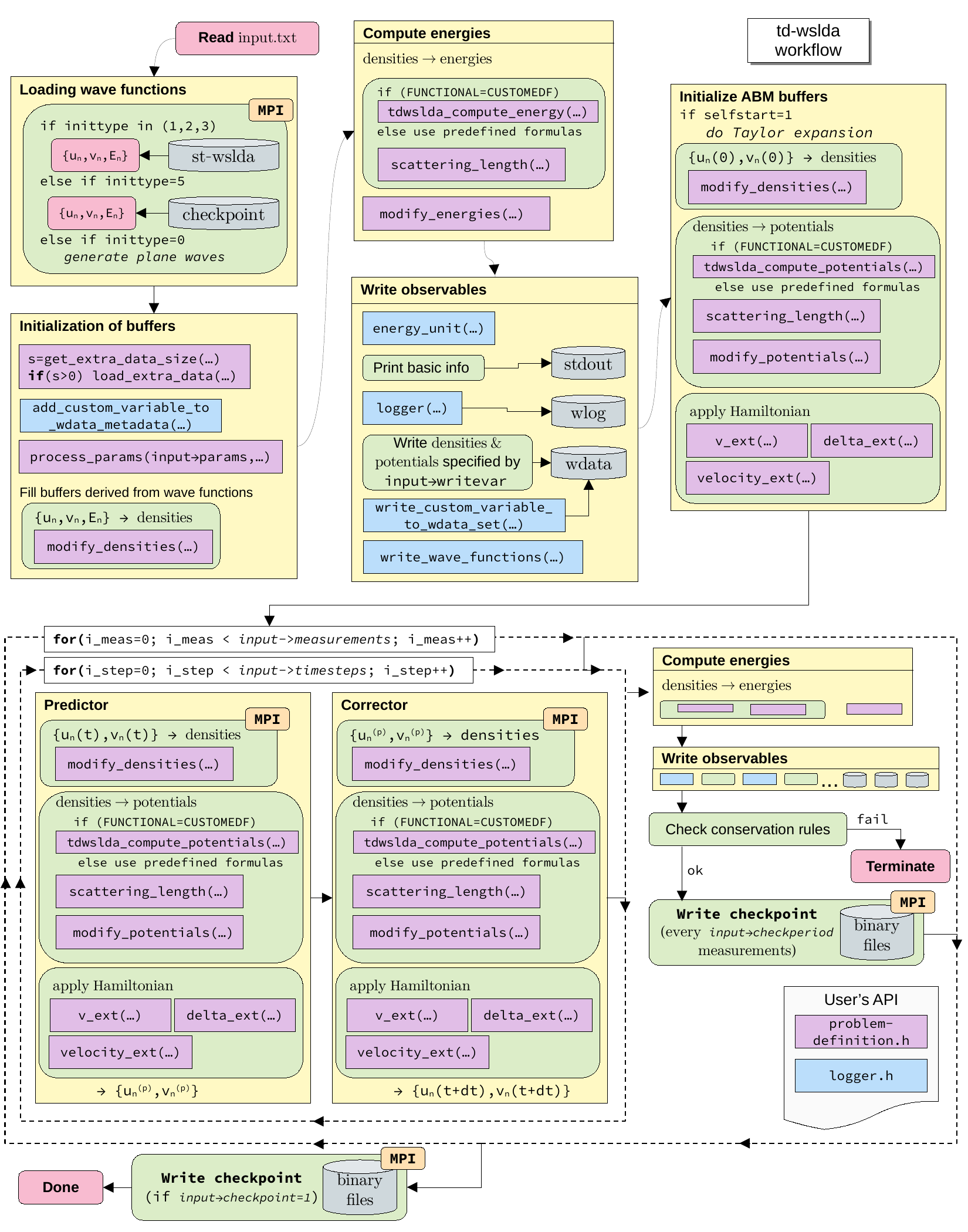}
	\caption{Workflow for the family of time-dependent codes.
    Yellow boxes indicate the logical components of the code, with arrows denoting the order of execution.
    The process begins with reading the \filename{input.txt} file.
    The sections enclosed by dashed rectangles correspond to time-evolution loops: the main loop (iterator \lstinline{i_meas}) and its nested loop (iterator \lstinline{i_step}).
    The computation terminates at the block labeled \textsf{Done}.
    Functions shown in purple and blue boxes are defined in the files \filename{problem-definition.h} and \filename{logger.h}, respectively, and can be customized by the user.
    The \textsf{MPI} icon marks blocks that involve communication between GPUs.}
	\label{fig:td-wslda-diagram}
\end{figure}
The computational process implemented in the family of the time-dependent codes \filename{td-wslda-?d} is illustrated schematically in \cref{fig:td-wslda-diagram}.
As in the static codes, the workflow can be customized by modifying routines defined in the \filename{problem-definition.h} and \filename{logger.h} files. These routines are similar in structure to those used in the static modules, but they differ in certain implementation details; therefore, the corresponding files are not interchangeable between the static and time-dependent codes.

The computation begins by reading the input file. The code requires a set of wave functions to evolve, which are either loaded from files produced by the static solver or from checkpoint files, depending on the input settings. For testing or benchmarking purposes, a uniform-system solution can also be used, as its wave functions can be generated quickly. Once the wave functions are loaded, all buffers are initialized with quantities derived from them. After the initialization is complete, basic observables (such as total energy) are evaluated and printed, allowing the user to verify that the wave functions were loaded correctly.  
Next, the Adams-Bashforth-Moulton integration scheme is initialized. Depending on the selected option in the input file (tag \lstinline{selfstart}), the code either assumes that the wave functions represent a stationary state (so previous steps can be reconstructed analytically) or, if this assumption is not valid, the Taylor expansion~\cref{eq:Taylor-exp} is performed to initialize the scheme. This step concludes the preparation for the time evolution.

The main time-evolution loop (enumerated by the iterator \lstinline{i_meas}) consists of a nested integration loop (\lstinline{i_step}) and a reporting block responsible for writing observables and monitoring numerical stability. Inside the nested loop, the evolution proceeds through alternating predictor and corrector steps:
\begin{description}
\item[Predictor] The predictor step uses formulas~\crefrange{eq:AB3}{eq:AB5}, depending on the selected integration order (set in the \filename{predefines.h} file). In this logical block, densities are first computed, which involves MPI communication. From these densities, the potentials are constructed to form the instantaneous Hamiltonian operator. The Hamiltonian is then applied to the wave functions to evaluate $f(\vb{\varphi},t)$ defined in \cref{eq:ABM-f}. This step is the most computationally expensive part of the algorithm, as it requires the calculation of derivatives, corresponding to multiple Fourier transforms.

\item[Corrector] The corrector step is constructed analogously to the predictor. Depending on the integration order, either formula \cref{eq:AM4} or \cref{eq:AM5} is applied. As in the predictor stage, this step requires the evaluation of densities (involving MPI communication) and the application of the Hamiltonian operator, which again entails performing Fourier transforms.
\end{description}
Each pair of predictor and corrector executions advances the wave functions in time by a single time step $dt$. Once the system is propagated by the desired interval (corresponding to one measurement block), the code computes observables. Since densities and potentials have already been computed during the preceding corrector step, only additional quantities, such as total energy, need to be evaluated. These observables are then appended to the output files.  

At this stage, diagnostic tests are performed to monitor the stability of the time integration. The tests verify the conservation of key physical quantities, such as total energy and particle number, with the corresponding tolerances defined in the input file. If any of these checks fail, the code terminates automatically.  
The code also supports periodic checkpointing (controlled by the \lstinline{checkperiod} parameter), allowing long simulations to be safely resumed. Once the trajectory of the requested total length has been generated, the computation concludes and the program exits.

\subsubsection{Parallelization scheme}

\begin{figure}[tb]
	\centering
	\includegraphics[width=0.8\linewidth,trim={0cm 0cm 0cm 0cm},clip]{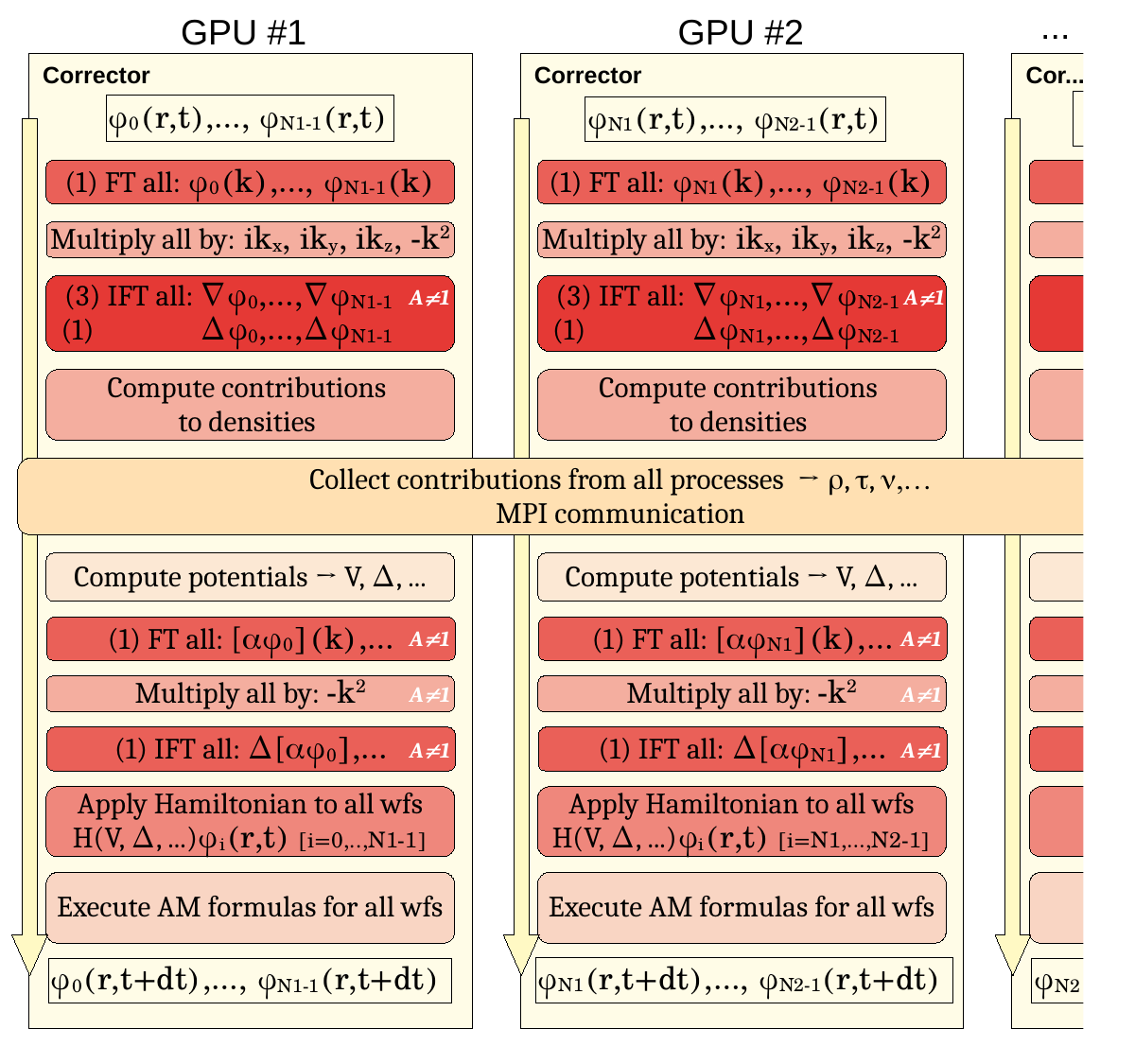}
	\caption{Parallelization scheme applied in the time-dependent codes.
    The total set of wave functions is distributed evenly among GPUs, so that each process evolves only a subset of them.
    Each computational block applies the same operation to all wave functions simultaneously to maximize performance.
    The color intensity of the blocks reflects the relative computational cost: the most time-consuming operations are the inverse (IFT) and forward (FT) Fourier transforms, whereas the least expensive is the construction of the potentials.
    The numbers in parentheses denote the number of times each transform is performed within the block.
    In total, the wave functions are transformed seven times (1+3+1+1+1) for $A_{\sigma} \neq 1$, and only twice (1+1) otherwise.
    The wide orange block spanning all compute units represents MPI communication between GPUs.}
	\label{fig:td-parallelization-scheme}
\end{figure}
The time-dependent codes perform all computations on GPUs. The total number of wave functions (\texttt{nwf}) to be evolved in time is evenly distributed among the available GPUs, so that each GPU handles approximately \texttt{nwf/np}, where \texttt{np} is the total number of computational units participating in the simulation.
To maximize performance, the toolkit adopts a strategy of massive parallel processing. Rather than evolving each wave function independently, operations are applied collectively to all wave functions in a single batch. This approach is illustrated in \cref{fig:td-parallelization-scheme}, which shows the sequence of operations performed during the corrector step. The predictor step follows the same logical structure.  

In each iteration, all wave functions are first transformed to momentum space using the Fourier transform. Next, element-wise multiplications by factors such as $ik_x$, $ik_y$, $ik_z$, and $-k^2$ are performed, and the resulting quantities are stored in working buffers. These buffers are then transformed back to coordinate space using the inverse Fourier transform, producing the gradients and Laplacians of the wave functions. 
The wave-functions together with derivatives are required to compute local densities ($\rho_{\sigma}$, $\tau_{\sigma}$, ...), the corresponding potentials ($V_\sigma$, $\alpha_\sigma$, $\Delta$, ...), and, ultimately, to apply the Hamiltonian operator. If the computation is performed with $A_{\sigma}\neq 1$, additional Fourier transforms are needed to evaluate the Laplacian of the product $\alpha(\vbr)\varphi(\vbr)$, as required for numerical stability\footnote{The Laplacian of $\alpha(\vbr)$ itself is also computed, but this step is omitted from the diagram since its cost is negligible compared to the processing of the quasiparticle wave functions.}; see \cref{eq:kinop-alpha}.
The computation of gradients and Laplacians (hence, the repeated execution of FFTs) constitutes the most time-consuming part of the time-dependent simulations. MPI communication occurs only during the construction of densities, where density fields are exchanged between GPUs. The total communication volume depends on the code variant but, in practical cases, does not exceed several tens of megabytes.  
Once the densities are assembled, the corresponding potentials are computed, defining the instantaneous Hamiltonian matrix. The Hamiltonian is then applied to all wave functions, and the subtraction of the instantaneous quasiparticle energy is carried out. Finally, the formulas defining the integration scheme are evaluated, completing the propagation block. 

\subsubsection{Memory consumption}

The time-dependent code achieves high performance at the cost of increased memory demand. Memory is required to store the solution vector $\vb{\varphi}$, historical states $\vb{f}_{k-1}, \vb{f}_{k-2}, \dots$ (whose number depends on the order of the integration scheme and typically ranges from 3 to 5), buffers for the gradients and Laplacians of the wave functions (four buffers for the 3D code, with one fewer buffer for each reduction in dimensionality), and additional working buffer used during intermediate steps.  
Each buffer has a size given by
\begin{equation}
  \texttt{buffer} = 2\,N_{\textrm{wf}}\,M^{(d)} \times \texttt{sizeof(double complex)},
\end{equation}
where $N_{\textrm{wf}}$ is the number of evolved wave functions (typically comparable to the number of lattice points), and the factor of 2 accounts for the $u$ and $v$ components of each quasiparticle orbital. For perspective, in 3D simulations on a $100^3$ lattice, a single buffer requires approximately $15$~TB of memory.  

In the most demanding case (3D calculations performed with the AB5AM5 integration scheme) the total memory consumption can be estimated as
\begin{equation}
  \texttt{mem}_{\textrm{td}} \approx 11 \times \texttt{buffer} \; [\texttt{Bytes}].
\end{equation}
The W-SLDA Toolkit provides a dedicated script, located at \filename{$WSLDA/tools/td-memory.py}, that allows users to estimate the total memory requirement for a given setup and determine the minimal number of GPUs necessary to execute the simulation efficiently.  
For optimal performance, it is recommended that the memory utilization on each GPU reaches at least 50\% of its available capacity. 

\subsubsection{Scaling}\label{sec:Scaling-td}

\begin{figure}[htbp]
	\centering
	\includegraphics[width=0.9\linewidth,trim={0cm 0cm 0cm 0cm},clip]{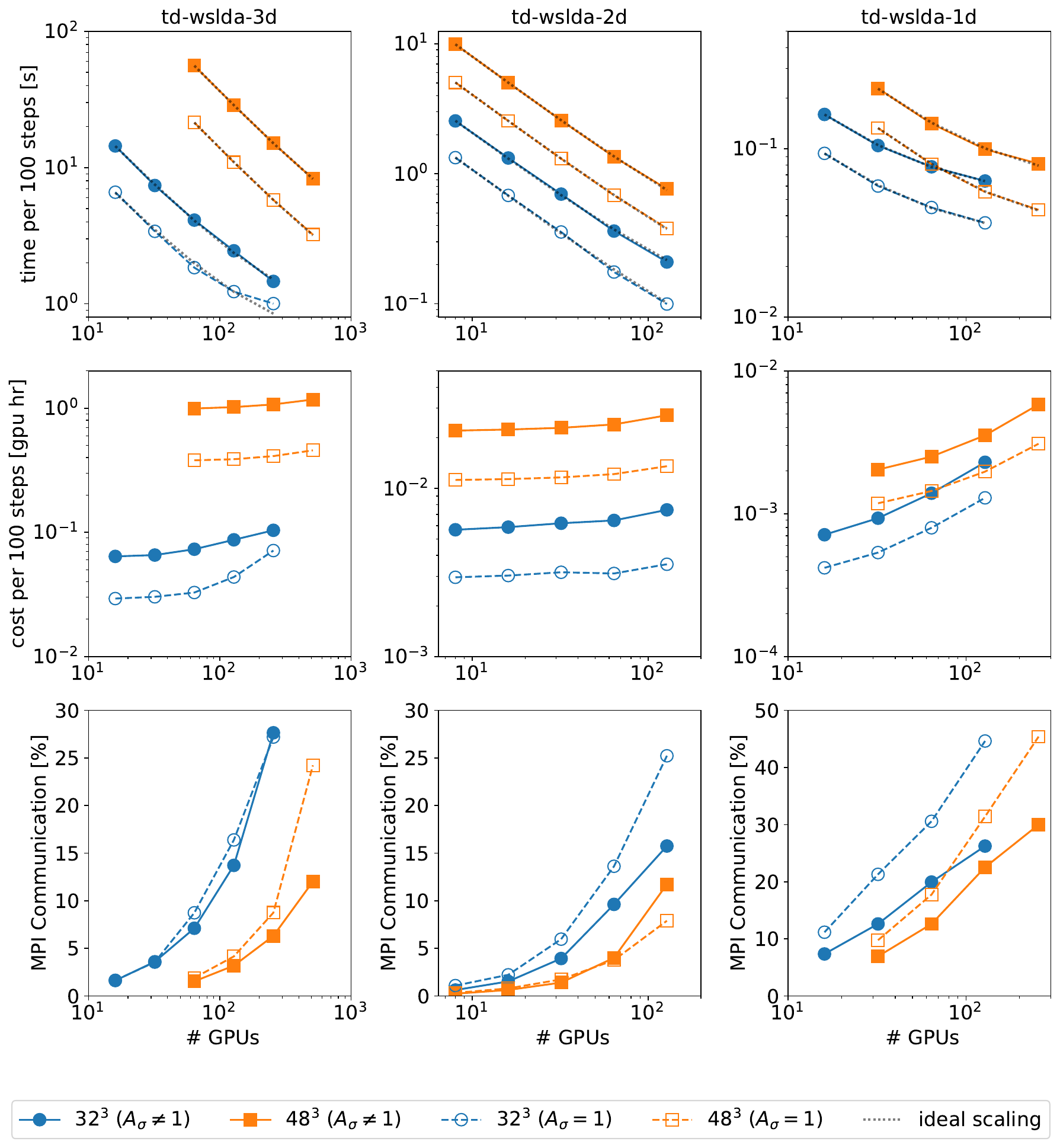}
	\caption{Scaling plots for the family of time-dependent codes. The top row shows the time per 100 integration steps (in seconds) versus the number of utilized GPUs. Two lattice sizes were considered, $32^3$ and $48^3$, and two cases were analyzed: the effective mass equal to the bare mass ($A_{\sigma}=1$) and the effective mass different from the bare one ($A_{\sigma}\neq 1$). The gray dotted line indicates the ideal scaling, i.e., the computation time decreases by a factor of two when the number of GPUs is doubled. The middle row presents the computation cost per 100 time steps, measured in GPU hours, defined as the product of run time and the number of GPUs. For ideal scaling, this quantity should remain constant with respect to the number of GPUs. \respch{The bottom row presents the percentage of runtime spent on MPI communication.} All tests were performed on the LUMI supercomputer (EuroHPC/CSC, Finland), equipped with AMD MI250X GPUs.}
	\label{fig:td-scaling}
\end{figure}

\Cref{fig:td-scaling} presents scaling results for the family of time-dependent codes. Two representative lattice sizes, $32^3$ and $48^3$, were considered. The plots show the time required to perform 100 integration time steps, including the evaluation of observables and data output. Two variants of the energy functional were tested: the general functional \cref{eq:sldae-functional-generic}, denoted by $A_{\sigma} \neq 1$, and the simplified form assuming $A_{\sigma}=1$.

The simplified functional yields a performance gain of approximately a factor of three, in agreement with theoretical expectations. As discussed in \cref{subsec:integation-td}, the general mode requires seven Fourier transforms per predictor (or corrector) step, whereas this number is reduced to only two when $A_{\sigma} = 1$. Consequently, the total number of Fourier transforms decreases by roughly a factor of 3.5, which directly translates into the observed speed-up. This result confirms that the overall computational time is indeed dominated by FFT operations. 

As in the static case, exploiting translational symmetries has a significant impact on the computational cost. The time-dependent codes demonstrate good strong-scaling properties, especially for the 3D and 2D variants. This is evident in the plots of computation cost versus the number of GPUs, which show only weak dependence, indicating that scaling is primarily limited by computation rather than communication. In contrast, the \respch{quasi-}1D variant is more affected by MPI communication overheads. This behavior is expected, since in the \respch{quasi-}1D case the computation involves a large number of short vectors (each of size $N_x$), leading to low computation time but a relatively larger share of communication cost. 
\respch{Indeed, the time percentage spent on MPI communication, for \respch{quasi-}1D code is significantly larger than in the case of \respch{quasi-}2D and \respch{3}D codes, as seen from the bottom row of \Cref{fig:td-scaling}.}
However, this effect is of minor practical importance because \respch{quasi-}1D calculations are computationally inexpensive compared to the higher-dimensional cases.  

The overall computational scaling per time step follows
\[
  \texttt{cost}_{\textrm{td}} \propto N_{\textrm{wf}} \cdot M^{(d)}\log M^{(d)},
\]
where $N_{\textrm{wf}}$ is the number of evolved wave functions and $M^{(d)}$ is the total number of lattice points. The first factor depends on the chosen energy cutoff. For the recommended case $E_c \approx \frac{\pi^2}{2\Delta x^2}$, one can estimate
\[
  N_{\textrm{wf}} \approx \frac{1}{2} N_x N_y N_z,
\]
assuming a spin-symmetric system ($N_a = N_b$). For spin-imbalanced systems, the number of wave functions doubles. The second factor $M^{(d)}$ depends on dimensionality. This scaling behavior originates from the FFT routines, which dominate the computational workload. For example, increasing the lattice size from $32^3$ to $48^3$ should increase the computational cost by a factor of
$
\frac{48^6 \log 48^3}{32^6 \log 32^3} \approx 13,
$
which is consistent with the measured results.

\subsection{Interoperability between static and time-dependent codes}\label{sec:dependencies}
\begin{figure}[t]
	\centering
	\includegraphics[width=0.9\linewidth]{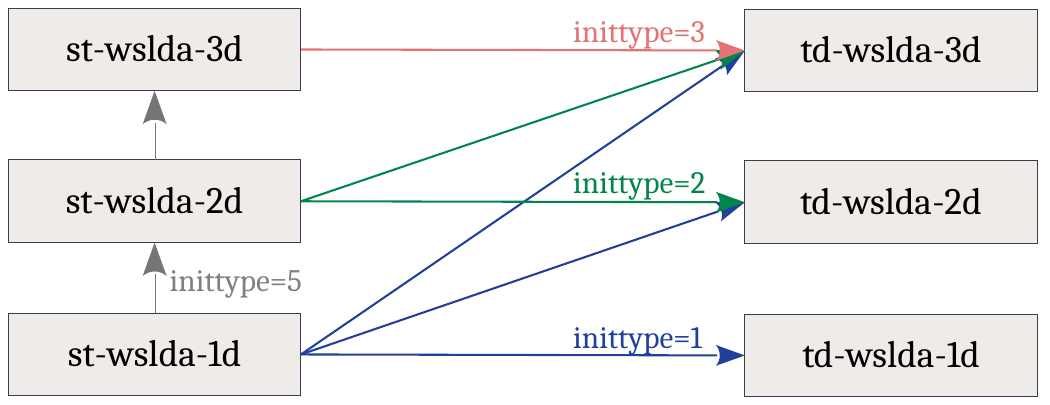}
  \caption{Dependencies between static and time-dependent codes.
    Arrows indicate which code outputs can serve as inputs for others.
    The tag \lstinline{inittype} specifies the value that should be used in the input file to properly initialize a given calculation.}
	\label{fig:inittypes}
\end{figure}
All codes (\filename{st-wslda-?d} and \filename{td-wslda-?d}) are fully compatible with each other, meaning that the output generated by one can be used as input for another.  
Typically, static codes are employed to generate initial states for subsequent time-dependent simulations.  
\Cref{fig:inittypes} illustrates these relationships, showing how the output of each code can serve as the starting point for simulations performed by others.  
For example, \filename{td-wslda-1d} can evolve only wave functions produced by \filename{st-wslda-1d}, whereas the three-dimensional variant \filename{td-wslda-3d} can be initialized using the output from any of the static codes.  
Furthermore, the output of a lower-dimensional static code can be used as an initial guess for self-consistent iterations of a higher-dimensional solver.  
This capability is particularly useful for improving convergence efficiency and is described in detail in \cref{subsec:conv-releasing-symmetry}.  

The interoperability between the static and time-dependent codes provides users with significant flexibility when designing efficient computational workflows.

\subsection{Storage format of results}\label{sec:storage-format}

\subsubsection{Format}\label{subsec:format}
The results generated by the W-SLDA Toolkit are stored in a unified format. These are either text files or binary files. The format is designed to be conceptually easy to understand, which turns out to be an important issue for academic applications where students are typically involved. All produced files share a common prefix, controlled by the \lstinline{outprefix} tag in the input file. For example, if \lstinline{outprefix=test}, the produced files will be:
\begin{lstlisting}[style=Input]
test.stdout          # Text file containing a copy of the standard output
                     #   (all messages printed on the screen) 
test.wlog            # text file produced by the logger(...) function
test.wtxt            # text file describing all generated binary files
                     #   (use this file when reading data with Python or VisIt)
test_rho_a.wdat      # binary file containing the density field
test_delta.wdat      # binary file containing the delta field
test_j_a.wdat        # binary file containing the current field
...
\end{lstlisting}
The data storage for functions $f(\vbr)$, like densities or potential, is   organized as follows: each variable is stored in a separate binary file. These binary files store consecutive arrays (for each measurement), with no extra header. The header is instead placed in a separate text file. On the list presented above, the header file is \filename{test.wtxt}, and its content may look like
\begin{lstlisting}[style=Input]
# W-DATA 0.3.0
# Comments with additional  info about the data set
# Comments are ignored when reading by the parser  
nx            24   # lattice
ny            28   # lattice
nz            32   # lattice
dx             1   # spacing
dy             1   # spacing
dz             1   # spacing
datadim        3   # dimension of block size: 1=nx, 2=nx*ny, 3=nx*ny*nz
prefix      test   # prefix for files belonging to this data set
cycles        10   # number of cycles (measurements)
t0             0   # time value for the first cycle
dt             1   # time interval between cycles
# variables
# tag       name     type    unit  format
var        rho_a     real    none    wdat
var        delta  complex    none    wdat
var          j_a   vector    none    wdat
\end{lstlisting}
The content is self-explanatory. In this case, for a 3D simulation (\lstinline{datadim}) preformed on a lattice $24\times 28\times 32$, ten measurements (\lstinline{cycles}) was generated for quantities: density $\rho_a(\vbr)$, delta $\Delta(\vbr)$ and current $\vbj_a(\vbr)$.
These measurements correspond to times $\texttt{t0}+k\texttt{dt}$, where $k=0,1,\dots,9$.
The format we use for storing binary data is known as the W-data format~\cite{WDataFormat}.
Its simplicity assures that it can be read with a variety of tools, without the need to rely on external libraries.
The details on this format can be found on webpage~\cite{WDataFormat}, or in folder \filename{$WSLDA/lib/wdata/doc}. 

The form of output is not strictly imposed. The user can customize the output format and the content of files using the provided interface through the \filename{logger.h} file. However, we recommend maintaining the integrity of the format we used, as the toolkit provides many extensions that can be used for results processing, and all of them assume that results are stored in a format as implemented in the default form of \filename{logger.h}. 

\subsubsection{Reproducibility packs}\label{subsec:repropacks}

Result reproducibility is a cornerstone of modern computational science. It has been widely recognized that, in many cases, reproducing one’s own results after some time (for example, after a year) can be challenging~\cite{Perkel2020,Ziemann2023,costa2025}. In most situations, having access to the same code version is not sufficient; one also needs precise knowledge of all input parameters and access to the exact input data.  

Since scientific research often relies on a trial-and-error methodology, researchers typically generate numerous datasets during a project. Only a subset of these results is ultimately published, while the remainder serve as exploratory or test runs. Under such conditions, tracking every modification introduced during the research process becomes difficult.  
To address this issue, the W-SLDA Toolkit implements an automated mechanism for result reproducibility. Each produced data set is accompanied by a \textit{reproducibility pack} that contains all information required to exactly reproduce the results (up to machine precision).  

Each output file generated by the toolkit includes in its header the version of the code used. During execution, the toolkit automatically saves copies of all user-definable files and attaches them to the output data set. For example, the W-data set (as described in the preceding section) is supplemented by the following files:
\begin{lstlisting}[style=Input]
...
test_input.txt            # input file used for the calculation
test_predefines.h         # compile-time predefines
test_problem-definition.h # user-defined problem setup 
test_logger.h             # customized logging configuration
test_machine.h            # machine configuration used in the computation
test_checkpoint.dat.init  # checkpoint file used as input (st codes only)
test_extra_data.dat       # binary file containing the extra_data array (if provided)
test_reprowf.tar          # reproducibility pack for restoring input wave functions (td codes only)
\end{lstlisting}
Together, these files provide all the information necessary to reproduce the results exactly, assuming the same version of the toolkit is used.  
In general, the W-SLDA Toolkit also supports backward compatibility, allowing newer versions of the code to reproduce results obtained with earlier releases.

\subsubsection{Integration with Python and VisIt}

The W-data format used by the W-SLDA Toolkit is intentionally designed to be simple and lightweight.  
It can be accessed directly using standard programming tools without the need for any dedicated libraries.  
However, for convenience and to enhance research productivity, the toolkit provides dedicated support for this format through the \texttt{wdata} Python library, which can be easily installed via \texttt{pip}:
\begin{lstlisting}[style=Commands]
pip install wdata
\end{lstlisting}
This library offers a user-friendly interface for reading, writing, and processing W-data files. It also includes high-level utilities for handling time-dependent datasets, computing derived quantities, and producing quick visualizations (with the help of \texttt{matplotlib}). See \cref{subsec:example-usage} for a demonstration of its usage.   

In addition to Python integration, the W-data format is natively supported by the VisIt visualization software~\cite{HPVVisIt,VisItweb}.  VisIt provides a powerful environment for data exploration, especially in the case of 2D and 3D simulations.  
It allows users to perform volume rendering, contour plotting, vector field visualization, and interactive data analysis.  
For example, several visualizations shown in \cref{fig:mesh-dim} were produced using VisIt.  

Furthermore, both C and Python libraries supporting the W-data format are available in the \filename{$WSLDA/lib/wdata} directory.  
These can be used to build custom data analysis or post-processing tools that integrate directly with the toolkit’s output files.
Comprehensive documentation and example scripts demonstrating their use are provided online in the W-SLDA Toolkit wiki pages and code repository.

\section{Toolkit usage}

\subsection{Example: dynamics of a Josephson junction}\label{subsec:example-usage}

In this section, we present an example of a complete workflow: from setting up the calculations to executing them and analyzing the results. We consider one of the most fundamental phenomena in superfluids and superconductors: the Josephson effect~\cite{JOSEPHSON1962251,RevModPhys.36.216}. This effect has been observed experimentally in fermionic superfluids using configurations that split an atomic cloud into two parts through a narrow potential barrier~\cite{Valtolina2015,PhysRevLett.120.025302,PhysRevLett.124.045301,Kwon2020,Luick_2020}. It has also been investigated theoretically~\cite{PhysRevLett.99.040401,Spuntarelli2010}, including studies within the W-SLDA Toolkit framework~\cite{Wlazlowski2023}.  

To keep the example light enough for execution even on a small personal computer (equipped with either NVIDIA or AMD graphics cards), we restrict the problem to a quasi-1D geometry. The complete set of files for this example is also available via Zenodo~\cite{zenodo_example}. 

\paragraph{Step 1: Creating working folders}  
Assuming that the toolkit is located in the path stored in the variable \lstinline{$WSLDA}, we begin by creating working folders for the project. We will first generate an initial state (solving the static problem) and then evolve it in time (using the time-dependent code). Thus, we create two working directories, \filename{st-jj-example} and \filename{td-jj-example}, by copying the corresponding template folders:
\begin{lstlisting}[style=Commands]
cp -r $WSLDA/st-project-template ./st-jj-example
cp -r $WSLDA/td-project-template ./td-jj-example
\end{lstlisting}
All file modifications related to the static part of the calculation should be performed within the \filename{st-jj-example} directory (Steps 2-4), while those associated with the time-dependent calculations should be made within the \filename{td-jj-example} directory (Steps 5-7).

\paragraph{Step 2: Editing files for the static calculations}  
In this step, we prepare the configuration for producing the initial state, which represents a unitary Fermi gas confined in a one-dimensional external trap of the form
\begin{equation}
  \label{eq:st-jj-trap}
  \Vext{\sigma}(x) = \frac{1}{2}\omega_x^2 x^2 + V_0e^{-2x^2/w^2} + f_{\textrm{tilt}} x,
\end{equation}
where the first term corresponds to harmonic confinement, the second describes the potential barrier separating the two reservoirs, and the last introduces a small linear tilt to generate a population imbalance between the left and right wells in the initial state.

\begin{figure}[tb]
	\centering
	\includegraphics[width=0.9\linewidth,trim={0cm 0.6cm 0cm 0.6cm},clip]{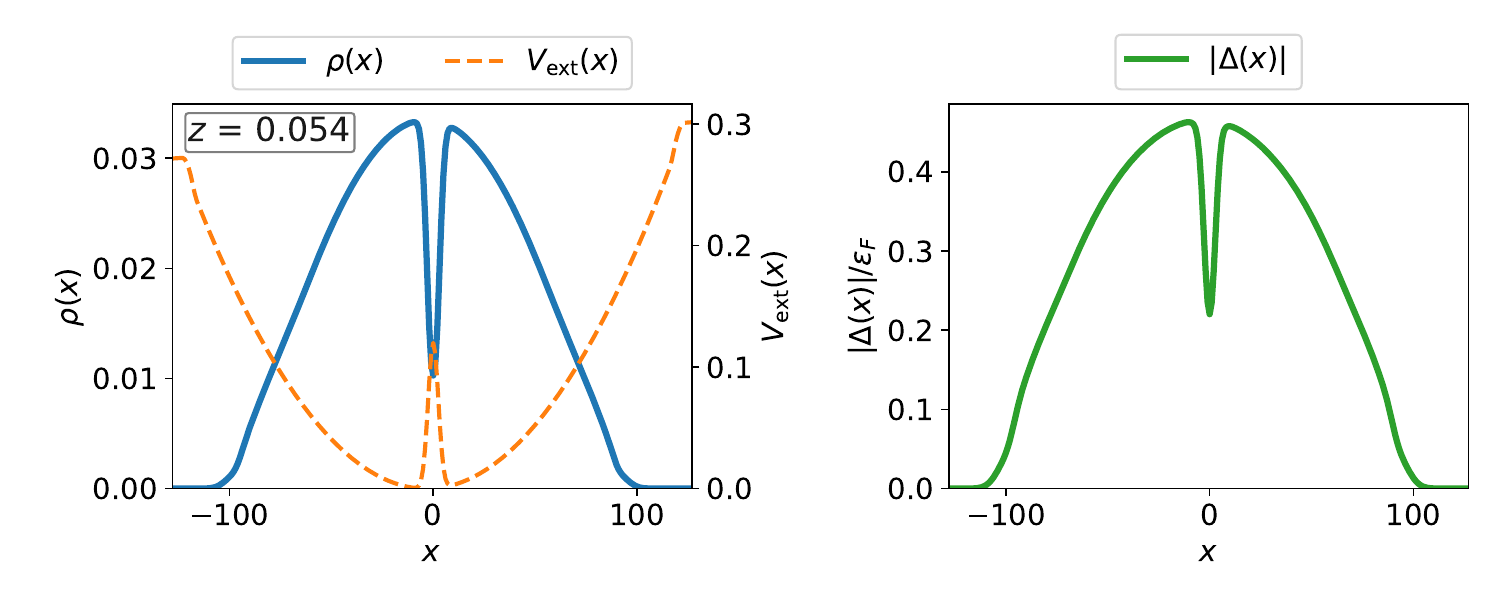}
	\caption{Visualization of the self-consistent solution for a unitary Fermi gas confined in the potential~\cref{eq:st-jj-trap}. 
	Left: total density (solid line) and external potential (dashed line). 
	Right: spatial profile of the magnitude of the pairing field $\abs{\Delta(x)}$. 
	The complete Python script used to generate this visualization is provided in the supplementary material.}
	\label{fig:st-jj-example}
\end{figure}

First, we select the lattice size and functional by editing the \filename{predefines.h} file. For brevity, only the modified fragments are listed below:
\begin{lstlisting}
// in predefines.h
/**
 * Define lattice size and lattice spacing.
 * */
#define NX (256)
#define NY (16)
#define NZ (16)
// ...
#define FUNCTIONAL SLDAE
/**
 * Sets the effective mass to be equal 1.
 * */
#define SLDA_FORCE_A1
// ...
\end{lstlisting}
Here, we enforce calculations with the effective mass equal to the bare mass ($A_{\sigma} = 1$) to accelerate convergence.  
Next, we define the physical problem in the \filename{problem-definition.h} file:
\begin{lstlisting}
// in problem-definition.h
#include "../extensions/wslda_utils.h"
// ...
void process_params(double *params, double *kF, double *mu, size_t extra_data_size, void *extra_data)
{
    // if akF is active, then overwrite the value of input->sclgth
    // value of kF is generated by the function referencekF(...)
    if(input->akF!=0.0) input->sclgth = input->akF/kF[0];

    // PROCESS INPUT FILE PARAMETERS
    // Definition of R_TF:  0.5*m*omega^2 * R_TF^2 = mu
    //                      => omega = sqrt(2*mu/m)/R_TF
    params[11] = sqrt(2.*mu[SPINA])/ (params[1]*LX*0.5); // omega_x
    double eF = kF[0]*kF[0]/2.0;
    params[4] = params[4]*eF; // convert barrier height to internal units
    params[5] = params[5]/kF[0]; // convert barrier width to internal units
}
// ...
double v_ext(int ix, int iy, int iz, int it, int spin, double *params, size_t extra_data_size, void *extra_data)
{
    double x = DX*(ix-NX/2);

    double V_ho = harmonic_oscillator_smooth_edges(x, params[11], params[2]*LX*0.5, params[3]*LX*0.5, 1.0);
    double V_barrier = params[4] * exp(-2.0*pow(x/params[5],2));
    double V_tilt = params[6]*x/(LX/2); // linear tilt potential
    return V_ho + V_barrier + V_tilt;
}
// ...
\end{lstlisting}
The function \lstinline{process_params(...)} converts user-provided parameters (in human-readable units) into internal units suitable for computation. For example, rather than specifying the trapping frequency $\omega_x$ directly, user provides the desired Thomas-Fermi radius $x_{\mathrm{TF}}$, which is related to the chemical potential through
\begin{equation}
    \frac{1}{2}\omega_x^2 x_{\mathrm{TF}}^2 = \mu
    \quad \Rightarrow \quad
    \omega_x = \frac{\sqrt{2\mu}}{x_{\mathrm{TF}}}.
\end{equation}
Since $\mu$ is not known a priori, \lstinline{process_params(...)} provides the necessary interface to adjust user input to internal code parameters.  
Another example concerns the barrier height $V_0$ and width $w$, which are more conveniently expressed in terms of the Fermi energy $\eF$ and Fermi wave vector $\kF$. The function \lstinline{process_params(...)} performs these conversions automatically. The parameters, in turn, are provided through the \filename{input.txt} file, for example:
\begin{lstlisting}[style=Input]
# -------------- USER DEFINED PARAMETERS --------------
# Data flow: [Read params from input file] -> [execute process_params( )] 
#                    -> [pass params to functions]
params[1] = 0.80 # Thomas-Fermi radius for x direction, in units of LX/2
params[2] = 0.90 # Smoothing parameter x1 for harmonic_oscillator_smooth_edges
params[3] = 0.97 # Smoothing parameter x2 for harmonic_oscillator_smooth_edges
params[4] = 0.24 # height of the barrier, in eF units
params[5] = 5.0  # width of barrier, in 1/kF units
params[6] = 0.015# tilt potential (value at the edge of the trap)
# ...
\end{lstlisting}
Each function in the toolkit accesses the \lstinline{params[]} array, containing parameters converted by \lstinline{process_params(...)}. The external potential formula is defined in \lstinline{v_ext(...)}. Here we employ a predefined smooth-edged harmonic potential to ensure continuity and differentiability at the boundaries of the simulation box (see \cref{fig:st-jj-example}). The function also uses the parameter \lstinline{params[11]}, which stores $\omega_x$ computed in \lstinline{process_params(...)}.

Since no customization of the output is required (which could otherwise be defined in \filename{logger.h}), only the input file remains to be prepared. The other relevant parameters for this example are:
\begin{lstlisting}[style=Input]
# ...
# ------------------- INITIALIZATION ------------------
inittype     0       # create uniform solution and start from it,
# ...
# ------------------- INPUT/OUTPUT ------------------
outprefix    jj-r1   # all output files will start with this prefix
writewf      1       # write wf at the end of computation: yes=1, no=0
                     # we will evolve them by td-wslda-1d code
# variables to write
writevar    rho delta j V_ext 
sclgth      -10000.0 # scattering length in units of lattice spacing
referencekF 1.0      # in this case, I can define the value of kF=1
# ...
# ----------------- SELF-CONSISTENT LOOP -----------------
energyconveps 1.0e-6 # energy convergence epsilon
npartconveps  1.0e+6 # large number means that this condition will always be satisfied
muchange      0.0    # do not change chemical potential, 
                     # so computation will be for fixed chemical potential
maxiters     200     # maximum number of iterations
spinsymmetry 1       # to impose Na=Nb
# ...
# ----------- IN CASE OF: inittype=0 ----------------
# See: Wiki -> Initialization of the solver
init0na      0.01689 # required density for component a
init0nb      0.01689 # and component b, so corresponding kF=1
init0muchange 0.5    # change rate of chemical potentials, default muchange
init0eps      1.0e-9 # epsilon for convergence for uniform solution
init0maxiter  1000   # maximum number of iterations when searching for a  uniform solution
# ...
\end{lstlisting}
With this configuration, the solver starts from a uniform solution corresponding to $\kF = 1$. For this reason, we can define the value of the reference scale in advance (tag \lstinline{referencekF}). The computation proceeds at fixed chemical potential (\lstinline{muchange = 0}), while the particle-number convergence condition is disabled (\lstinline{npartconveps = 1.0e+6}).

\paragraph{Step 3: Compiling and running the static code}  
To generate the binary for a quasi-1D simulation, execute the command:
\begin{lstlisting}[style=Commands]
make 1d    
\end{lstlisting}
Depending on your system configuration, you may need to modify the \filename{Makefile} to ensure compatibility with your local environment, and set environment variables such as \lstinline{LD_LIBRARY_PATH}. It may also be necessary to edit the \filename{machine.h} file, which stores information about the computing environment and installed libraries. In particular, this file allows you to choose the diagonalization routine (by default, ScaLAPACK is used as the diagonalization engine). The database of examples of Makefiles and environmental settings for various machines is located in the folder \filename{$WSLDA/templates}.

If the compilation completes successfully, a binary named \filename{st-wslda-1d} will be created.
This program should be executed within an MPI environment.
The following example illustrates a run using four MPI processes:
\begin{lstlisting}[style=Commands]
mpirun -np 4 ./st-wslda-1d input.txt
# START OF THE MAIN FUNCTION
# CODE: ST-WSLDA-1D
...
\end{lstlisting}
As a result, a set of output files will be generated, all starting with the string specified by \lstinline{outprefix} (in this example, \lstinline{jj-r1}). Some of these files are text-based (\filename{*.stdout}, \filename{*.wtxt}, \filename{*.wlog}, \filename{*.h}), while others contain binary data (\filename{*.wdat}).  

\paragraph{Step 4: Analyzing the results of the static run}  
We now demonstrate a basic analysis workflow using Python. As a simple example, we will plot the density profile and compute the population imbalance, defined as $z=(N_L-N_R)/(N_L+N_R)$, where $N_L$ and $N_R$ are particle numbers in left and right reservoirs:
\begin{lstlisting}[style=Python]
import numpy as np
import matplotlib.pyplot as plt
# to get wdata lib use: pip install wdata
from wdata.io import WData, Var

# Load data
data = WData.load("./st-jj-example/jj-r1.wtxt")
x=data.xyz[0]
rho=data.rho_a[-1,:]+data.rho_b[-1,:] # take last iteration
# compute initial population imbalance
NL = np.sum(rho[x<0])
NR = np.sum(rho[x>0])
z=(NL-NR)/(NL+NR) # initial population imbalance
print(z)

# Plot rho(x)
plt.plot(x,rho)
plt.show()
\end{lstlisting}
This script yields an initial population imbalance of $z = 0.054$.  
A more detailed visualization of the static solution for the settings defined in Step~2 is shown in \cref{fig:st-jj-example}.

\paragraph{Step 5: Editing files for the time-dependent calculations}  
The workflow for time-dependent runs is very similar to that for the static case. One should modify the \filename{predefines.h} and \filename{problem-definition.h} files in an analogous manner, but now editing the files located in the \filename{td-jj-example} directory. These files are not interchangeable between the static and time-dependent codes; however, most of the functions they contain, such as \lstinline{process_params(...)} and \lstinline{v_ext(...)}, can be reused directly by copy and paste. For this reason, their listings are omitted here.  

In the input file, the key parameters for this example are:
\begin{lstlisting}[style=Input]
# -------------- USER DEFINED PARAMETERS --------------
# ...
params[6] = 0.0  # no tilt
# ...
# ------------------- INITIALIZATION ------------------
inittype     1  # start from st-wslda-1d solution, 
                # inprefix points to data from static calculations
# ...
# ------------------- INPUT/OUTPUT ------------------
outprefix    jj-r2   # all output files will start with this prefix
inprefix  ../st-jj-example/jj-r1 # prefix of static result
# variables to write
writevar    rho delta j V_ext 
sclgth      -10000.0 # scattering length in units of lattice spacing
# ...
# ------------------- TIME EVOLUTION ----------------
dt        0.0035 # integration time step, in units of eF^-1
measurements 200 # number of requested measurements
timesteps   1430 # number of time steps between each measurement
                 # measurements are written with time resolution: timesteps*dt
                 # the total trajectory length is: measurements*timesteps*dt
selfstart      1 # use Taylor expansion for first 5 steps? 
# ...
\end{lstlisting}
Since the goal is to observe Josephson dynamics, the tilting potential is deactivated (\lstinline{params[6]=0.0}), so the initial state no longer corresponds to a stationary configuration. The code reads the wave functions generated by the static solver (specified via \lstinline{inprefix}) and writes results under a new prefix (\lstinline{outprefix}).  
The integration time step is controlled by the tag \lstinline{dt}. The number of measurements (\lstinline{measurements}) determines how many times observables will be computed and written to output files, while the parameter \lstinline{timesteps} specifies the number of time steps between successive measurements.  
In this case, $1430$ time steps are executed between each measurement, corresponding to a sampling interval of $1430 \times 0.0035 = 5.005$ (in $\eF^{-1}$ units). The total trajectory length is therefore $200 \times 5.005 = 1001$.  

\paragraph{Step 6: Compiling and running the time-dependent code}  
The compilation and execution procedure is analogous to the static case, except that the resulting executable is now named \filename{td-wslda-1d}. This program requires access to a GPU, so a compatible GPU software stack must be installed (CUDA Toolkit for NVIDIA GPUs or ROCm for AMD GPUs).  
Compilation and execution can be performed as follows:
\begin{lstlisting}[style=Commands]
make 1d
...
mpirun -np 1 ./td-wslda-1d input.txt
# START OF THE MAIN FUNCTION
# CODE: TD-WSLDA-1D
...
\end{lstlisting}
As before, the program must be launched within an MPI environment. Here, the argument \texttt{np} specifies the number of GPUs to be used. In this example, a single GPU is used, making the code easily executable on a workstation or laptop equipped with a single GPU.  

\paragraph{Step 7: Analyzing the results of the time-dependent run}  
The output format of the time-dependent code is identical to that of the static code. Therefore, the same Python scripts can be used to compute quantities such as the population imbalance and to visualize the density distribution. The only difference is that, instead of iterations, the time-dependent results are stored as successive measurements.  

\begin{figure}[tb]
	\centering
	\includegraphics[width=0.9\linewidth,trim={0cm 0.6cm 0cm 0.6cm},clip]{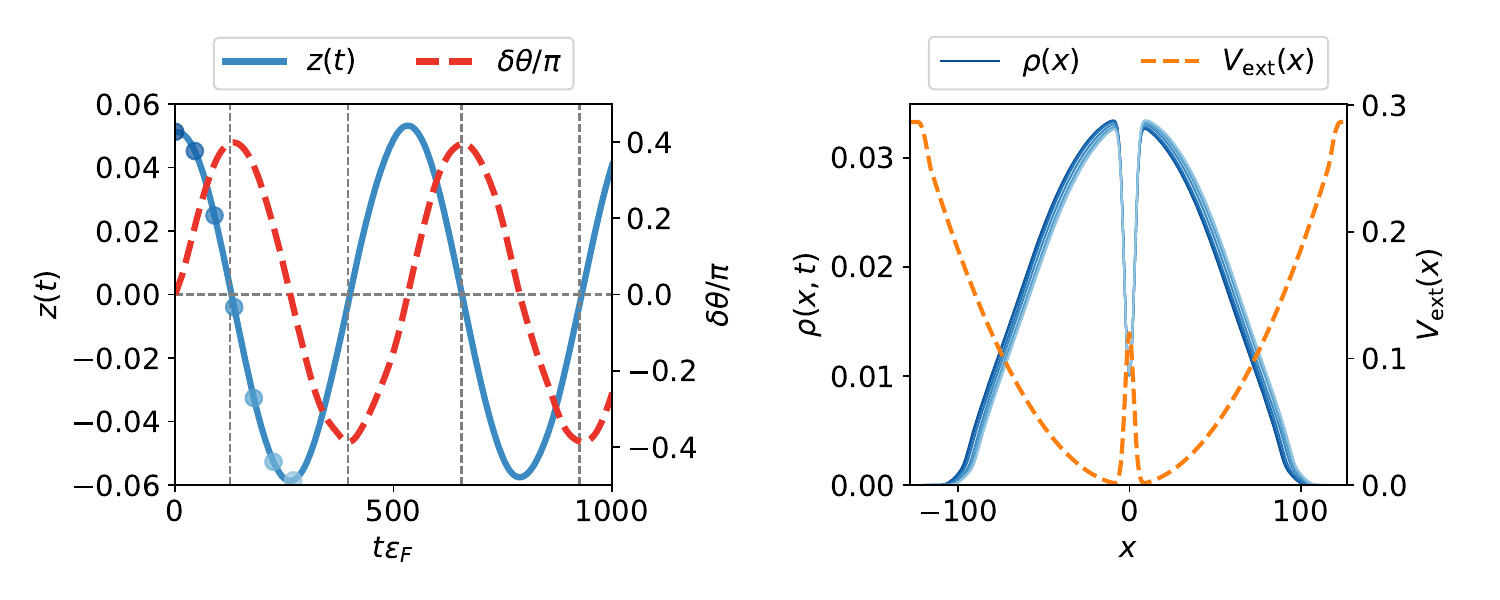}
	\caption{Time evolution of the atomic Josephson dynamics in a unitary Fermi gas.
	Left: time dependence of the population imbalance (solid line) and the phase difference across the junction (dashed line). 
	Right: spatial profile of the density distribution $\rhoP(x)$ for selected time moments, indicated by circles in the left panel. The dashed line shows the external potential.  
	The full Python script used to generate this visualization is provided in the supplementary material.}
	\label{fig:td-jj-example}
\end{figure}

\Cref{fig:td-jj-example} presents the results of the Josephson dynamics obtained from the simulation.
The figure shows two key quantities: the population imbalance $z(t)$ and the phase difference $\delta\theta(t)$ of the order parameter across the junction. Both exhibit periodic oscillations in time, with their phases shifted by $\pi/2$, as expected for an ideal Josephson junction.

\respch{
\paragraph{Checking the stability of the results with respect to the algorithm parameters}

Finally, we demonstrate the stability of the results with respect to the choice of algorithmic parameters. We note that it is the user's responsibility to verify whether the obtained results are trustworthy for a given physical setup. Here, we provide basic stability checks for two control parameters: the lattice resolution $\Delta x \kF$ and the integration time step $\Delta t\eF$ (both expressed in dimensionless units).

In the left panel of Fig.~\ref{fig:jj-example-stability}, we show how the energy of the initial state changes as a function of the lattice resolution in the static calculations. Specifically, we keep the physical size of the box fixed, $L_x=N_x\Delta x$, and vary the lattice spacing $\Delta x$. The figure shows that, for the resolution used in the example, $\Delta x\kF=1$, the ground-state energy is reproduced with a relative accuracy of about $1\%$. This is typically smaller than the uncertainty of the quantum Monte Carlo data used to calibrate the functional.
In the right panel of Fig.~\ref{fig:jj-example-stability}, we show the total energy of the system during the Josephson oscillations. This quantity should be conserved, and numerically this is achieved with a relative accuracy of about $10^{-4}$. This accuracy threshold is set by the regularization procedure, as discussed in Sec.~\ref{sec:reg_and_energy_conservation}. We also observe that, for time steps $\Delta t\eF\lesssim 0.014$, the integration remains numerically stable, whereas larger time steps lead to numerical instabilities.

\begin{figure}[tb]
	\centering
	\includegraphics[width=0.9\linewidth]{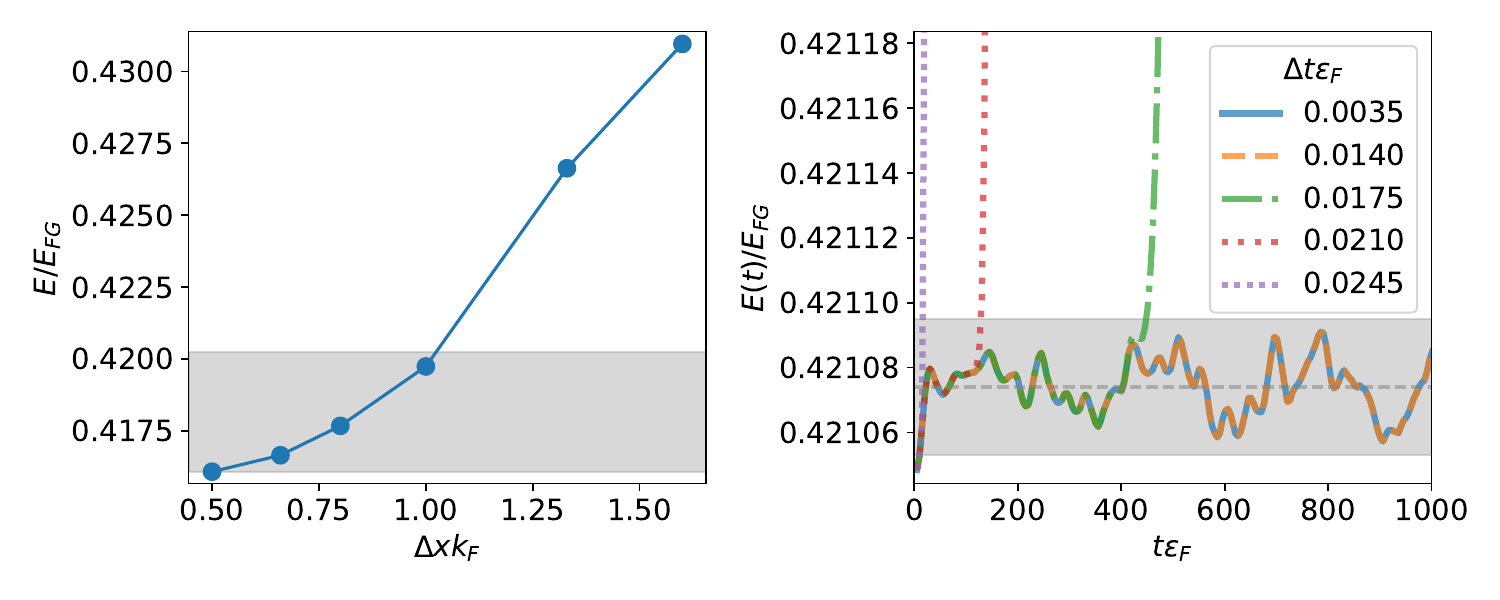}
	\caption{\respch{Stability checks with respect to the basic algorithmic parameters. Left: dependence of the static-solution energy on the lattice resolution. The gray band indicates a relative variation of $1\%$ with respect to the highest resolution case. Right: evolution of the total energy for different values of the integration time step $\Delta t$. For sufficiently small time steps, $\Delta t\eF\lesssim 0.014$, the accuracy of energy conservation is limited by the regularization scheme, while for larger time steps the integration becomes unstable. The gray band indicates relative variations of $10^{-4}$.}}
	\label{fig:jj-example-stability}
\end{figure}
}

\subsection{Obtaining ground and initial states for time evolution}

The example presented above illustrates a relatively simple case in which the initial state can be generated without difficulty. However, complications arise when attempting to construct more complex configurations, particularly in full three-dimensional geometries.  

The accuracy and stability of superfluid dynamics simulations strongly depend on the ability to prepare suitable initial states for time evolution.
Ultracold atomic systems provide an exceptionally versatile platform, where parameters such as the interaction strength can be tuned across different regimes, the system can be confined in external potentials of arbitrary geometry, and its effective dimensionality can vary.
Moreover, many other experimental parameters can be precisely controlled in modern setups.
This broad versatility also poses a significant challenge for computational approaches that aim to offer general-purpose functionality.
The main difficulty stems from the highly nonlinear character of the SLDA equations.
As is well established in the DFT community, there exists no universal procedure that guarantees convergence of the self-consistent cycle in static problems.
Consequently, various methods and convergence strategies have been developed over time and are applied depending on the specific physical system under investigation~\cite{Woods2019}.

As a general guiding principle, the following rule can be formulated:
\epigraph{The closer the initial guess is to the true solution, the higher the chance of convergence, and the fewer iterations are required to achieve it.}%
This principle motivates different strategies for constructing the initial guess. The W-SLDA Toolkit supports multiple methodologies for generating ground or initial states for subsequent time evolution.

\subsubsection{Starting from custom or predefined initial states}
A common strategy is to begin from a predefined or user-specified starting configuration. The following initialization modes are currently supported (selected in the \filename{input.txt} file):
\begin{lstlisting}[style=Input]
# ------------------- INITIALIZATION ------------------
# See: Wiki -> Initialization of the solver
inittype                0       #  0 - create uniform solution and start from it
                                #  5 - start from st-wslda checkpoint
                                # -1 - custom initialization

\end{lstlisting}
The toolkit provides an internal solver for a uniform system (i.e., in the absence of external potentials), which serves as the default initialization mode. Alternatively, the user may define a custom starting configuration by implementing the body of the function \lstinline{modify_potentials(...)}. This function is called prior to the first diagonalization step and allows the user to modify densities and potentials directly. An example of such custom initialization, defined in the \filename{problem-definition.h} file, is shown below:
\begin{lstlisting}
void modify_potentials(int it, wslda_density h_densities, wslda_potential h_potentials, double *params, size_t extra_data_size, void *extra_data)
{    
    // DETERMINE LOCAL SIZES OF ARRAYS (CODE DIMENSIONALITY DEPENDENT)
    int lNX=h_densities.nx, lNY=h_densities.ny, lNZ=h_densities.nz; // local sizes
    int ix, iy, iz, ixyz;
    
    if(it==-1 && wsldapid==0) wprintf("# SETTING MY VALUES FOR STARTING POINT\n");
    
    // ITERATE OVER ALL POINTS
    ixyz=0;
    if(it==-1) for(ix=0; ix<lNX; ix++) for(iy=0; iy<lNY; iy++) for(iz=0; iz<lNZ; iz++)
    {
        double x = DX*(ix-lNX/2);
        double y = DY*(iy-lNY/2); // for 1d code y will be always 0
        double z = DZ*(iz-lNZ/2); // for 1d and 2d codes z will be always 0
        
        // my custom initialization of the solver for point (x,y,z)
        h_densities.rho_a[ixyz]=h_densities.rho_b[ixyz]=0.018; // density guess
        
        // all potentials must be filled
        // since they are used to construct HFB matrix
        h_potentials.delta[ixyz]=0.1 + I*0.0; // guess for delta
        h_potentials.alpha_a[ixyz]=h_potentials.alpha_b[ixyz]=1.0; // bare mass
        h_potentials.V_a[ixyz]=h_potentials.V_b[ixyz]=0.0; // no mean-field
        h_potentials.A_a_x[ixyz]=h_potentials.A_b_x[ixyz]=0.0; 
        h_potentials.A_a_y[ixyz]=h_potentials.A_b_y[ixyz]=0.0; 
        h_potentials.A_a_z[ixyz]=h_potentials.A_b_z[ixyz]=0.0; 
        ixyz++; // go to next point
    }
    
    // my guess for chemical potentials 
    if(it==-1)
    {
        h_potentials.mu[SPINA]=0.5;
        h_potentials.mu[SPINB]=0.5;
    }
}
\end{lstlisting}
The initialization step is identified by the iteration index \texttt{it==-1}, and an appropriate conditional statement ensures that the initialization is executed only once.  

It should be noted that starting from either a predefined uniform solution or a custom configuration is not always optimal. In some cases, the number of iterations required for convergence can be large, and in others, the self-consistent algorithm may fail to converge if the initial guess is too far from the physical solution. This is not a limitation of the present implementation but a general feature of all self-consistent HFB-type solvers.

\subsubsection{Converging via releasing symmetry constraints}\label{subsec:conv-releasing-symmetry}

Following the general guiding principle, an effective alternative strategy consists in the gradual release of symmetry constraints. This approach is particularly useful for systems confined in external potentials, where solving the fully three-dimensional problem directly may be computationally demanding or numerically unstable.

As an illustrative example, consider a system of ultracold atoms confined in a three-dimensional harmonic trap,
\begin{equation}
\Vext{\sigma}(x,y,z)=\frac{1}{2}\bigl(\omega_x^2 x^2+\omega_y^2 y^2 +\omega_z^2 z^2 \bigr).
\end{equation}
A self-consistent solution for this system can be obtained in several stages.  
First, we impose translational symmetry in two directions by setting $\omega_y=\omega_z=0$, thereby reducing the problem to a quasi-1D case. Suppose we label this run as \texttt{ho-1d}. Once convergence is achieved, we use the obtained solution as a starting point for the next step, in which the symmetry is partially released. Specifically, we restore the correct value of $\omega_y$ while still keeping $\omega_z=0$, and solve the problem using the quasi-2D mode, which we tag by \texttt{ho-2d}. This can be done by specifying the following options in the \filename{input.txt} file:
\begin{lstlisting}[style=Input]
# ------------------- INITIALIZATION ------------------
# See: Wiki -> Initialization of the solver
inittype                5     # start from st-wslda checkpoint
                              # inprefix points to the checkpoint file
# ------------------- INPUT/OUTPUT ------------------
outprefix               ho-2d    # all output files with start with this prefix
inprefix                ho-1d    # checkpoint name
...
\end{lstlisting}
The code automatically detects the change in the dimensionality mode and performs the necessary conversion of the stored data. This operation is reported in the \texttt{stdout} file, for example:
\begin{lstlisting}[style=Input, numbers=none]
...
# INSPECTING CHECKPOINT FILE `ho-1d_checkpoint.dat`
# CHECKPOINT FOR 1D LATTICE: [NX,NY,NZ]=[128,32,32], [DX,DY,DZ]=[1.000,1.000,1.000]...
# !!! --- WARNING --- WARNING --- WARNING --- WARNING --- WARNING --- WARNING --- !!!
#   Dimensionality of the lattice has changed!
#   The code will change the dimensionality of the given checkpoint data to the new lattice.
# !!! --- ------- --- ------- --- ------- --- ------- --- ------- --- ------- --- !!!
...
\end{lstlisting}
Subsequently, the same procedure can be repeated by using the quasi-2D solution as input for the full 3D calculation. This iterative approach, progressively relaxing the imposed symmetry constraints, was used to obtain the 3D solution shown in the bottom row of \cref{fig:mesh-dim}.  

This methodology not only stabilizes the convergence process but also significantly reduces the overall computational cost. For instance, in the presented example, obtaining the 3D solution directly from a uniform initial state requires approximately 100 iterations. In contrast, initializing the 3D solver with the quasi-2D solution reduces this number to about 50 iterations. Moreover, since the computational cost decreases drastically with lower effective dimensionality (as discussed in \cref{sec:Scaling-st}), the combined cost of the preceding quasi-1D and quasi-2D runs is lower than that of a single iteration in 3D. In other words, by adopting this convergence strategy based on the gradual release of symmetry constraints, the total computational effort, in this example, was reduced by roughly a factor of two. 

\subsubsection{Converging via refining lattice resolution}
\begin{figure}[t]
	\centering
	\includegraphics[width=1.0\linewidth]{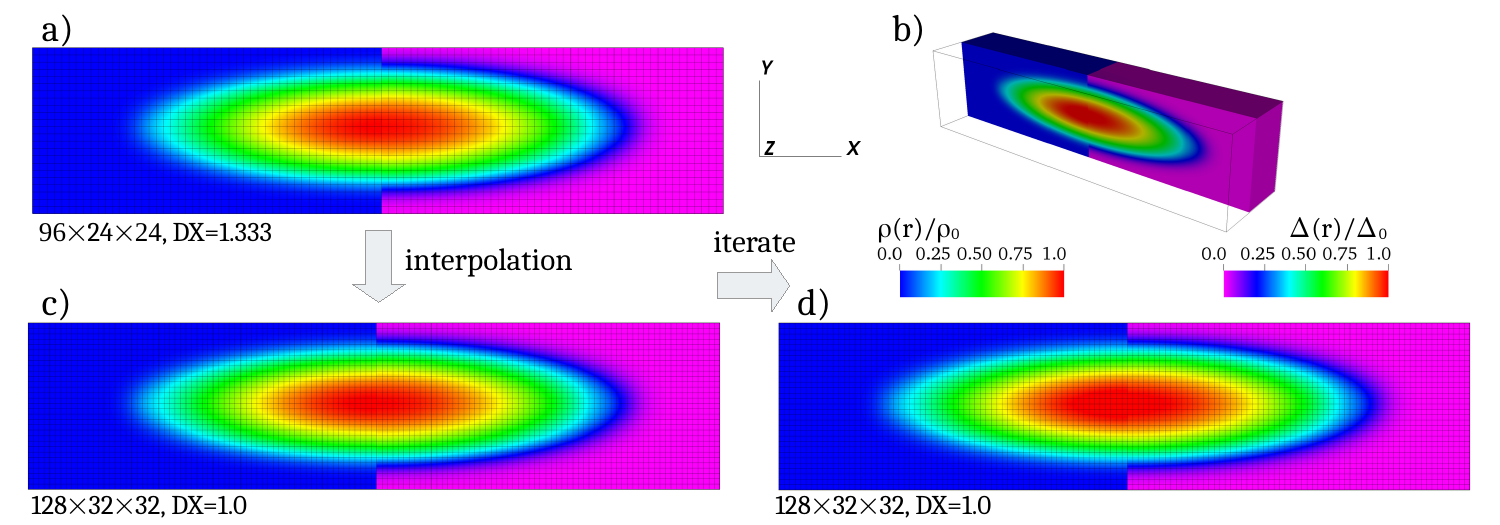}
    \caption{Schematic representation of the convergence strategy based on refining lattice resolution. A self-consistent solution obtained on a coarse lattice (panel a) is interpolated onto a finer lattice (panel c) and used as the initial state for subsequent iterations. The final converged solution is shown in panel d. Panels a, c, and d display cross-sections through the $y$-$z$ plane, while panel b presents a 3D view of the solution. Each panel is divided into two halves: the left side shows the particle density normalized to its central value, and the right side shows the pairing field normalized in the same way.}
	\label{fig:automatic-interpolation}
\end{figure}
Another effective convergence strategy consists in using a lattice of lower resolution to obtain an approximate solution, which is then interpolated to a finer (target) lattice and used as the starting point for a new round of iterations. The idea is illustrated in \cref{fig:automatic-interpolation}, where we again consider the example of an atomic gas confined in a harmonic oscillator potential.

In the first stage, the 3D problem was solved on a lattice of size $96\times 24^2$ with a lattice spacing of $dx = 128/96 = 1.333$. This calculation required about 100 iterations to converge. The resulting solution was then used as the initial state for a calculation on a refined lattice of size $128\times 32^2$ with a lattice spacing of $dx = 1$, so that the total physical volume was preserved.  
The toolkit includes routines for automatic interpolation between lattices of different sizes. When a checkpoint file from a lattice with a different resolution is provided, the code detects this mismatch and performs the interpolation automatically. This process is reported in the \texttt{stdout} file, for example:
\begin{lstlisting}[style=Input, numbers=none]
# LATTICE: 128 x 32 x 32
# SPACING: 1.000000 x 1.000000 x 1.000000
...
# INSPECTING CHECKPOINT FILE `ho-3d-dx1.3-r1_checkpoint.dat`
# CHECKPOINT FOR 3D LATTICE: [NX,NY,NZ]=[96,24,24], [DX,DY,DZ]=[1.333,1.333,1.333],...
# !!! --- WARNING --- WARNING --- WARNING --- WARNING --- WARNING --- WARNING --- !!!
#   Resolution of the lattice has changed!
#   The code will interpolate given checkpoint data to the new resolution.
# !!! --- ------- --- ------- --- ------- --- ------- --- ------- --- ------- --- !!!   
...
\end{lstlisting}
In this example, the refined-lattice calculation required about 25 iterations to converge. However, the cost per iteration increased by a factor of
$\bigl[(128\times 32^2)/(96\times 24^2)\bigr]^3\approx 13$,
meaning that one iteration on the refined lattice is equivalent to roughly 13 iterations on the coarse one. The total computational cost can therefore be estimated as
$\frac{100}{13} + 25\approx 33$
in units of refined-lattice iterations. Thus, in this particular case, obtaining the ground state through successive lattice refinements proved even more efficient than using the convergence strategy based on gradually releasing symmetry constraints.
 
The interpolation routines are implemented in the spirit of the Discrete Variable Representation (DVR) method~\cite{Littlejohn2002-1,Littlejohn2002-2,PhysRevC.87.051301}. Specifically, the value of a function at an arbitrary coordinate $(x,y,z)$ is evaluated as
\begin{equation}
f(x,y,z) = \frac{1}{L_x L_y L_z}\sum_{lmn} \tilde{f}_{lmn}\exp\left[i(k_l^{(x)} x+k_m^{(y)}y+k_n^{(z)}z)\right],
\end{equation}
where $\tilde{f}_{lmn}$ are the Fourier coefficients obtained via FFT, and the sum is computed numerically in a direct fashion. This formulation ensures that if $(x,y,z)$ coincides with a lattice point, the expression exactly reproduces the stored value (as it reduces to the inverse FFT). For off-grid points, it yields a smooth and physically consistent interpolation. The lower-dimensional variants are constructed analogously to the 3D case: in the \respch{quasi-}2D version the summation runs over two indices $(l,m)$, while in the \respch{quasi-}1D version it is reduced to a single index $l$.

The interpolation library is provided as a separate module within the toolkit, located in \filename{$WSLDA/lib/winterp}. It is implemented in C for performance and also includes a Python interface for user convenience, enabling its use in both production runs and post-processing analyses.

\subsubsection{Correcting the initial state via time-dependent approach}

The strategies discussed above (based on releasing symmetry constraints or refining the lattice resolution) allow for substantial savings in computational time, particularly when simulating full 3D systems. However, in some cases the cost of obtaining a fully converged static solution may still be prohibitively high. In such situations, an alternative approach can be employed: first, generate a partially converged solution, and then evolve it in time using a dissipative extension of the time-dependent code~\cite{bulgac2013f,Alba-Arroyo2025}, whose role is to remove residual excitation energy.  

The time-dependent module of the toolkit is capable of starting from an arbitrary initial state, so providing it with a not fully converged configuration poses no difficulty. To illustrate this procedure, let us again consider the example of an atomic gas confined in a 3D harmonic oscillator potential. In the previous section, we discussed a case where, after interpolating the solution from a coarse lattice to a finer one, the static solver required about 25 iterations to reach full convergence. As shown in \cref{fig:automatic-interpolation}, the interpolated (initial) and fully converged (final) states are visually almost indistinguishable: the differences are barely noticeable even when inspecting cross-sectional density maps.  Therefore, instead of performing all static iterations, one may execute only a single iteration to generate the wave functions on the refined grid and then use this configuration as the starting point for a time-dependent simulation. Although such a state contains some residual excitation energy, it already lies close to the minimum-energy configuration. By enabling the dissipative terms in the time-dependent evolution, this excess energy can be gradually reduced, effectively relaxing the system toward the true stationary state. 

A detailed derivation of the dissipative terms, as well as examples of their use in eliminating residual excitations, can be found in Ref.~\cite{Alba-Arroyo2025}. Here, we summarize the final expressions implemented in the toolkit. The dissipative mean field is given by
\begin{equation}
  \label{cooling_alpha}
  \Udiss_\sigma(\vbr,t) = - \frac{\alpha_{\text{qf}}}{\rho_0}\vb{\nabla}\cdot \vb{j}_\sigma(\vbr,t),
\end{equation}
where $\alpha_{\text{qf}}$ is a control parameter adjustable via the input file. When the dynamics are performed with this additional contribution, i.e.\ $V_{\sigma}\rightarrow V_{\sigma}+\Udiss_\sigma$,  
the system undergoes dissipative evolution in which the total energy decreases whenever nonzero currents $\vb{j}_\sigma$ are present. This form of the dissipative mean field was referred to in Ref.~\cite{bulgac2013f} as \textit{quantum friction}, representing a force that opposes irrotational currents and damps them. Numerical tests indicate that $\alpha_{\text{qf}} \approx 5$-$10$ provides optimal damping efficiency.  

The dissipative contribution to the pairing field is expressed as
\begin{align}\label{eq:Deltadiss2}
    \Ddiss(\vbr,t)&=  \Delta(\vbr,t)\Bigr[\beta_{\text{qf}}\sin\bigl[\theta_\nu(\vbr,t)-\theta_{\mathcal{B}'}(\vbr,t)\bigr] + \I\tilde{\gamma}_{\text{qf}}\Bigl],\\
    \theta_{\mathcal{B}'}(\vbr,t)&=\arg\left[ -\frac{1}{2m}\sum_n\left[u_{n,b}(\vbr,t)\nabla^2v_{n,a}^*(\vbr,t) + v_{n,a}^*(\vbr,t)\nabla^2u_{n,b}(\vbr,t) \right] \right],\\
    \theta_{\nu}(\vbr,t)&=\arg\left[\nu(\vbr,t)\right],\\
    \tilde{\gamma}_{\text{qf}} &= \gamma_{\text{qf}}\frac{N(t)-\Nreq(t)}{\Nreq(t)}.
\end{align}
Here, $\beta_{\text{qf}}$ and $\gamma_{\text{qf}}$ are control parameters that can be adjusted via the input file. When the system is evolved with the pairing field modified as $\Delta \rightarrow \Delta +\Ddiss$,  
the term proportional to $\beta_{\text{qf}}$ damps excitations in the pairing field, while the term proportional to $\gamma_{\text{qf}}$ drives the system toward the desired particle number $N_{\mathrm{req.}}$. The latter correction is useful when the initial configuration has a particle number different from the target value, typically occurring when the static code has not fully converged and the chemical potential is slightly misadjusted. Numerical tests show that $\beta_{\text{qf}} \approx 5$ and $\gamma_{\text{qf}} \approx 0.1$ yield optimal results for most applications.  

This hybrid approach, which combines static initialization with dissipative time evolution, offers a practical and efficient method for correcting imperfections in the initial state and achieving a well-converged configuration at a significantly lower computational cost.

\section{Ensuring high-quality results} 
\label{sec:CodeResultsQuality}

Ensuring the correctness and reliability of computational results is one of the most important challenges in modern computational science.  
The rapid increase in the complexity of numerical simulations and the use of large-scale computing infrastructures have made it increasingly difficult to guarantee that results are free from implementation or methodological errors.  
As emphasized in several studies~\cite{Merali2010,Soergel2015,Perkel2022}, reproducibility and verification are now recognized as fundamental pillars of scientific credibility.  
Computational physics, in particular, requires a combination of careful validation, reproducibility, and transparency to maintain a high level of confidence in numerical predictions.

Within the W-SLDA Toolkit, we apply a multilayered strategy to ensure the highest possible standard of result correctness.  
This strategy is organized into three complementary levels, each addressing a different aspect of code and result validation.

\paragraph{Level 1: Inspection of the results}
The first and most traditional approach to verifying computational correctness is direct inspection of the results, often summarized by the informal rule:
\begin{center}
    \textit{Results are correct if they look to be correct.}
\end{center}
Although this method may appear subjective, it remains a very powerful and widely used practice in computational physics.  
Researchers typically possess strong physical intuition and expectations regarding the qualitative behavior of a system.  
Comparing simulation outcomes with known analytical solutions, experimental data, or theoretical trends often allows one to identify discrepancies that may reveal implementation or setup issues.  

From an implementation perspective, the W-SLDA Toolkit provides extensive functionality to facilitate such result inspections and qualitative validations:  
\begin{itemize}
    \item \textbf{W-data format}: a conceptually simple and open data format that enables convenient post-processing using various tools and programming languages.  
    \item \textbf{Integration with external tools}: built-in support for the VisIt visualization platform, which allows advanced graphical analysis and interactive exploration of 2D and 3D simulation data, as well as seamless interoperability with Python.  
    Python integration enables the use of scientific frameworks such as \texttt{NumPy}, \texttt{SciPy}, and \texttt{Matplotlib} for data analysis and visualization.  
    \item \textbf{Auxiliary tools and extensions}: a collection of validated analysis scripts and examples demonstrating how to correctly read, interpret, and process data produced by the toolkit.  
    These resources are provided in the \filename{$WSLDA/extensions} directory.  
    \item \textbf{Customization of output}: the toolkit allows users to tailor the generated output to their specific needs through the \filename{logger.h} interface. Users can define additional observables, introduce custom diagnostics, and adapt the output format to ensure compatibility with external data-analysis or verification tools.
\end{itemize}
Together, these tools make qualitative verification efficient and accessible, ensuring that users can easily inspect and validate the physical plausibility of their results.

\paragraph{Level 2: Internal tests}
While Level~1 inspection ensures the qualitative plausibility of results, it is not sufficient to guarantee quantitative correctness.  
As Lubarsky’s Law humorously states,
\begin{center}
    \textit{every program has at least one more bug},
\end{center}
emphasizing that testing must be a continuous and systematic process~\cite{arxiv-guidelines}.  
To address this, the developers of the W-SLDA Toolkit maintain an extensive suite of automated internal tests that verify the correctness of computations across multiple scenarios and code configurations.  

The automated testing framework, called \texttt{testsuite}, executes a large collection of predefined test cases designed to assess numerical accuracy and code stability.  
All tests are automatically run after each new commit to the main code repository to ensure consistency and to detect potential regressions.  
Currently, the test suite includes more than 330 individual test cases covering all major modules and functionalities of the toolkit.  
These tests verify:  
(i) whether the code successfully compiles for a given selection of compilation flags;  
(ii) whether the executable runs correctly without errors; and  
(iii) whether the produced results agree with the corresponding reference values within specified tolerances.  

Over the years of toolkit development, a comprehensive database of benchmark cases has been accumulated.  
It includes both uniform systems and trapped configurations (typically under harmonic confinement), which serve as reference datasets for validation.  
This collection of trusted results is continuously expanded, ensuring that the test suite remains robust and representative of the diverse physical scenarios supported by the W-SLDA Toolkit.

\paragraph{Level 3: Open-source and transparency}
The third level of quality assurance leverages community engagement and transparency.  
External review, long established in the context of scientific publishing, is equally valuable for scientific software.  
The growing open-source movement in computational physics follows Linus’s Law: 
\epigraph[author={Eric S. Raymond},
          source={Linus's Law: The Cathedral and the Bazaar}
         ]{Given enough eyeballs, all bugs are shallow.}%
By making the code openly available, the W-SLDA Toolkit invites external verification, peer review, and community-driven improvement.

Actions supporting Level~3 quality control include:
\begin{itemize}
    \item \textbf{Open-source development}: the W-SLDA Toolkit is distributed under an open-source license, enabling full inspection and independent verification by the scientific community.
    \item \textbf{Result reproducibility}: since 2022, all published results have been accompanied by dedicated \textit{reproducibility packs} (see \cref{subsec:repropacks}) containing complete input files, configuration parameters, and setup definitions required to replicate computations from start to finish. This guarantees full transparency of the research workflow and enables independent validation of numerical findings.
\end{itemize}

Through the combined operation of these three levels, qualitative inspection, systematic internal testing, and open scientific transparency, the W-SLDA Toolkit maintains a high standard of computational reliability and scientific integrity.

\section{Applications}
\label{sec:applications}

The W-SLDA Toolkit has reached a mature stage of development and has been successfully applied in a wide range of research contexts.
\Cref{tab:Applications} summarizes representative applications, together with references to the corresponding publications reporting the obtained results.  
For cases where reproducibility packs are available, references to them are also provided.  
These reproducibility packs not only serve as complete documentation of the computational process but also constitute a valuable database of practical examples that users can adopt as starting points for their own projects.  

\begin{table}[htbp]
  \centering
  \caption{List of applications of the W-SLDA Toolkit. The second column lists the publications reporting the results obtained with the toolkit, while the third column indicates the availability of corresponding reproducibility packs.}
  \label{tab:Applications}
  \begin{tabular}{p{12cm}ll}%
    \toprule
    \textbf{Application} & \textbf{Reference} & \textbf{Repro. pack}\\
    \midrule
3D dynamics of the decay of a dark soliton in a spin-imbalanced unitary Fermi gas. The work reproduces an experimental setup of Ref.~\cite{PhysRevLett.116.045304}.  
    & \cite{Wlazlowski2018} & \\
3D dynamics and stability of spin-polarized droplets (\textit{ferrons}) in a unitary Fermi gas.  
    & \cite{Magierski2019,Tuzemen2020,Magierski2021} & \\
3D dynamics of reconnecting quantum vortices across the BCS-BEC crossover.  
    & \cite{Tylutki2021} & \\
Static properties of the Abrikosov lattice in a spin-imbalanced unitary Fermi gas. The work reproduces the experimental setup of Ref.~\cite{Zwierlein2006}.   
    & \cite{Kopycinski2021} & \\
Investigation of the sensitivity of unitary Fermi gas dynamics to initial conditions and quantification of Lyapunov exponents. 
    & \cite{Bulgac2022} & \\
3D dynamics of rotating quantum turbulence in the unitary Fermi gas, including spin-polarized configurations.  
    & \cite{Hossain2022} & \\
Structure of a quantum vortex as a function of the scattering length.  
    & \cite{Boulet2022} & \cite{Boulet2022-RP}\\
Structure of a quantum vortex in spin-imbalanced Fermi gases in the BCS and UFG regimes.  
    & \cite{Magierski2022} & \\
Dynamics of colliding vortex dipoles in an ultracold Fermi gas (BCS and UFG regimes) confined in a 2D cylindrical trap. The work reproduces the experimental setup of Ref.~\cite{Kwon2021}.  
    & \cite{Barresi2023a} & \cite{Barresi2023a-RP}\\
Simulations of Josephson dynamics in harmonically trapped atomic gases in the BCS and UFG regimes. The work corresponds to the experiment described in Ref.~\cite{PhysRevLett.120.025302}.  
    & \cite{Wlazlowski2023} & \cite{Wlazlowski2023-RP}\\
Ground-state properties of spin-imbalanced Fermi gases.  
    & \cite{Tuzemen2023,Alba-Arroyo2025} & \cite{Tuzemen2023-RP,Alba-Arroyo2025-RP}\\
Dynamics and stability of the Higgs mode in the unitary Fermi gas as a function of the system's dimensionality.  
    & \cite{Barresi2023b} & \cite{Barresi2023b-RP}\\
Dynamics of 3D isotropic quantum turbulence in the BCS and UFG regimes. This work reports the largest simulations performed within the W-SLDA Toolkit to date.  
    & \cite{Wlazlowski2024} & \cite{Wlazlowski2024-RP}\\
Thermalization processes in the unitary Fermi gas.  
    & \cite{Bulgac2024} & \cite{Bulgac2024-RP}\\
Stability of persistent currents in superfluid fermionic rings. The work reproduces the experimental setup of Ref.~\cite{PhysRevX.12.041037}.  
    & \cite{Xhani2025,Tuzemen2025} & \cite{Xhani2025-RP,Tuzemen2025-RP}\\
Extraction of the vortex mass as a function of scattering length and temperature.  
    & \cite{Richaud2025} & \cite{Richaud2025-RP}\\
Dissipative dynamics of a quantum vortex confined in a 2D cylindrical trap. The simulations directly benchmark experimental measurements.  
    & \cite{Grani2025} & \cite{Grani2025-RP}\\
Vortex dynamics in spin-imbalanced unitary Fermi gases.  
    & \cite{Barresi2025} & \cite{Barresi2025-RP}\\
    \bottomrule
  \end{tabular}
\end{table}
The list includes only works in which the software presented in this publication was employed (works generated by the toolkit's predecessors are not listed).  
Up to the submission of this paper, the toolkit has contributed to results reported in more than twenty scientific publications.  
This record demonstrates that the W-SLDA Toolkit is now a well-tested, reliable, and versatile framework capable of simulating diverse setups and physical phenomena related to ultracold Fermi gases. The framework has been successfully employed to study systems ranging from small quasi-1D configurations to large-scale 3D turbulent dynamics, covering diverse physical regimes from the BCS limit to unitarity and beyond.  
Its demonstrated scalability, both in terms of computational performance and physical applicability, establishes the W-SLDA Toolkit as a state-of-the-art platform for quantitative and predictive simulations of strongly correlated fermionic superfluids.  

\section{Conclusions}
The W-SLDA Toolkit constitutes a mature, general-purpose simulation platform for studying the static and dynamical properties of ultracold Fermi gases within the framework of density functional theory for superfluids. It provides a comprehensive environment for modeling a broad spectrum of physical phenomena, from vortex dynamics, solitonic excitations, and superfluid transport to collective oscillations, in both spin-balanced and spin-polarized systems. Owing to its modular structure and consistent treatment of both static and time-dependent formalisms, the toolkit can also serve as a versatile research infrastructure for testing and developing new energy density functionals.

The software is continually developed and maintained to ensure compatibility with evolving HPC environments. Its architecture is designed to exploit the capabilities of modern leadership-class supercomputers, enabling large-scale three-dimensional simulations with microscopic spatial resolution. These developments make it possible to study strongly correlated superfluids in domains that were previously inaccessible to microscopic approaches, offering quantitative insight into their real-time dynamics. The code is routinely tested and benchmarked on state-of-the-art HPC systems, guaranteeing its scalability and robustness across diverse computational platforms.

In its current form, the W-SLDA Toolkit allows simulations involving on the order of $10^5$ atoms in three dimensions~\cite{Wlazlowski2024}, approaching the scale of real experiments with ultracold atomic gases, which typically confine $10^5-10^6$ atoms in optical traps. This capability bridges the gap between microscopic \textit{ab initio} simulations and laboratory realizations, providing a direct avenue for comparison with experimental observations and for exploring regimes beyond the reach of mean-field or hydrodynamic descriptions.

Moreover, the toolkit has served as the foundation for the W-BSk Toolkit~\cite{pecak2024WBSkMeff}, an extension specifically optimized for the study of nuclear matter in neutron-star environments. This development extends the reach of the framework beyond ultracold-atom physics, enabling numerical experiments on dense, strongly interacting Fermi systems that cannot be probed directly in the laboratory. In this context, the W-SLDA and W-BSk toolkits form a coherent ecosystem for exploring the universal behavior of superfluid fermions across vastly different energy, density, and temperature scales: from atomic to astrophysical systems.

In summary, the W-SLDA Toolkit represents a robust, extensible, and HPC-ready platform for microscopic investigations of Fermi superfluids. Its continuous evolution ensures long-term adaptability to future computing architectures and emerging research directions, including applications to superconducting systems, neutron-star matter, and quantum-technology platforms based on superfluid atomic circuits.

\section{Further information}
Comprehensive documentation, including additional examples, detailed descriptions of advanced functionalities, and guidelines for implementing custom extensions, is available on the toolkit’s official Wiki, accessible through the Git repository~\cite{WSLDAToolkitGit}. The repository also provides up-to-date installation instructions, release notes, and developer guidelines. The mirrors of the main repository are also available at GitLab~\cite{GitLab-mirror} and GitHub~\cite{GitHub-mirror}. Links to related resources, such as reproducibility packs accompanying published works, are available on the official website~\cite{WSLDAToolkit}.
Users seeking assistance, reporting issues, or wishing to contribute to the project are encouraged to contact the development team directly at \email{wslda@fizyka.pw.edu.pl}.

\section{Acknowledgments}
We gratefully acknowledge all contributors who, over the years, have supported the development of W-SLDA Toolkit, the maintenance of its webpage, and the management of its public repository: Antoine Boulet,  Daniel P{\k{e}}cak, Jose Ernesto Alba Arroyo, Maciej Marchwiany, Maciej Szpindler, Wojciech Pudełko, Andrzej Makowski, Janusz Oleniacz, Maksymilian Odziemczyk, Bartosz Ruszczak, Andrea Barresi, Stanisław Tabisz, Agata Zdanowicz, Szymon Sieradzki, Michał Śliwiński, Krzysztof Calik. We express special thanks to Kenneth J. Roche for his invaluable assistance during the parallelization stage of the early predecessors of the toolkit, which laid the groundwork for the current high-performance implementation. We thank the Polish National Science Center (NCN) for continuous support of the W-SLDA project since 2017. 

The present work was financially supported by the (Polish) National Science Center Grant No.~2022/45/B/ST2/00358.
We acknowledge the Polish high-performance computing infrastructure, PLGrid, for awarding this project access to the LUMI supercomputer, owned by the EuroHPC Joint Undertaking, hosted by CSC (Finland), and the LUMI consortium through PLL/2024/07/017603.
P.M. was supported by the Polish National Science Center under Grant No. UMO-2021/43/B/ST2/01191.
A.B. greatly appreciates the funding from the Office of Science, Grant No. DE-FG02-97ER41014    and also the partial support provided by NNSA cooperative Agreement DE-NA0003841.
M.M.F.\ acknowledges \supportfromNSFgrant[PHY]{2309322}.
We thank the Institute for Nuclear Theory at the University of Washington for its kind hospitality and stimulating research environment.
This research was supported in part by the INT's U.S. \fundrefDOE[Department of Energy]{} grant No.\ DE-FG02-00ER41132.

\section{CRediT authorship contribution statement}
\begin{description*}[mode=unboxed]
  \item[Gabriel Wlazłowski] Conceptualization (code architecture; SLDAE framework), Methodology (workflow design; algorithms), Software (static and time-dependent solvers; extensions; W-data), Validation (implementation; framework; algorithms), Supervision (implementation), Writing -- original draft.
\item[Piotr Magierski] Conceptualization (SLDAE framework), Methodology (foundational theory), Validation (framework), Writing -- review \& editing.
\item[Michael McNeil Forbes] Conceptualization (ASLDA framework), Methodology (foundational theory; algorithms), Software (Python interface; W-data), Validation (implementation; framework; algorithms), Writing -- review \& editing.
\item[Aurel Bulgac] Conceptualization (SLDA/ASLDA framework; regularization scheme), Methodology (foundational theory;  algorithms), Validation (framework; algorithms), Writing -- review \& editing.
\end{description*}

\bibliographystyle{elsarticle-num}
\bibliography{biblio,wslda}







\end{document}